\documentclass[a4paper,11pt]{article}

\usepackage{jheppubMOD} 
\usepackage{lipsum}
\usepackage{graphicx}  
\usepackage{multirow}
\usepackage[x11names,dvipsnames]{xcolor}

\usepackage{mciteplus}

\hypersetup{
    colorlinks,
    linkcolor={jpac-blue},
    citecolor={jpac-blue},
    urlcolor={jpac-blue}
}

\usepackage{lmodern}
\usepackage{amsmath,amssymb}
\usepackage{mathtools}
\usepackage[numbers,sort&compress]{natbib}
\usepackage[numbers,sort&compress]{natbib}

\usepackage{tikz}
\usepackage{tikz-3dplot}
\usepackage{pgfplots,pgfplotstable}
\usetikzlibrary{pgfplots.groupplots}
\usetikzlibrary{decorations.markings}
\usetikzlibrary{shapes}
\usetikzlibrary{shapes.misc}
\usetikzlibrary{patterns}
\usepgfplotslibrary{fillbetween}
\pgfplotsset{compat=newest}
\usetikzlibrary{positioning}
\usetikzlibrary{calc}

\usepackage{lipsum}
\usepackage{stackrel}

\usepackage[compat=1.1.0]{tikz-feynman}

\usepackage{cleveref}
\usepackage{ifthen}
\usepackage{xspace}
\usepackage[shortlabels]{enumitem}

\definecolor{jpac-blue}{RGB}{ 31,119,180}
\definecolor{jpac-red}{RGB}{214,39, 40}
\definecolor{jpac-green}{RGB}{ 44,160, 44}
\definecolor{jpac-orange}{RGB}{255,127, 14}
\definecolor{jpac-purple}{RGB}{148,103,189}
\definecolor{jpac-brown}{RGB}{140, 86, 75}
\definecolor{jpac-pink}{RGB}{227,119,194}
\definecolor{jpac-gold}{RGB}{188,189, 34}
\definecolor{jpac-aqua}{RGB}{ 23,190,207}
\definecolor{jpac-grey}{RGB}{127,127,127}

\newcommand{\DLR}{\smash{\overset{\text{\small$\leftrightarrow$}}{\smash{D}
\vphantom{+}}}}
\newcommand{\DL}{\smash{\overset{\text{\small$\leftarrow$}}{\smash{D}\vphantom{+}}}}
\newcommand{\vp}{\varphi}
\newcommand{\eq}[1]{Eq.~\eqref{eq:#1}}
\newcommand{\nn}{\nonumber}

\begin{document}

\title{\boldmath Global analysis of $\mu \rightarrow e$ interactions in the SMEFT}

\author[a]{Filippo Delzanno,}
\emailAdd{fdelzanno@ucsb.edu}

\author[b]{Kaori Fuyuto,}
\emailAdd{kfuyuto@lanl.gov}

\author[b,c,d]{Sergi Gonz\`{a}lez-Sol\'is,}
\emailAdd{sergig@icc.ub.edu}

\author[b]{and Emanuele Mereghetti}
\emailAdd{emereghetti@lanl.gov}

\affiliation[a]{Department of Physics, University of California Santa Barbara, Santa Barbara, CA 93106, USA}
\affiliation[b]{Theoretical Division, Los Alamos National Laboratory, Los Alamos, NM 87545, USA}
\affiliation[c]{Departament de F\'isica Qu\`{a}ntica i Astrof\'isica (FQA), 
Universitat de Barcelona (UB), Mart\'i i Franqu\`{e}s, 1, 08028 Barcelona, Spain}
\affiliation[d]{Institut de Ci\`{e}ncies del Cosmos (ICCUB), 
Universitat de Barcelona (UB), Mart\'i i Franqu\`{e}s, 1, 08028 Barcelona, Spain}

\preprint{LA-UR-24-32279}

\abstract{We study current experimental bounds on charged lepton flavor violating (CLFV) $\mu- e$ interactions in the model-independent framework of the Standard Model Effective Field Theory (SMEFT). Assuming a generic flavor structure in the quark sector, we consider the contributions of CLFV operators to low-energy observables, including $\mu\to e\gamma$ and $\mu\to e$ conversion for quark-flavor conserving operators and CLFV meson decays for quark-flavor violating operators.  At high energy, we
consider limits on CLFV decays of the Higgs and $Z$ bosons and of the top quark, and obtain bounds on operators with light quarks by recasting searches for production of $e\mu$ pairs in $pp$ collisions at the Large Hadron Collider (LHC).
We connect observables at low- and high-energy by taking into account renormalization group running and matching between CLFV operators. 
We also discuss the sensitivity of the future Electron-Ion Collider, where the prospective bounds are derived by imposing simple cuts on final state particles. 
We find that, in a single operator scenario, bounds 
on purely leptonic operators are dominated by $\mu \rightarrow e \gamma$ and $\mu \rightarrow e$ conversion. 
Semileptonic operators with down-type quarks are also dominantly constrained by low-energy observables, while LHC searches lead the bounds on up-type quark-flavor violating operators. Taking simplified multiple-coupling scenarios, we show that it is easy to evade the strongest low-energy bounds from spin-independent $\mu \rightarrow e$ conversion, and that collider searches are competitive and complementary to constraints from spin-dependent $\mu \rightarrow e$ conversion and other low-energy probes.
}

\frenchspacing
\maketitle
\newpage



\section{Introduction}
Lepton flavor is exactly conserved in the 
Standard Model (SM) with massless neutrinos. 
The observation of neutrino oscillations 
\cite{Super-Kamiokande:1998kpq,KamLAND:2002uet,SNO:2001kpb}
has showed that lepton flavor is not conserved in the neutrino sector. 
While the non-conservation of neutrino flavor implies that charged lepton flavor is not conserved, 
the smallness of the active neutrino masses  causes 
an enormous suppression of charged lepton flavor violation (CLFV) in minimal extensions of the SM with massive neutrinos
\cite{Lee:1977tib,Lee:1977qz,Marciano:1977wx,Petcov:1976ff}. 
On the other hand, flavor violation in the neutrino and charged lepton sectors are often intimately related, especially in models
in which the particles responsible for the generation of the active neutrino masses have other interactions with SM particles.
Examples are Left Right Symmetric Models, 
in which heavy sterile neutrinos can interact with quarks and charged leptons via the exchange of new massive bosons \cite{Mohapatra:1980yp,Mohapatra:1979ia}, or supersymmetric see-saw models \cite{Borzumati:1986qx,Ilakovac:2012sh}.
The observation of CLFV processes in the next generation of experiments would thus play a crucial role in probing SM extensions connected to the generation of neutrino mass \cite{Ilakovac:1994kj,Abada:2007ux,Alonso:2012ji,Cirigliano:2005ck,Raidal:2008jk,deGouvea:2013zba,Bernstein:2013hba,Calibbi:2017uvl}.
For this reason, there is a rich experimental program aiming at improving constraints on $e \leftrightarrow \tau$, 
$\mu \leftrightarrow \tau$,
and $e \leftrightarrow \mu$ transitions \cite{Banerjee:2022xuw,Davidson:2022jai,CGroup:2022tli}.
 
Searches for $e \leftrightarrow \mu$ transitions provide for some of the strongest constraints on beyond the Standard Model (BSM) physics.
The very strong bounds on the $\mu \rightarrow e \gamma$ and
$\mu \rightarrow e e e$ branching ratios, and
on $\mu \rightarrow e$ conversion in nuclei
can naively be translated into BSM scales of hundreds of TeV, far from the reach of present and future colliders. 
Experiments such as Mu2e at Fermilab \cite{Mu2e:2014fns,Bernstein:2019fyh} and COMET at JPARC \cite{Kuno:2013mha,COMET:2018auw} will further improve the constraints on the $\mu \rightarrow e$ branching ratio by four orders of magnitude, apparently putting the direct observation of new particles with CLFV couplings even more out of reach. On the other hand, the number of highly-sensitive low-energy observables is very limited. It is thus important to assess which 
directions in the CLFV parameters space remain unconstrained, and which experiments are in the best position or would be required to close them. The goal of this paper is to address this question by presenting a global picture of $e \leftrightarrow \mu$ CLFV interactions, in the framework of the Standard Model Effective Field Theory (SMEFT)
\cite{Buchmuller:1985jz,Grzadkowski:2010es},
which allows us to parameterize CLFV interactions with minimal model dependence and to study the correlations between low-energy and present and future collider experiments.
Our work builds on the large body of studies of $e\leftrightarrow \mu$ transitions in EFT frameworks, at both high and low energy \cite{Raidal:1997hq,Kuno:1999jp,Kitano:2002mt, Cirigliano:2009bz,Crivellin:2013hpa,Crivellin:2014cta,Crivellin:2017rmk, Cirigliano:2017azj,Calibbi:2017uvl,Angelescu:2020uug, Calibbi:2021pyh,Calibbi:2022ddo, Davidson:2022jai,Davidson:2022nnl,Ardu:2021koz,Ardu:2022pzk,Cirigliano:2022ekw, Rule:2021oxe,Ardu:2024bua,Hoferichter:2022mna, Haxton:2022piv,Haxton:2024lyc,Coloma:2024ict, Plakias:2023esq,Becirevic:2024vwy,Redigolo:2024ztw}, which we extend to carry out a detailed analysis of the sensitivity of the upcoming Electron Ion Collider (EIC) to $e \rightarrow \mu$ transitions.

The paper is organized as follows. In Section \ref{sec:formalism} we introduce the SMEFT formalism, and discuss renormalization group effects, that play an important role in linking experiments at the energy and precision frontiers. In Section \ref{sec:mu2e} we discuss the constraints from $\mu \rightarrow e \gamma$ and $\mu \rightarrow e$ conversion in nuclei,
including both spin-independent and spin-dependent contributions and using up-to-date nucleon form factors and nuclear matrix elements \cite{Hoferichter:2020osn,Hoferichter:2022mna,Rule:2021oxe,Haxton:2024lyc}.  
Section \ref{sec:Flavor} is dedicated to low-energy bounds on quark flavor changing operators, via the study of the decays of $K$, $D$ and $B$ mesons.  In addition to obtaining bounds from existing experimental limits, we  discuss how differential observables, such as decay distributions and Dalitz plots, could disentangle the contributions of different SMEFT operators.
Section \ref{sec:lhc} studies the constraints coming from experiments at the Large Hadron Collider (LHC).  These include searches for CLFV decays of the Higgs and $Z$ bosons and of the top quark, and searches for $e \mu$ production in $pp$ collisions. We update the bounds from $pp \rightarrow e\mu$ by including the latest CMS data  \cite{CMS:2022fsw}, and by performing the analysis at next-to-leading order in QCD. In Section \ref{sec:eic} we perform a study of the sensitivity of the Electron-Ion Collider.
We carry out a detailed study of SM background, and identify an optimal set of analysis parameter to maximize the EIC's sensitivity.
In Section \ref{sec:summary} we summarize 
the bounds on SMEFT operators and compare the reach of different high- and low-energy experiments.
In Section \ref{sec:single} we assume that only one operator is turned on at the scale of new physics. In this case, we confirm the expectation that quark flavor diagonal couplings are overwhelmingly constrained by low-energy processes. We notice that this is true for both quark-flavor diagonal operators with valence quarks ($u$ and $d$), and with strange or heavy quarks ($s$, $c$, $b$ and $t$). Similarly, quark-flavor-changing operators in the \textit{down} sector are very well constrained by $K$ and $B$ decays, while flavor-changing operators in the \textit{up} sector are less constrained. In Section \ref{sec8} we study simplified scenarios with multiple couplings, inspired by the quark flavor symmetries of the SM. We thus illustrate that it is relatively easy to realize scenarios in which the strongest limits coming from spin-independent $\mu \rightarrow e$ conversion are evaded, and in which CLFV searches at the EIC and LHC can play an important role, complementary to low-energy. We conclude in Section \ref{sec:conclusions}.


\section{$\mu \rightarrow e$ interactions in the Standard Model Effective Field Theory}\label{sec:formalism}

We study $\mu \rightarrow e$ transitions in the framework of SMEFT \cite{Buchmuller:1985jz,Grzadkowski:2010es}. Similar EFT-based approaches to CLFV were used in Refs. 
\cite{Crivellin:2013hpa,Crivellin:2017rmk,Cirigliano:2017azj,Husek:2021isa,Cirigliano:2021img,Calibbi:2021pyh,Ardu:2021koz,Ardu:2022pzk,Ardu:2023yyw,Ardu:2024bua,Haxton:2024lyc,Fernandez-Martinez:2024bxg,Redigolo:2024ztw,Coloma:2024ict,Kumar:2021yod}. 
We follow the conventions of Ref. \cite{Cirigliano:2021img}, which are summarized in Appendix \ref{app:basis}. We 
work with dimensionless matching coefficients, by factoring out powers of the weak scale, see Eqs. \eqref{eq:Z}, \eqref{eq:dipole}, \eqref{eq:Y} and \eqref{eq:fourfermion}. The Wilson coefficients in this work thus scale as
\begin{equation}
    C = \mathcal O\left(\frac{v^2}{\Lambda^2} \right),
\end{equation}
where $v =246$ GeV is the Higgs vacuum expectation value, and $\Lambda$ the scale of BSM physics.
At dimension-six in the SMEFT, there are two classes of CLFV operators, lepton bilinears and  four-fermion operators. 
The former include the operators $c^{(1)}_{L \varphi}$ and $c^{(3)}_{L \varphi}$, which, after electroweak symmetry breaking, induce couplings of the $Z$ boson to
left-handed $e_L$ and $\mu_L$,
$c^{}_{e \varphi}$, which couples the $Z$ boson to right-handed $e_R$ and $\mu_R$,
the photon and $Z$ dipole couplings
$\Gamma^{e}_{\gamma}$ and $\Gamma^e_Z$, 
and the operator $Y^\prime_{e}$, which induces CLFV couplings of the Higgs to $e$ and $\mu$.
The hermiticity of the SMEFT Lagrangian implies that $\left[ C\right]_{e\mu} = \left[ C\right]^*_{\mu e}$, for $C \in \left\{ c^{(1)}_{L\varphi}, c^{(3)}_{L\varphi}, c_{e\varphi}  \right\}$.  On the other hand,  $[Y^{\prime}]_{\mu e}$, $[Y^{\prime}]_{e \mu}$,
$[\Gamma^{e}_{Z(\gamma)}]_{\mu e}$, $[\Gamma^{e}_{Z(\gamma)}]_{e \mu}$
are independent complex couplings. In the lepton bilinear sector, we therefore have 9 independent complex couplings. 
As an example of our normalization choices, we give here the dipole and Yukawa Lagrangian
\begin{equation}
\mathcal L_{\psi^2 X \varphi} + \mathcal L_{\psi^2 \varphi^3} =  -\frac{1}{\sqrt{2}}\bar \ell_L \sigma^{\mu\nu}(g_1\Gamma_{B}^e B_{\mu\nu}+g_2\Gamma_{W}^e {\tau}^I  {W}^I_{\mu\nu})\frac{\varphi}{v^2}  e_R 
 - \sqrt{2} \frac{\varphi^{\dagger} \varphi}{v^2} \bar \ell_L Y^\prime_e \varphi e_R  + \textrm{h.c.}, \label{eq:chiral}
 \ , \end{equation}
where $\varphi$ and $\ell_L$ denote the
Higgs and
left-handed lepton doublets, $e_R$
the right-handed charged
lepton field, $W_{\mu\nu}$ and $B_{\mu\nu}$ the field strengths of the $SU(2)_L$ and $U(1)_Y$
gauge fields.
Our conventions for the fields and covariant derivatives are defined in Appendix \ref{app:basis}.
$\Gamma^{e}_{B, W}$ and $Y^\prime_e$ are matrices in flavor space, and we will here consider the $\mu e$ and $e\mu$ components.

The semileptonic four-fermion operators include
7 vector operators, $C_{LQ, U}$, $C_{LQ, D}$, $C_{Lu}$, $C_{Ld}$, $C_{eu}$, $C_{ed}$ and $C_{Qe}$, two scalar operators, $C^{(1)}_{LeQu}$
and $C_{LedQ}$, and one tensor operator $C^{(3)}_{LeQu}$.
These operators encode all possible $SU(2)_L$-invariant couplings of left-handed quark and lepton doublets ($Q$ and $L$) 
and right-handed electrons, up and down quarks ($e$, $u$, $d$).  We find it convenient to define the four-fermion operators in the 
mass basis for both the $u$ and $d$ quarks, as explained in Ref. \cite{Cirigliano:2021img}. The simplification of working in the mass basis comes at the cost of a slight complication of the renormalization group equations.
We will keep the quark flavor structure of the operators generic. For vector operators, we have $\left[C \right]_{e \mu i j} = \left[C \right]^*_{\mu e j i}$. For scalar and tensor operators, the $e \mu$ and $\mu e$ components are independent. The semileptonic sector therefore contains 63 complex coefficients from the vector operators  and 54 complex coefficients from scalar and tensor operators.  Counting both leptonic and semileptonic operators, we study a total of 126 coefficients.
Finally there are three purely leptonic operators.
In our analysis we focus mostly on the operators that can be probed at the EIC and LHC. Purely leptonic operators affect our analysis since they are generated via renormalization group evolution. 
In our analysis, we truncate the SMEFT expansion at dimension-6, which, being CLFV forbidden in the SM, imply that 
CLFV observables receive their first contributions at 
$\mathcal O(v^4/\Lambda^4)$. 
The next SMEFT corrections will arise at
$\mathcal O(v^6/\Lambda^6)$ from the interference of dimension-6 and dimension-8 operators,
and at $\mathcal O(v^8/\Lambda^8)$
from the square of dimension-8 contributions or from the interference of dimension-6 and dimension-10 operators.
Ref. \cite{Ardu:2021koz} pointed out
that at dimension-8 there arise new semileptonic operators that are forbidden by $SU(2)_L \times U(1)_Y$ gauge invariance at dimension-6, and 
that $\mu \rightarrow e \gamma$,
$\mu \rightarrow 3e$ and $\mu \rightarrow e$ conversion in nuclei could probe their scale
up to 50 TeV. In this work, we work in the power counting assumption that, in a given quark flavor sector, the scale of dimension-6 and dimension-8 operators are not completely uncorrelated, to that dimension-6 operators are always dominant. It would be interesting to relax this assumption in future work. 

In this work, we study constraints on $\mu \leftrightarrow e$ interactions from a variety of low- and high-energy probes, and with very different sensitivities. To describe low-energy probes, we run the SMEFT coefficients from the scale of new physics to the electroweak scale, at the electroweak scale we
match the SMEFT basis onto the LEFT basis 
of Refs. \cite{Jenkins:2017dyc,Jenkins:2017jig,Dekens:2019ept}.
In this basis, semileptonic neutral current operators are denoted by $C^{f_1 f_2}_{\rm X Y Z}$, with $X$ denoting the Lorentz structure of the operator, $X \in \{V, S, T\}$ for vector, scalar and tensor,
while $Y, Z \in \{ L,R\}$ denote the chiralities of fermions $f_1$ and $f_2$, respectively.
We then further evolve the coefficients to the scales relevant to low-energy experiments.
The importance of renormalization group effects in analyses of CLFV was pointed out in Refs. \cite{Cirigliano:2017azj,Crivellin:2017rmk,Kumar:2021yod}.
Renormalization group evolution (RGE) induces several effects,
which become important when performing a global analysis:
\begin{itemize}
    \item the heavy flavor components of the vector operators run at one loop into purely leptonic operators
    and semileptonic operators with light flavors,
    \item vector operators with strange quarks generate operators with $u$ and $d$ quarks at one loop, 
    \item for light flavor operators, axial and vector linear combinations of SMEFT operators mix under RGE, 
    \item tensor operators mix strongly onto dipoles. 
    \end{itemize}
Some select results of the RGEs from the scale of new physics $\Lambda$ to $\mu = 2$ GeV
are shown in Table \ref{topRGE} and \ref{lightRGE}.
All the SMEFT renormalization group equations can be extracted from Refs. \cite{Alonso:2013hga,Jenkins:2013wua,Jenkins:2013zja},
while the LEFT renormalization group equations are given in Ref. \cite{Jenkins:2017dyc}.
For a complete discussion of RGE effects we refer to Ref. \cite{Cirigliano:2021img}.
We work here at leading logarithmic accuracy in both SMEFT and LEFT. 
That is, we start with the SMEFT coefficients at a scale $\Lambda$ much larger than the electroweak scale.
We run the SMEFT coefficients from $\mu \sim \Lambda$ to $\mu \sim v$ 
using the one-loop anomalous dimension. This allows one to sum the leading logarithmic series, so that the matching coefficients at the electroweak scale are, very schematically, given by 
\begin{equation}
    C_j(\mu_{\rm ew}) = \left[\sum_n  \left(\frac{g^2}{(4\pi)^2} \log \frac{\mu_{\rm ew}}{\mu_{\rm high}}\right)^n\right]_{ij} C_j(\mu_{\rm high}),
\end{equation}
where $g$ here denotes any gauge and Yukawa couplings in the SMEFT, and the notation $[\ldots]_{ij}$ denotes the operator mixing induced by the SMEFT anomalous dimension.
In practice, we take $\mu_{\rm high}= 1$ TeV and $\mu_{\rm ew} = m_t$. At $\mu_{\rm ew}$ we integrate out heavy degrees of freedom and match onto the LEFT. We carry out the matching at tree level. At tree level,  semileptonic LEFT four-fermion  operators receive contributions from semileptonic SMEFT operators, and from tree level Higgs and $Z$ boson exchanges. The LEFT photon dipole operator receives a contribution from a linear combination of $\Gamma^e_B(\mu_{\rm ew})$ and 
$\Gamma^e_W(\mu_{\rm ew})$.
The tree level matching coefficients between SMEFT and LEFT are given in Appendix \ref{sec:tree}.
The only case in which we include higher loop corrections to the matching coefficients is for the contributions
of the LFV Yukawa couplings to the dipole operator, for which we consider two-loop Barr-Zee diagrams with virtual top and $W$ running in the loop \cite{Barr:1990vd}. These diagrams are finite and do not induce scale or scheme dependence. They can thus be considered an improvement of the tree level matching coefficient \cite{Buras:1998raa}.
Finally, we take the LEFT coefficients at the electroweak scale, and evolve them down to the scale relevant to low-energy experiments using the LEFT RGEs.
Again, schematically, the low-energy coefficient are given by
\begin{equation}
    C^{\rm LEFT}_j(\mu_{\rm low}) = \left[\sum_n \left(\frac{g^2}{(4\pi)^2} \log\frac{\mu_{\rm low}}{\mu_{\rm ew}}\right)^n\right]_{ij} C^{\rm LEFT}_j(\mu_{\rm ew}),
\end{equation}
where $g$ now is the electromagnetic or 
strong coupling, and
$\mu_{low}$ is the typical scale for the low-energy experiment under consideration.
This framework can be improved by performing matching and running in SMEFT and LEFT at higher order \cite{Dekens:2019ept,Aebischer:2025hsx}. In particular, 
two-loop mixing between vector and dipole operators has been shown to be important \cite{Ardu:2021koz,Crivellin:2017rmk,EliasMiro:2021jgu}.
As the two-loop SMEFT RGEs are not yet known,
we do not include two-loop LEFT and SMEFT running in our baseline analysis, but discuss the impact of some known contributions in Appendix \ref{app:twoloop}.

\begin{table}[t]
\small
\centering
\begin{tabular}{||c|| c c c c ||}
\hline
                      &  $\left[{C}_{LQ,D}+{C}_{Ld}\right]_{bb}$ 
                      &  $\left[{C}_{LQ,D}-{C}_{Ld}\right]_{bb}$
                      &  $\left[{C}_{LQ,U}+{C}_{Lu}\right]_{tt}$
                      &  $\left[{C}_{LQ,U}-{C}_{Lu}\right]_{tt}$\\
                      \hline \hline
$[C^{ed}_{\rm VLL} +  C^{ed}_{\rm VLR}]_{kk}$ & $-3.1  $ & $-0.22$  & $2.3$ & $37$        \\               
$[C^{ed}_{\rm VLL} -  C^{ed}_{\rm VLR}]_{kk}$ & $-0.38$  & $-0.42$  &    $0.57$ & $55$   \\  
$[C^{eu}_{\rm VLL} +  C^{eu}_{\rm VLR}]_{uu}$ &  $5.9 $ & $0.14$      & $-4.1$ & $-21$  \\        $[C^{eu}_{\rm VLL} -  C^{eu}_{\rm VLR}]_{uu}$ &  $0.60$ & $0.52$   & $-0.79$ & $-56$    \\ 
\hline
$[C^{ee}_{\rm VLL}]_{\mu e ee}$ & $-4.3$ & $-0.14$ & $3.1$ & $28$ \\ 
$[C^{ee}_{\rm VLR}]_{\mu e ee}$ & $-4.1$ & $0.30$ & $2.7$ & $-28$ \\
\hline\hline
& $[C_{Qe}]_{bb}$ & $[C_{ed}]_{bb}$ & $ [C_{eu}]_{tt} $ &
 \\
\hline \hline
$[C^{ed}_{\rm VRR} +  C^{de}_{\rm VLR}]_{kk}$          & $37$  & $-2.8$ & $-36$ &  \\               
$[C^{ed}_{\rm VRR} -  C^{de}_{\rm VLR}]_{kk}$         & $-55$ & $3.5 \cdot 10^{-3}$  & $56$ & \\  
$[C^{eu}_{\rm VRR} +  C^{ue}_{\rm VLR}]_{uu}$         & $-17$ & $5.7$ & $15 $ & \\               
$[C^{eu}_{\rm VRR} -  C^{ue}_{\rm VLR}]_{uu}$         & $54 $& $0.12$ & $-55$ & \\ 
\hline 
$[C^{ee}_{\rm VRR}]_{\mu e ee}$  & $-27$ & $-4.2$  & $29$ & \\
$[C^{ee}_{\rm VLR}]_{ee \mu e}$  & $28$  &  $-4.3$  & $-27$ & \\
\hline
\end{tabular}
\caption{The Wilson coefficients at $\mu=2$ GeV induced by nonzero $b$- and $t$-quark operators
at the scale $\Lambda =1$ TeV, in units of $10^{-3}$. The indices  
$k$ denote either $d$ or $s$ quarks. We organize low-energy semileptonic operators into vector and axial combinations (the first column).
To highlight linear combinations not constrained by low-energy,
we also reorganize SMEFT operators with left-handed leptons into vector and axial combinations (the first and third lines). For operators with right-handed leptons, 
the fact that $C_{Qe}$ contributes to both $u$- and $d$-type operators, and large effects from the top Yukawa couplings are such that no simple linear combination gives small corrections to the vector couplings. 
}
\label{topRGE}
\end{table}

Table \ref{topRGE} shows the effects of the mixing of
semileptonic vector-like operators with 
heavy flavors onto low-energy  light flavor operators. 
Because of the parity symmetry of the strong interactions, low-energy observables are more naturally expressed in terms of operators with well defined parity, namely vector, axial, scalar, pseudoscalar and tensor operators. For SMEFT operators with $L$ leptons, we can construct similar linear combinations. 
From Table \ref{topRGE} we see that
the vector and axial combinations of top quark operators $\left[C_{LQ, U} \pm C_{Lu}\right]_{tt}$ 
 generate the light-flavor vector operators
 $[C^{ed}_{\rm VLL} +C^{ed}_{\rm VLR} ]_{dd} $ and
 $[C^{eu}_{\rm VLL} +C^{eu}_{\rm VLR} ]_{uu} $
 at the $10^{-2}$ level.  These vector operators are strongly constrained by $\mu \rightarrow e$ conversion in nuclei, resulting in extremely strong constraints on operators with top quarks. 
In the case of bottom quarks, on the other hand,
the axial combination
$\left[C_{LQ, D} - C_{Ld}\right]_{bb}$ generates low-energy vector operators only at the level of $10^{-4}$, resulting in much weaker bounds
from $\mu \rightarrow e$ conversion.

\begin{table}[t]
\small
\centering
\begin{tabular}{||c|| c c c c ||}
\hline
                      &  $\left[{C}_{LQ,D}+{C}_{Ld}\right]_{dd}$ 
                      &  $\left[{C}_{LQ,D}-{C}_{Ld}\right]_{dd}$
                      &  $\left[{C}_{LQ,U}+{C}_{Lu}\right]_{uu}$    & $\left[{C}_{LQ,U}-{C}_{Lu}\right]_{uu}$\\
                      \hline \hline
$[C^{ed}_{\rm VLL} +  C^{ed}_{\rm VLR}]_{dd}$ & $1.0  $ & $-1.8 \cdot 10^{-2}$          &     $6.5 \cdot 10^{-3}$    & $7.4 \cdot 10^{-5}$ \\               
$[C^{ed}_{\rm VLL} -  C^{ed}_{\rm VLR}]_{dd}$ & $-1.8 \cdot 10^{-2}$  & $1.0$   &     $-1.6 \cdot 10^{-4}$    & $-1.1\cdot 10^{-4}$    \\  
$[C^{eu}_{\rm VLL} +  C^{eu}_{\rm VLR}]_{uu}$ &  $-7.4 \cdot 10^{-3} $ & $-1.4 \cdot 10^{-2} $   &     $1.0$    & $4.7 \cdot 10^{-2}$     \\        $[C^{eu}_{\rm VLL} -  C^{eu}_{\rm VLR}]_{uu}$ &  $-1.4 \cdot 10^{-2}$ & $-1.4 \cdot 10^{-2}$  &     $4.7 \cdot 10^{-2}$    & $1.0$     \\  
$[C^{ed}_{\rm VLL} +  C^{ed}_{\rm VLR}]_{ss}$ & $-3.2 \cdot 10^{-3}  $ & $7.5 \cdot 10^{-5}$          &     $6.5 \cdot 10^{-3}$    & $7.4 \cdot 10^{-5}$ \\               
$[C^{ed}_{\rm VLL} -  C^{ed}_{\rm VLR}]_{ss}$ & $1.3 \cdot 10^{-4}$  & $1.0\cdot 10^{-4}$   &     $-1.6 \cdot 10^{-4}$    & $-1.1\cdot 10^{-4}$    \\  
\hline \hline
          &  $\left[{C}_{LQ,D}+{C}_{Ld}\right]_{ss}$ 
                      &  $\left[{C}_{LQ,D}-{C}_{Ld}\right]_{ss}$
                      &      & \\
                      \hline \hline
 $[C^{ed}_{\rm VLL} +  C^{ed}_{\rm VLR}]_{dd}$ & $-3.2 \cdot 10^{-3}  $ & $7.4 \cdot 10^{-5}$  &       &  \\               
$[C^{ed}_{\rm VLL} -  C^{ed}_{\rm VLR}]_{dd}$ &  $1.3 \cdot 10^{-4}$  & $1.0\cdot 10^{-4}$   &      &    \\  
$[C^{eu}_{\rm VLL} +  C^{eu}_{\rm VLR}]_{uu}$ &  $5.8 \cdot 10^{-3}$  & $-7.0 \cdot 10^{-4} $   &         &      \\     $[C^{eu}_{\rm VLL} -  C^{eu}_{\rm VLR}]_{uu}$ &  $5.4 \cdot 10^{-5}$ & $-7.6 \cdot 10^{-4}$  &     &     \\  
$[C^{ed}_{\rm VLL} +  C^{ed}_{\rm VLR}]_{ss}$ & $0.99  $ & $-1.8 \cdot 10^{-2}$     &    & \\               
$[C^{ed}_{\rm VLL} -  C^{ed}_{\rm VLR}]_{ss}$ & $-1.8 \cdot 10^{-2}$  &      $1.0 $ &    &     \\  \hline 
\end{tabular}
\caption{The Wilson coefficients at $\mu=2$ GeV induced by nonzero four-fermion operators with light quarks and left-handed leptons at the scale $\Lambda =1$ TeV. }
\label{lightRGE}
\end{table}

Table \ref{lightRGE} shows RGE effects 
on light quark operators. Since spin-independent $\mu \rightarrow e$ conversion constraints are much stronger than spin-dependent, the percent level one-loop electroweak mixing between axial and vector operators is relevant, as pointed out in Ref. \cite{Cirigliano:2017azj}.
In the case of strange operators, the tree level contribution to $\mu \rightarrow e$ conversion mediated by the nucleon strange matrix elements is usually quite suppressed, as we discuss in more detail in the following section. Thus, penguin running into $d$ and $u$ quarks also plays an important role.


\section{Low-energy bounds: $\mu \rightarrow e \gamma$ and $\mu \rightarrow e$ conversion in nuclei}
\label{sec:mu2e}
\begin{table}[t]
\centering
\begin{tabular}{l l l}
\hline
Decay mode & Upper limit on BR ($90\%$ C.L.)&Reference  \\ 
\hline\hline
$\mu \to e\gamma$  & $<3.1\times 10^{-13}$&MEG II~\cite{MEGII:2023ltw}   \\
$\mu+{\rm Au} \to e+{\rm Au}$ & $<7\times 10^{-13}$ & SINDRUM II~\cite{SINDRUMII:2006dvw}\\
$\mu+{\rm Ti} \to e+{\rm Ti}$ & $<6.1\times 10^{-13}$ & SINDRUM II~\cite{Wintz:1998rp}\\
$\mu\to 3e$ & $<1.0\times 10^{-12}$& SINDRUM~\cite{SINDRUM:1987nra} \\
\hline 
\end{tabular}
\caption{Current experimental limits on branching ratios of $\mu \to e$ transition.}
\label{Table:mutoe}
\end{table}

In this section, we will consider constraints from $\mu \to e$ transitions.
The experimental bounds 
on the $\mu \rightarrow e \gamma$ and
$\mu \rightarrow e e e$
branching ratios, 
and on $\mu \rightarrow e$ conversion with Au and Ti targets
are presented in Table \ref{Table:mutoe}. 
The branching ratio of $\mu \to e \gamma$ can be given in terms of dipole operators at low energy as
\begin{align}
{\rm BR}(\mu \to e \gamma)&=\tau_{\mu}\frac{m^3_{\mu}\alpha_{\rm em}}{4v^2}\left[\left|\left(\Gamma^e_{\gamma}\right)_{e\mu}\right|^2+\left|\left(\Gamma^e_{\gamma}\right)_{\mu e}\right|^2\right],
\end{align}
with $\tau_{\mu}^{-1}=G^2_Fm^5_{\mu}/(192\pi^3)$
and $\alpha_{\rm em}$ the fine structure constant. The prefactor in front of the bracket yields BR$=2\times 10^{7}~|\Gamma^e_{\gamma}|^2$. 
The expression of the $\mu \rightarrow 3e $ branching ratio in terms of SMEFT operators in the conventions used in this paper can be read from Eq. (D.4) in Ref. \cite{Cirigliano:2021img}, upon replacing $\tau \rightarrow \mu$. In our analysis we use the bound on the branching ratio given in Table \ref{Table:mutoe}. It has been noted that the CP asymmetry in $\mu\rightarrow 3e$ could probe new CP-violation in the lepton sector not constrained by electric dipole moments \cite{Kuno:1999jp,Bolton:2022lrg,Redigolo:2024ztw}.

The $\mu \rightarrow e$ conversion process is generally dominated by spin-independent (SI) interactions due to coherence of nucleons, while spin-dependent (SD) contributions are relatively suppressed. The SI process has been extensively studied in the EFT framework  \cite{Weinberg:1959zz, Shanker:1979ap, Czarnecki:1998iz, Kitano:2002mt, Cirigliano:2009bz, Rule:2021oxe, Davidson:2022nnl, Heeck:2022wer, Cirigliano:2022ekw, Haxton:2022piv}, while nuclear many-body calculations dedicated to the conversion process up to medium mass nuclei ($Z=29$) have been conducted with shell model over the last few years \cite{Haxton:2022piv, Hoferichter:2022mna}.
In our analysis we use the formalism and nuclear matrix elements calculated in Ref. \cite{Haxton:2022piv, Haxton:2024lyc}.
A detailed mapping between the LEFT operators and the non-relativistic operators used in Ref. \cite{Haxton:2022piv, Haxton:2024lyc}, together with the list of nuclear response functions and of the nucleon form factors used in this work, are provided in Appendix \ref{App:mutoe}.
We give bounds on LFV operators  mainly using the experimental result with titanium targets \cite{Wintz:1998rp}, for which accurate nuclear response functions were recently computed with the nuclear shell-model \cite{Haxton:2022piv, Hoferichter:2022mna}.
The branching ratio, expressed in terms of LEFT operators at the scale $\mu =2$ GeV and neglecting interference terms, is given by
\begin{align}
{\rm BR}(\mu\to e;~{\rm Ti})=&~2.8\times 10^3\left|C_{\rm SRR}^{eu}+C_{\rm SRL}^{eu} \right|^2_{\mu e uu}+2.7\times 10^3\left|C_{\rm SRR}^{ed}+C_{\rm SRL}^{ed} \right|^2_{\mu e dd}\nonumber\\
&+12.0\left|C_{\rm SRR}^{ed}+C_{\rm SRL}^{ed} \right|^2_{\mu e ss}\nonumber\\
&+1.1\times 10^2\left|C_{\rm VRR}^{eu}+C_{\rm VLR}^{ue} \right|^2_{\mu e uu}+2.3\times 10^2\left|C_{\rm VRR}^{ed}+C_{\rm VLR}^{de} \right|^2_{\mu e dd}\nonumber\\
&+6.7\times 10^{-7}\left|C_{\rm VRR}^{ed}+C_{\rm VLR}^{de} \right|^2_{\mu e ss}\nonumber\\
&+0.16\left|C_{\rm SRR}^{eu}-C_{\rm SRL}^{eu} \right|^2_{\mu e uu}+2.1\times 10^{-3}\left|C_{\rm SRR}^{ed}-C_{\rm SRL}^{ed} \right|^2_{\mu e dd}\nonumber\\
&+2.6\times 10^{-5}\left|C_{\rm SRR}^{ed}-C_{\rm SRL}^{ed} \right|^2_{\mu e ss}\nonumber\\
&+5.8\times 10^{-5}\left|C_{\rm VRR}^{eu}-C_{\rm VLR}^{ue} \right|^2_{\mu e uu}+5.0\times 10^{-4}\left|C_{\rm VRR}^{ed}-C_{\rm VLR}^{de} \right|^2_{\mu e dd}\nonumber\\
&+2.2\times 10^{-6}\left|C_{\rm VRR}^{ed}-C_{\rm VLR}^{de} \right|^2_{\mu e ss}, \label{BR_Mu2E}
\end{align}
where the first four lines correspond to SI contributions from scalar and vector operators while the rest of the lines are from SD process. Interestingly, in the case of strange quark, the contribution from the vector combination is suppressed due to the somewhat small form factor compared to the axial-vector one.
Eq. \eqref{BR_Mu2E} does not include the theoretical errors from the nucleon form factors or from nuclear matrix elements. The former are fairly well understood, but still affected by large uncertainties in the case of nucleon strange form factors \cite{Crivellin:2014cta,Hoferichter:2016nvd,FlavourLatticeAveragingGroupFLAG:2024oxs}. The estimate of nuclear theory errors is in a less developed state. For a discussion we refer to Refs.  \cite{Noel:2024swe,Noel:2024led}.

Table \ref{limit_dipole_yukawa_Z} shows upper bounds on lepton bilinear operators at $90\%$ confidence level (C.L.) where we assume that only one LFV operator is present at $\Lambda=1$ TeV. While the operators in the left column are bounded by $\mu\to e\gamma$, those in the right one are obtained by $\mu\to e$ conversion  in Ti (Au) target. Photon dipole operators directly generate the $\mu \to e\gamma$ process, leading to the severe limit of $|\Gamma_{\gamma}^e|< 6\times 10^{-11}$ while those from $\mu\to e$ conversion and $\mu\to 3e$ are ${\cal O}(10^{-9})$.
As discussed in Ref.~\cite{Cirigliano:2021img}, $Z$-boson dipole operators induce the photon dipole via renormalization group evolution between the scale of new physics and the electroweak scale, $\Gamma^{e}_{\gamma}(m_t)=-2\times 10^{-2}~\Gamma^{e}_Z(\Lambda)$. The limit on $Z$ dipoles is therefore two orders of magnitude weaker. The limit on the LFV Yukawa coupling is obtained by estimating one- and two-loop contributions to the photon dipole operator. Two-loop Barr-Zee diagrams give the dominant contribution due to the suppression from the electron or muon Yukawa couplings in the one-loop diagram. 
\begin{table}[t]
\centering
\begin{tabular}{|c c c| c c c | }
\hline
 $\left(\Gamma^e_{\gamma}\right)_{\mu e}$  & $\left(\Gamma^e_Z\right)_{\mu e}$ & $\left(Y_e^{\prime}\right)_{\mu e}$ &
 $\left(c_{L\varphi}^{(1)}\right)_{\mu e}$ &
 $\left(c_{L\varphi}^{(3)}\right)_{\mu e}$ &
 $\left(c_{e\varphi}\right)_{\mu e}$\\
\hline \hline
\multirow{2}{*}{$6.0\times 10^{-11}$}  & \multirow{2}{*}{$3.1\times 10^{-9}$} & \multirow{2}{*}{$1.2\times 10^{-6}$} & $4.3 \times 10^{-7}$ & $4.3\times 10^{-7}$ & $4.3\times 10^{-7}$ \\
 & & & $(2.9\times 10^{-7})$ & $(2.9\times 10^{-7})$ & $(2.9\times 10^{-7})$ \\
\hline
\end{tabular}
\caption{90\% C.L. upper
 limits on the dipole, Yukawa and $Z$ couplings operators assuming a single operator is turned on at $\Lambda=1$ TeV. The bounds on the dipole and Yukawa operators are originated from $\mu\to e\gamma$, while those on the LFV $Z$ couplings by $\mu \to e $ conversion in Ti (Au). }
\label{limit_dipole_yukawa_Z}
\vspace{2ex}
\renewcommand{\arraystretch}{1.1}
\begin{tabular}{||c|| c| c|| c | c || c | c |}
\hline
{$C_{LQ,U}$} & $uu$ & $6.9\times 10^{-8}$ & $cc$ & $1.1\times 10^{-5}$ & $tt$ & $4.6\times 10^{-6}$  \\
& & $(5.7\times 10^{-8})$ & & $(1.0\times 10^{-5})$ & & $(3.0\times 10^{-6})$ \\
{$C_{eu}$} & $uu$ & $7.2\times 10^{-8}$ & $cc$ & $1.2\times 10^{-5}$ & $tt$ & $3.2\times 10^{-6}$  \\
& & $(5.9\times 10^{-8})$ & & $(1.1\times 10^{-5})$ & & $(2.3\times 10^{-6})$ \\
{$C_{Lu}$} & $uu$ & $7.6\times 10^{-8}$ & $cc$ & $1.2\times 10^{-5}$  & $tt$ & $3.8\times 10^{-6}$   \\
& & $(6.2\times 10^{-8})$ & & $(1.1\times 10^{-5})$ & & $(2.6\times 10^{-6})$ \\
\hline \hline
$C_{LQ,D}$ & $dd$ & $7.4\times 10^{-8}$ & $ss$ & $3.7\times 10^{-5}$ & $bb$ & $2.8\times 10^{-5}$   \\
& & $(5.5\times 10^{-8})$ & & $(3.6\times 10^{-5})$ & & $(2.7\times 10^{-5})$ \\
$C_{ed}$ & $dd$ & $7.2\times 10^{-8}$ & $ss$ & $2.3\times 10^{-5}$ & $bb$ & $2.7\times 10^{-5}$ \\
& & $(5.3\times 10^{-8})$ & & $(2.1\times 10^{-5})$ & & $(2.5\times 10^{-5})$ \\
$C_{Ld}$ & $dd$ & $7.0\times 10^{-8}$  & $ss$ & $2.1\times 10^{-5}$  & $bb$ & $2.7\times 10^{-5}$  \\
& & $(5.2\times 10^{-8})$ & & $(1.8\times 10^{-5})$ & & $(2.4\times 10^{-5})$ \\
$C_{Qe}$ & $dd$ & $3.7\times 10^{-8}$ & $ss$ & $1.6\times 10^{-6}$  & $bb$ & $3.4\times 10^{-6}$   \\
& & $(2.9\times 10^{-8})$ & & $(1.3\times 10^{-6})$ & & $(2.4\times 10^{-6})$ \\
\hline\hline
$C_{LedQ}$ & $dd$ & $8.4\times 10^{-9}$ & $ss$ & $1.3\times 10^{-7}$ & $bb$ & $5.6\times 10^{-6}$  \\
& & ($9.4\times 10^{-9}$ ) & & ($3.2\times 10^{-7}$ ) & & ($5.4\times 10^{-6}$ ) \\
$C_{LeQu}^{(1)}$ & $uu$ & $8.2\times 10^{-9}$  & $cc$ & $1.3\times 10^{-6}$  & $tt$ & $9.3\times 10^{-8*}$ \\
& & $(9.8\times 10^{-9})$ & & $(1.3\times 10^{-6})$ & & \\
$C_{LeQu}^{(3)}$ & $uu$ & $3.1\times 10^{-8}$  & $cc$ & $7.6\times 10^{-9*}$ & $tt$ & $1.3\times 10^{-10*}$   \\
& & $(4.2\times 10^{-8})$ & & & & \\
 \hline
\end{tabular}
\caption{90\% C.L. upper limits on the quark-flavor-conserving semileptonic operators. $``*"$ represents that the bound is given by $\mu\to e \gamma$, while the rest of operators are constrained by $\mu\to e$ (Ti) conversion. The numbers in parentheses correspond to limits obtained by utilizing Au target in $\mu\to e$ conversion \cite{Cirigliano:2009bz}.}
\label{limit_semileptonic}
\end{table}

Table \ref{limit_semileptonic} presents bounds on semileptonic four-fermion operators. The first seven rows show the bounds on vector operators from $\mu\to e$ conversion, obtained by taking into account their renormalization group evolution from the scale of new physics $\Lambda =1$ TeV to low energy. As pointed out in Ref.~\cite{Cirigliano:2021img}, top-quark vector operators induce a  sizable mixing to the $Z$ couplings $c_{e\varphi, L\varphi}$, dominated by the top Yukawa coupling. This causes the bounds on the top coefficients to be  stronger than those on the $s$, $c$ and $b$ operators. 
For $b$ and $c$ operators, the dominant mixing is due to penguin diagrams.
One notable thing is that RGE effects are very important also for strange operators.
The limit on strange-quark vector operator is roughly a factor of 10 stronger than the one that would be obtained neglecting renormalization group evolution. This is because 
the evolution of
strange operators 
induces light-quark operators with coefficients
$C_{uu/dd}~(\mu=2~{\rm GeV})\sim {\cal O}(10^{-3})C_{ss}(\mu_0=1~{\rm TeV})$. As can be seen from Eq. \eqref{BR_Mu2E}, these contributions would dominate 
over ``direct''  contributions proportional to strange nucleon form factors. The last three rows correspond to limits on scalar and tensor operators, where we consider not only operator mixing but also their QCD self-running. Here, $\mu\to e \gamma$ gives the most stringent bounds, which is indicated by the symbol “$\ast$”, on tensor operators of top and charm quarks and on the top-quark scalar operator. 
While the tensor operators generate dipole operators at one-loop level, the top scalar operator does it in two steps, i.e., from scalar to tensor and then from tensor to dipole. Bounds on scalar operators for charm and bottom quarks are extracted from their threshold corrections to the gluonic operator $C_{GG}$. For reference, we also present limits in parentheses obtained by utilizing gold target in $\mu\to e$ conversion \cite{Cirigliano:2009bz}, which are in good agreement with the Ti case.

Finally, we note that pseudo-scalar and axial-vector operators become a source of SD $\mu\to e$ process, resulting in somewhat weaker bounds due to lack of coherent enhancement. Taking the results in Eq. (\ref{BR_Mu2E}), one can see
\begin{align}
\left|C^{eu}_{\rm SRR}-C^{eu}_{\rm SRL} \right|_{uu}&<1.9\times 10^{-6},\\
\left|C^{ed}_{\rm SRR}-C^{ed}_{\rm SRL} \right|_{dd}&<1.7\times 10^{-5},\\
\left|C^{eu}_{\rm SRR}-C^{eu}_{\rm SRL} \right|_{ss}&<1.5\times 10^{-4},
\end{align}
for pseudo-scalar operators, and
\begin{align}
\left|C^{eu}_{\rm VRR}-C^{ue}_{\rm VLR} \right|_{uu}&<1.0\times 10^{-4}, \label{SD_AVuu}\\
\left|C^{ed}_{\rm VRR}-C^{de}_{\rm VLR} \right|_{dd}&<3.5\times 10^{-5}, \label{SD_AVdd}\\
\left|C^{ed}_{\rm VRR}-C^{de}_{\rm VLR} \right|_{ss}&<5.2\times 10^{-4 }  \label{SD_AVss},
\end{align}
for axial-vector operators. It should be emphasized that the above bounds are obtained by employing nuclear response functions that take into account finite momentum transfer, which is, to the best of our knowledge, the up-to-date analysis for SD process. We find that the response functions for axial-vector (pseudo-scalar) operators in Ti can be reduced roughly by a factor of 10 $(3)$ compared to those assuming zero momentum transfer. This implies roughly a factor of 3 weaker bounds. In the case of Al target, the reduction for axial-vector operators is the similar size while pseudo-scalar ones receive a $\sim 50\%$ suppression.

SMEFT analyses of $\mu \rightarrow e$ conversion in nuclei have been recently carried out in Refs. \cite{Haxton:2024lyc,Fernandez-Martinez:2024bxg}. As we use the same nuclear matrix elements, our bounds agree with Ref. \cite{Haxton:2024lyc}.
We also checked that the above bounds as well as the results in Tables \ref{limit_dipole_yukawa_Z} and \ref{limit_semileptonic} are in good agreement with most of the bounds obtained  in Ref. \cite{Fernandez-Martinez:2024bxg}. 
These are based on tree level matching between SMEFT and LEFT, and do not include renormalization group evolution.
Our results indeed show that scale running effects based on the RGEs play a significant role in constraining more than half of LFV operators. One remarkable finding is
that, for strange vector operators,  
the contribution from the nucleon strange vector form factor is subleading with respect 
to the electroweak mixing of the $s$ with the $u$ and $d$ vector operators. 
Also, the large operator mixing of the tensor onto scalar operators, $C_{LeQu}^{(1)}(\mu=2~{\rm GeV})\sim {\cal O}(0.1)~C_{LeQu}^{(3)}(\mu_0=1~{\rm TeV})$, results in stronger bounds on the $u$ quark tensor operator than would be obtained without considering the RGEs. 
We also confirmed that our results are consistent with those obtained in Ref. \cite{Crivellin:2017rmk} except for $C^{\rm VRL}_{bb/cc}$. The $C_{qq}^{\rm VRL}$ operators
are induced by the SMEFT operator $C_{Qe}$, which links down-type and up-type components (see Eqs. \eqref{CQe}, \eqref{eq:cqe1} and \eqref{eq:cqe2}).   
$C^{\rm VRL}_{bb}$ is thus linked to operators with top quarks, which run strongly into CLFV $Z$ couplings and thus into operators with light-quarks. $C^{\rm VRL}_{cc}$ is on the other hand related to 
$C^{\rm VRL}_{dd}$ by factors of the CKM matrix, which are not too small. In both cases, the bounds on the SMEFT operators that induce $C^{\rm VRL}_{bb/cc}$
are about a factor of 10 stronger than the bounds based purely on RGE evolution in LEFT, as those obtained in Ref. \cite{Crivellin:2017rmk}.

\section{Low-energy bounds on quark-flavor changing operators}\label{sec:Flavor}

In this section we discuss the bounds on the quark-flavor-changing couplings that can be obtained from leptonic and semileptonic $K,D$ and $B$ meson decays.
Table~\ref{Table:ExperimentLimits} shows the decay channels that we analyze along with the present experimental upper bounds on the corresponding branching ratios at $90\%$ C.L., reported in the PDG~\cite{Workman:2022ynf}.
The only hadronic input required in the leptonic decays are the meson decay constants, which are well calculated in Lattice QCD~\cite{FlavourLatticeAveragingGroupFLAG:2021npn}, while the semileptonic decays require knowledge of the participating form factors.
For the functional form of these form factors we also use Lattice-QCD results.
As Lattice determines the form factors in a limited range of energies, these are presented as a formula from fits to a $z$-expansion parametrization to cover the whole kinematic range.
For our analysis we use the BCL representation \cite{Bourrely:2008za} with the results from the Fermilab Lattice and MILC collaboration (FNAL/MILC)~\cite{FermilabLattice:2015mwy,Bailey:2015dka,FermilabLattice:2022gku}.
The expressions for the BR in terms of the LEFT coefficients and the form factors are given in Appendix~\ref{App:BRandFF}. 
In the following we only give results for the most restrictive decay channels showing just one lepton chirality.

\begin{table}[t]
\centering
\begin{tabular}{l l l}
\hline
Decay mode & Upper limit on BR ($90\%$ C.L.)&Reference  \\ 
\hline\hline
$K^0_L \to e^{\pm}\mu^{\mp} $ & $<4.7\times 10^{-12}$&E871~\cite{BNL:1998apv}   \\
$K^0_L \to \pi^0 e^{\pm}\mu^{\mp} $ & $<7.6\times 10^{-11}$&KTeV~\cite{KTeV:2007cvy}   \\
$K^+ \to \pi^+ e^+ \mu^- $ & $<6.6\times 10^{-11}$&NA62~\cite{NA62:2021zxl}   \\
$K^+ \to \pi^+ e^- \mu^+ $ & $<1.3\times 10^{-11}$&E865,\,E777~\cite{Sher:2005sp}   \\
\hline 
$D^0 \to e^{\pm}\mu^{\mp} $ & $<1.3\times 10^{-8}$&LHCb~\cite{LHCb:2015pce}\\
$D^0 \to \pi^0 e^{\pm}\mu^{\mp} $ & $<8.0\times 10^{-7}$&BaBar~\cite{BaBar:2020faa} \\
$D^+ \to \pi^+ e^+ \mu^- $ & $<2.1\times 10^{-7}$&LHCb~\cite{LHCb:2020car}   \\
$D^+ \to \pi^+ e^- \mu^+ $ & $<2.2\times 10^{-7}$&LHCb~\cite{LHCb:2020car}   \\
$D^{+}_{s} \to K^{+} e^+ \mu^- $ & $<7.9\times 10^{-7}$&LHCb~\cite{LHCb:2020car}   \\
$D^{+}_{s}\to K^{+} e^- \mu^+ $ & $<5.6\times 10^{-7}$&LHCb~\cite{LHCb:2020car}   \\
\hline
$B^0 \to e^{\pm}\mu^{\mp} $ & $<1.0\times 10^{-9}$&LHCb~\cite{LHCb:2017hag}   \\
$B^+ \to \pi^{+} e^{\pm} \mu^{\mp} $ & $<1.7\times 10^{-7}$&BaBar~\cite{BaBar:2007xeb}    \\
$B^+ \to K^{+} e^{-} \mu^{+} $ & $<6.4\times 10^{-9}$&LHCb~\cite{LHCb:2019bix}    \\
$B^+ \to K^{+} e^{+} \mu^{-} $ & $<7.0\times 10^{-9}$&LHCb~\cite{LHCb:2019bix}    \\
$B_s \to e^\pm \mu^\mp $ & $<5.4\times 10^{-9}$&LHCb~\cite{LHCb:2017hag}    \\
\hline 
\end{tabular}
\caption{Summary of low-energy decay modes sensitive to CLFV quark-flavor-changing interactions,
and current experimental limits on their branching ratios~\cite{Workman:2022ynf}.}
\label{Table:ExperimentLimits}
\end{table}

The decays $D^{0}\to e^{\pm}\mu^{\mp},D^{0}\to\pi^{0}e^{\pm}\mu^{\mp},D^{+}\to\pi^{+}e\mu$ and $D^{+}_{s}\to K^{+}e\mu$ can be used to restrict the $cu$ and $uc$ elements of the LFV $up$-type operators.
The channels $D^{0}\to e^{\pm}\mu^{\mp}$ and $D^{+}\to\pi^{+}e^{+}\mu^{-}$ give the strongest bounds:
\begin{eqnarray}
\label{Eq.RestrictionsDdecays}
{\rm{BR}}(D^{0}\to e^{\pm}\mu^{\mp})&\simeq&0.0026\left|\left(C_{\rm{VLR}}^{eu}-C_{\rm{VLL}}^{eu}\right)_{\mu ecu}\right|^{2}+0.85\left|\left(C_{\rm{SRR}}^{eu}-C_{\rm{SRL}}^{eu}\right)_{\mu ecu}\right|^{2}\nonumber\\[1ex]
&+&c\leftrightarrow u\,,\\[1ex]
{\rm{BR}}(D^{+}\to\pi^{+}e^{+}\mu^{-})&\simeq&0.14\left|\left(C_{\rm{VLR}}^{eu}+C_{\rm{VLL}}^{eu}\right)_{\mu ecu}\right|^{2}+0.29\left|\left(C_{\rm{SRR}}^{eu}+C_{\rm{SRL}}^{eu}\right)_{\mu ecu}\right|^{2}\nonumber\\[1ex]&+&0.019\left|(C_{\rm{TRR}}^{eu})_{\mu ecu}\right|^{2}\,.
\end{eqnarray}
The mode $D^{+}\to\pi^{+}e^{-}\mu^{+}$ is described by replacing $u\leftrightarrow c$, and yields slightly weaker bounds due to its slightly weaker experimental limit.
The LEFT coefficients with opposite lepton chirality contribute to each mode with the same prefactor.

The decays $K_{L}^{0}\to e^{\pm}\mu^{\mp},K_{L}^{0}\to\pi^{0}e^{\pm}\mu^{\mp}$ and $K^{+}\to\pi^{+}e\mu$ can be used to restrict the $sd$ and $ds$ elements of the LFV $down$-type operators.
The more relevant channels are $K_{L}^{0}\to e^{\pm}\mu^{\mp}$ and $K^{+}\to\pi^{+}e^{-}\mu^{+}$, which yield:
\begin{eqnarray}
{\rm{BR}}(K_{L}^{0}\to e^{\pm}\mu^{\mp})&\simeq&53.9\left|\left(C_{\rm{VLR}}^{ed}-C_{\rm{VLL}}^{ed}\right)_{\mu esd}\right|^{2}+30.8\times10^{3}\left|\left(C_{\rm{SRR}}^{ed}-C_{\rm{SRL}}^{ed}\right)_{\mu esd}\right|^{2}\nonumber\\[1ex]
&+&s\leftrightarrow d\,,\\[1ex]
{\rm{BR}}(K^{+}\to\pi^{+}e^{-}\mu^{+})&\simeq&1.19\left|\left(C_{\rm{VLR}}^{ed}+C_{\rm{VLL}}^{ed}\right)_{\mu esd}\right|^{2}+17.66\left|\left(C_{\rm{SRR}}^{ed}+C_{\rm{SRL}}^{ed}\right)_{\mu esd}\right|^{2}\,.
\label{Eq.RestrictionsKdecays}
\end{eqnarray}
Similarly, the decay $K^{+}\to\pi^{+}e^{+}\mu^{-}$ is described by replacing $d\leftrightarrow s$, and yields weaker limits due to its weaker experimental bound. Again, only one lepton chirality is shown.

The LFV $down$-type components $bd$ and $db$ are constrained by the decays $B^{0}\to e^{\pm}\mu^{\mp}$ and $B^{+}\to\pi^{+}e^{\pm}\mu^{\mp}$, so that
\begin{eqnarray}
{\rm{BR}}(B^{0}\to e^{\pm}\mu^{\mp})&\simeq&0.027\left|\left(C_{\rm{VLR}}^{ed}-C_{\rm{VLL}}^{ed}\right)_{\mu ebd}\right|^{2}+107.8\left|\left(C_{\rm{SRR}}^{ed}-C_{\rm{SRL}}^{ed}\right)_{\mu ebd}\right|^{2}\nonumber\\[1ex]
&+&b\leftrightarrow d\,,\\[1ex]
{\rm{BR}}(B^{+}\to\pi^{+}e^{\pm}\mu^{\mp})&\simeq&11.32\left|\left(C_{\rm{VLR}}^{ed}+C_{\rm{VLL}}^{ed}\right)_{\mu ebd}\right|^{2}+15.39\left|\left(C_{\rm{SRR}}^{ed}+C_{\rm{SRL}}^{ed}\right)_{\mu ebd}\right|^{2}\nonumber\\[1ex]
&+&b\leftrightarrow d\,.
\label{Eq.RestrictionsBdecays}
\end{eqnarray}
Finally, the transitions $B_{s}\to e^{\pm}\mu^{\mp}$ and $B^{+}\to K^{+}e\mu$ can be used to set bounds on the $down$-type $sb$ and $bs$ elements:
\begin{eqnarray}
{\rm{BR}}(B_{s}\to e^{\pm}\mu^{\mp})&\simeq&0.028\left|\left(C_{\rm{VLR}}^{ed}-C_{\rm{VLL}}^{ed}\right)_{\mu ebs}\right|^{2}+117.6\left|\left(C_{\rm{SRR}}^{ed}-C_{\rm{SRL}}^{ed}\right)_{\mu ebs}\right|^{2}\nonumber\\[1ex]
&+&b\leftrightarrow s\,,\\[1ex]
{\rm{BR}}(B^{+}\to K^{+}e^{-}\mu^{+})&\simeq&17.18\left|\left(C_{\rm{VLR}}^{ed}+C_{\rm{VLL}}^{ed}\right)_{\mu esb}\right|^{2}+22.33\left|\left(C_{\rm{SRR}}^{ed}+C_{\rm{SRL}}^{ed}\right)_{\mu esb}\right|^{2}\,.\nonumber\\
\label{Eq.RestrictionsBsdecays}
\end{eqnarray}
The decay $B^{+}\to K^{+}e^{+}\mu^{-}$ is expressed by replacing $b\leftrightarrow s$ and gives similar bounds.

The resulting upper limits on the quark-flavor-changing operators are shown, respectively, in Tables~\ref{Table:UpTypeLimits_flavor} and \ref{Table:DownTypeLimits_flavor} for the $up$-and $down$-type operators. These have been obtained assuming one single operator is turned on at a time.
The bounds on the $up$-type $uc$ and $cu$ components are of $\mathcal{O}(10^{-3})$ for the vector and tensor operators and of $\mathcal{O}(10^{-4})$ for the scalar one, respectively.
The former come from $D^{+}\to\pi^{+}e\mu$ while the latter from $D^{0}\to e^{\pm}\mu^{\mp}$. 
The strongest bounds on the $down$-type $ds$ and $sd$ operators are obtained from $K_{L}^{0}\to e^{\pm}\mu^{\mp}$ and are of $\mathcal{O}(10^{-7})$ and $\mathcal{O}(10^{-8})$ for the vector and scalar operators, respectively.
The bounds on the $down$-type $db$ and $bd$ elements are of $\mathcal{O}(10^{-4})$ and $\mathcal{O}(10^{-6})$ for the vector and scalar operators, and come from $B^{+}\to\pi^{+}e^{\pm}\mu^{\mp}$ and $B^{0}\to e^{\pm}\mu^{\mp}$, respectively. 
The channel $B^{+}\to K^{+}e\mu$ constrains the $down$-type $sb$ and $bs$ vector operators at the $\mathcal{O}(10^{-5})$ level, while the limits on the scalar counterparts, which are slightly less than $\mathcal{O}(10^{-5})$, come from $B_{s}\to e^{\pm}\mu^{\mp}$.

\begin{table}[t]
\centering
\begin{tabular}{|c| c| c| c | c || c | c |}
\hline
$C_{LQ,U}$ & $uc$ & $1.3\times 10^{-3\S}$   \\
		& $cu$ & $1.2\times 10^{-3\S}$   \\
$C_{eu}$ & $uc$ & $1.3\times 10^{-3\S}$  \\
		& $cu$ & $1.2\times 10^{-3\S}$ \\
$C_{Lu}$ & $uc$ & $1.3\times 10^{-3\S}$   \\
		& $cu$ & $1.2\times 10^{-3\S}$  \\
$C_{Qe}$ & $uc$ & $2.8\times 10^{-3\S}$   \\
		& $cu$ & $2.8\times 10^{-3\S}$  \\
\hline
$C_{LeQu}^{(1)}$ & $uc$ & $1.2\times 10^{-4\ddag}$   \\
		& $cu$ & $1.2\times 10^{-4\ddag}$  \\
$C_{LeQu}^{(3)}$ & $uc$ & $3.4\times 10^{-3\S}$   \\
		& $cu$ & $1.7\times 10^{-3\S}$  \\
\hline
\end{tabular}
\caption{
90\% C.L. upper limits on the $up$-type quark-flavor-changing semileptonic operators assuming a single operator is turned on.
The strongest bounds come from (${\S}$) $D^+\to \pi^+e^{\pm}\mu^{\mp}$ and ($\ddag$) $D^0\to e^{\pm}\mu^{\mp}$. }
\label{Table:UpTypeLimits_flavor}
\end{table}

\begin{table}[t]
\centering
\begin{tabular}{|c|| c| c|| c | c || c | c |}
\hline
$C_{LQ,D}$ & $ds$ & $3.0\times 10^{-7\diamondsuit}$ & $db$ & $1.2\times 10^{-4\#}$ & $sb$ & $1.9\times 10^{-5\natural}$   \\
		& $sd$ & $3.0\times 10^{-7\diamondsuit}$ & $bd$ & $1.2\times 10^{-4\#}$ & $bs$ & $2.0\times 10^{-5\natural}$   \\
$C_{ed}$ & $ds$ & $3.0\times 10^{-7\diamondsuit}$& $db$ & $1.2\times 10^{-4\#}$ & $sb$ & $1.9\times 10^{-5\natural}$ \\
	  	& $sd$ &$3.0\times 10^{-7\diamondsuit}$ & $bd$ & $1.2\times 10^{-4\#}$ & $bs$ & $2.0\times 10^{-5\natural}$ \\
$C_{Ld}$ & $ds$ & $3.0\times 10^{-7\diamondsuit}$ & $db$ & $1.2\times 10^{-4\#}$  & $sb$ & $1.9\times 10^{-5\natural}$  \\
		& $sd$ &  $3.0\times 10^{-7\diamondsuit}$ & $bd$ & $1.2\times 10^{-4\#}$  & $bs$ & $2.0\times 10^{-5\natural}$  \\
$C_{Qe}$ & $ds$ & $3.0\times 10^{-7\diamondsuit}$& $db$ & $1.2\times 10^{-4\#}$  & $sb$ & $1.9\times 10^{-5\natural}$   \\
		 & $sd$ & $3.0\times 10^{-7\diamondsuit}$ & $bd$ & $1.2\times 10^{-4\#}$  & $sb$ & $2.0\times 10^{-5\natural}$   \\
\hline
$C_{LedQ}$ & $ds$ & $1.3\times 10^{-8\diamondsuit}$& $db$ & $3.1\times 10^{-6\flat}$  & $sb$ & $6.8\times 10^{-6\star}$   \\
		 & $sd$ & $1.3\times 10^{-8\diamondsuit}$ & $bd$ & $3.1\times 10^{-6\flat}$  & $sb$ & $6.8\times 10^{-6\star}$   \\
\hline
\end{tabular}
\caption{
90\% C.L. upper limits on the $down$-type quark-flavor-changing semileptonic operators assuming a single operator is turned on. The superscripts represent that the limit is imposed by decay modes (${\diamondsuit}$) $K_L^{0}\to e^{\pm}\mu^{\mp}$, (${\#}$) $B^+\to \pi^+ e^{\pm}\mu^{\mp}$, (${\flat}$) $B^0\to e^{\pm}\mu^{\mp}$, (${\natural}$) $B^+\to K^+ e^{\pm}\mu^{\mp}$, (${\star}$) $B_s\to e^{\pm}\mu^{\mp}$.}
\label{Table:DownTypeLimits_flavor}
\end{table}
For $B$ decays, our numerical results in Eqs. (\ref{Eq.RestrictionsDdecays})-(\ref{Eq.RestrictionsBsdecays}) agree with those obtained recently in a similar EFT approach in Refs. \cite{Plakias:2023esq,Becirevic:2024vwy}.
While we cannot directly compare bounds on SMEFT coefficients, the agreement on the semileptonic form factors implies that our results are consistent.

\begin{figure}[t]
\includegraphics[scale=0.45]{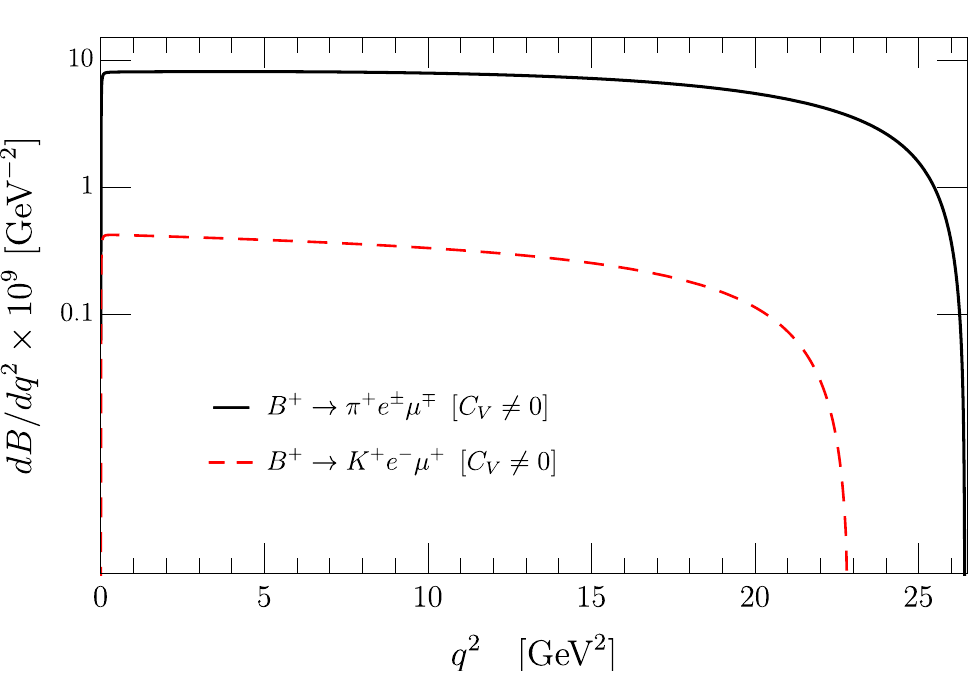}\quad\includegraphics[scale=0.45]{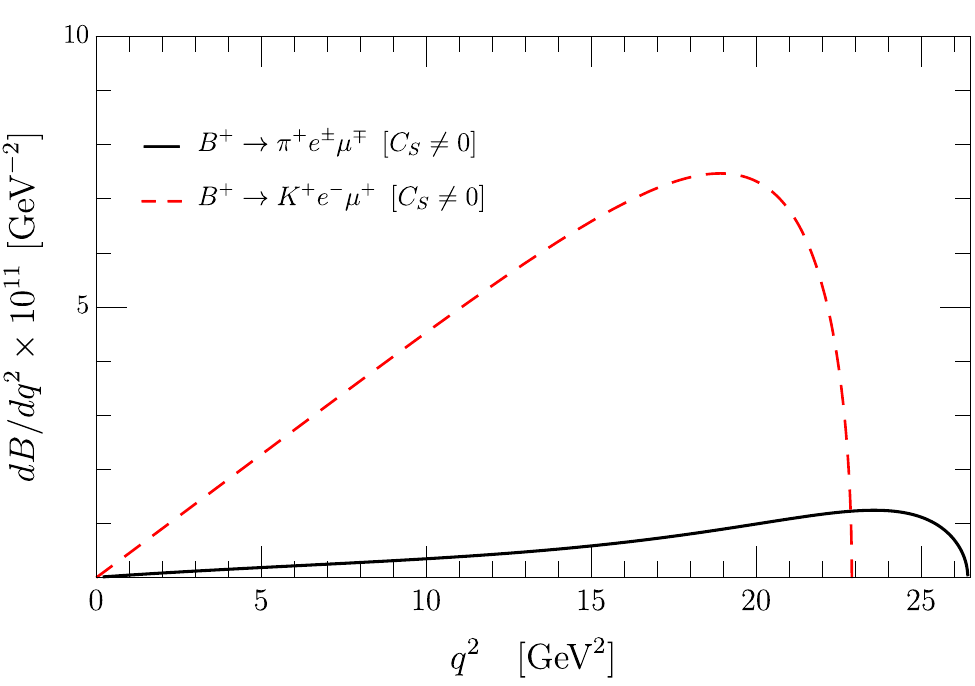}
\caption{\label{Fig:DistributionsBdecays}Differential branching ratio distributions for the decays $B^{+}\to\pi^{+}e^{\pm}\mu^{\mp}$ (solid black) and $B^{+}\to K^{+}e^{-}\mu^{+}$ (dashed red) as a function of the dilepton invariant mass ($q^{2}$) assuming one single vector (left) and scalar (right) operator from Table \ref{Table:DownTypeLimits_flavor} is turned on, with the others set to zero.
Here $C_V$ and $C_S$ denote the combination $C^{ed}_{\rm VLR} + C^{ed}_{\rm VLL}$
and $C^{ed}_{\rm SRR} + C^{ed}_{\rm SRL}$, respectively.
}
\end{figure}

In the following, we use the bounds given in Tables \ref{Table:UpTypeLimits_flavor} and \ref{Table:DownTypeLimits_flavor} to study different decay observables that can be useful to discriminate operators contributing to these channels in experimental searches. These observables
include decay distributions, Dalitz plots and the forward-backward asymmetries.
In Fig. \ref{Fig:DistributionsBdecays}, we show the differential branching ratio distributions for the decays $B^{+}\to\pi^{+}e^{\pm}\mu^{\mp}$ and $B^{+}\to K^{+}e^{-}\mu^{+}$  as a function of the $e\mu$ invariant mass $(q^{2})$ for vector (left) and scalar (right) operators, assuming that only one operator is turned on with the others set to zero.
We note that the $q^{2}$-shapes for the vector operators follow the same trend for the two decays and are $1-2$ orders of magnitude larger than the scalar ones, which peak at the high $q^{2}$ region of the spectra. 
The latter is also observed in Ref. \cite{Becirevic:2024vwy} for the $B$ decays studied here. 
Our results, however, contrasts Ref. \cite{Becirevic:2024vwy} in the size of the corresponding contributions. 
While in this reference the standalone vector and scalar contributions result from setting a single effective coefficient to 1 and both are of the same order, we use the actual bounds from Tables \ref{Table:UpTypeLimits_flavor} and \ref{Table:DownTypeLimits_flavor} to find that the vector one largely dominates.
This is due to the fact that we use the most stringent scalar couplings, which come from the purely leptonic decays $B^{0}\to e^{\pm}\mu^{\mp}$ and $B_{s}\to e^{\pm}\mu^{\mp}$ that are not considered in Ref. \cite{Becirevic:2024vwy}.

\begin{figure}[h!]\centering
\includegraphics[scale=0.65]{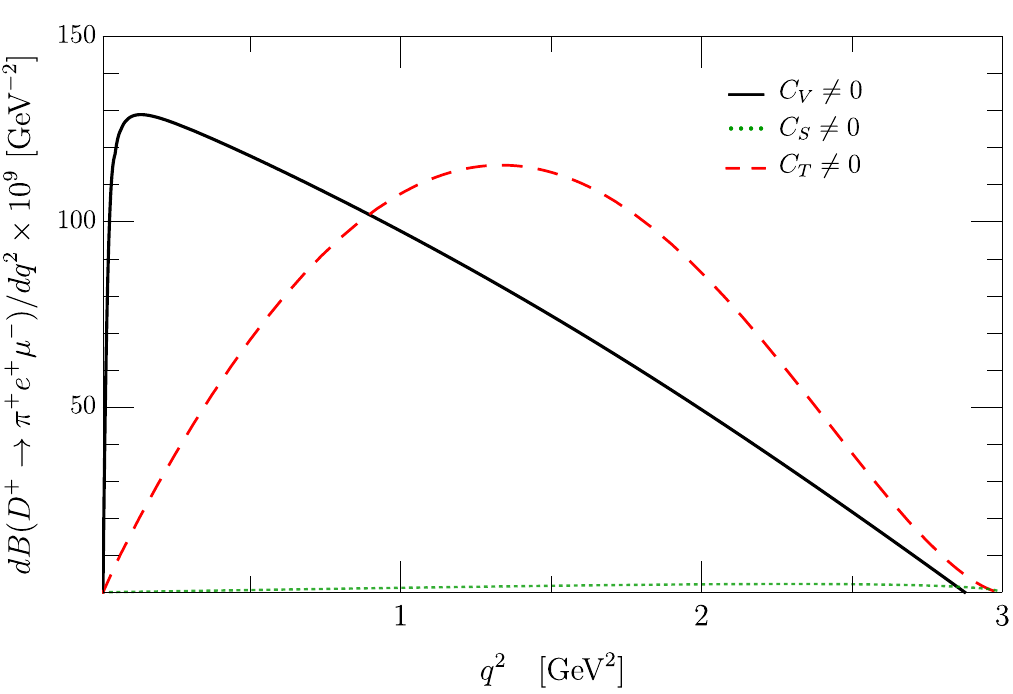}\\
\includegraphics[scale=0.5]{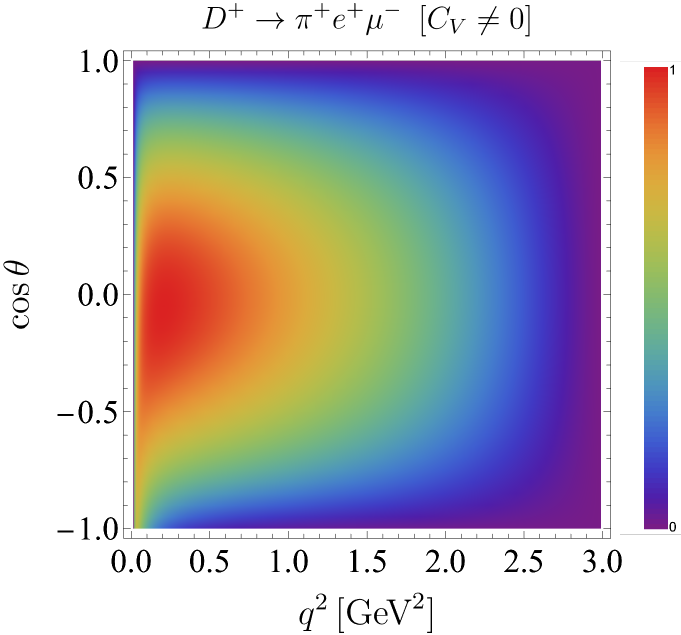}\quad
\includegraphics[scale=0.5]{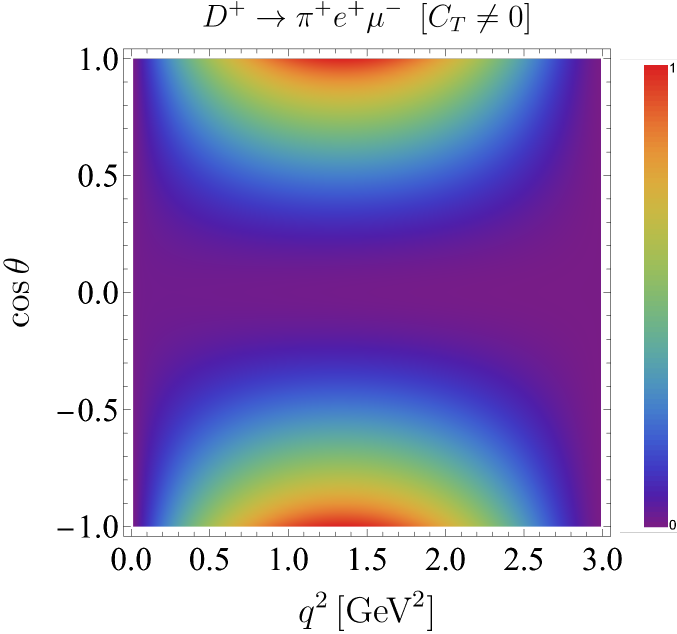}
\caption{\label{Fig:DmesonDecay} Differential branching ratio distribution (top) and Dalitz plot (in arbitrary units) in the $(q^{2},\cos\theta)$ variables (bottom) for the decay $D^{+}\to\pi^{+}e^{+}\mu^{-}$ assuming one single vector, scalar or tensor operator from Table \ref{Table:UpTypeLimits_flavor} is turned on, with the others set to zero.
Here $C_V$,  $C_S$ and $C_T$ denote the combinations $C^{eu}_{\rm VLR} + C^{eu}_{\rm VLL}$,
 $C^{eu}_{\rm SRR} + C^{eu}_{\rm SRL}$ and $C^{eu}_{\rm TRR}$, respectively.}
\end{figure}

The differential branching ratio distribution for the decay $D^{+}\to\pi^{+}e^{+}\mu^{-}$ is shown in the upper panel of Fig. \ref{Fig:DmesonDecay} assuming one single vector (solid black), scalar (dotted green) and tensor (dashed red) operator from Table (\ref{Table:UpTypeLimits_flavor}) is turned on at a time.
In this case, we note that the standalone tensor contribution is found to be of the same order of magnitude as the vector one but with the peak shifted to the central part of the decay spectrum, whereas the scalar contribution is again constrained by the purely leptonic $D^{0}\to e^{\pm}\mu^{\mp}$ decay, and hence it is small.
Another result of this is work is given 
in the bottom panel of Fig. \ref{Fig:DmesonDecay}, where we show the Dalitz plot distributions in the $(q^{2},\cos\theta)$ variables, where $\theta$ is the angle between the meson in the final state and the lepton $e$ in the $e-\mu$ rest frame (we follow Refs. \cite{Gratrex:2015hna,Duraisamy:2016gsd} for the expression of the decay distribution in the $(q^{2},\cos\theta)$ variables). 
As seen, the vector distribution (left) is visibly different from the tensor distribution (right).
Therefore, if LFV was observed in these decays, all of these differences could serve to discriminate among the operators that contribute to these channels.\footnote{
The Dalitz plot for the $B$ decays and the predictions for the $q^{2}$ variation of the forward-backward asymmetry are relegated to Appendix \ref{App:BRandFF}.}

Our predictions for these observables with SMEFT
can be compared with some of those in Refs. \cite{Duraisamy:2016gsd,deBoer:2015boa} resulting from specific leptoquark models.

\section{High-energy limits on CLFV operators}
\label{sec:lhc}

In addition to $\mu \rightarrow e \gamma$, $\mu \rightarrow e$ conversion in nuclei and the meson decay channels discussed so far, $\mu \leftrightarrow e$ transitions can be looked for in collider processes. 
At colliders, one can directly look for LFV decays of the $Z$, $H$ and top quark.
$Z \rightarrow e \mu$
has been looked for at the LHC and LEP. The best bound is now from the LHC,
with, at the 95\% C.L., \cite{ATLAS:2022uhq,ParticleDataGroup:2024cfk}
\begin{equation}\label{eq:LHC_Z}
    \textrm{BR}(Z \rightarrow e \mu ) < 2.62 \cdot 10^{-7}.
\end{equation}
The bound on the Higgs branching ratio, at the 95\% C.L., is
\cite{CMS:2023pte,ParticleDataGroup:2024cfk}
\begin{equation}
     \mathcal B_e\equiv     \textrm{BR}(H \rightarrow e \mu )   < 4.4  \cdot 10^{-5}.   
\end{equation}
Finally, the CMS collaboration provides the strongest limits on the lepton and quark flavor violating decays of the top quark   \cite{CMS:2022ztx,ParticleDataGroup:2024cfk}
\begin{equation}
    \textrm{BR}(t \rightarrow e \mu u ) < 7 \cdot 10^{-8}, \qquad \textrm{BR}(t \rightarrow e \mu c ) < 8.9 \cdot 10^{-7}.
\end{equation}
In SMEFT, the $Z\rightarrow e \mu$ branching ratio can 
be expressed as \cite{Cirigliano:2021img}
\begin{equation}\label{Zpre}
\textrm{BR}(Z\rightarrow e\mu) = \frac{1}{4\widehat\Gamma_Z} \left(  \left| \left[c^{(1)}_{L\varphi} + c^{(3)}_{L\varphi}\right]_{\mu e}\right|^2 
+  |\left[c_{e\varphi}\right]_{\mu e}|^2  + \frac{2 m_Z^2}{v^2} \left( |\left[ \Gamma^e_Z\right]_{e\mu} |^2 + \left[ \Gamma^e_Z\right]_{\mu e} |^2\right)
\right),
\end{equation}
with $\widehat \Gamma_Z = 3.76$. The branching ratio includes both channels $e^\pm \mu^\mp$.
Using Eq. \eqref{eq:LHC_Z},
the $90\%$ C.L. limits on the SMEFT operators are
\begin{eqnarray}\label{Zbound}
|c_{e\varphi}| < 1.8 \cdot 10^{-3}, \quad  |c^{(1)}_{L\varphi} + c^{(3)}_{L\varphi}| < 1.8 \cdot 10^{-3}, \quad 
|\left[\Gamma^e_{Z}\right]_{e\mu, \, \mu e}| < 3.4 \cdot 10^{-3}.
\end{eqnarray}
The Higgs decay width into $\mu e$ is mostly sensitive to non-standard Yukawa couplings   \cite{Harnik:2012pb}
\begin{equation}
\Gamma(H\rightarrow e^- \mu^+ + \mu^- e^+) = \frac{m_H}{8\pi} \left( \left[Y_e^\prime \right]_{\mu e}^2 + \left[Y_e^\prime \right]_{e \mu}^2  \right).
\end{equation}
Assuming no other new Higgs decay channel, the total width can be written as
\begin{equation}
    \Gamma_{\rm tot} = \Gamma_{\rm SM} + \Gamma(H \rightarrow e^- \mu^+ + \mu^- e^+),
\end{equation}
which leads to
\begin{equation}
\left( \left[Y_e^\prime \right]_{\mu e}^2 + \left[Y_e^\prime \right]_{e \mu}^2  \right) = \frac{8 \pi}{m_H}  \frac{\mathcal B_e}{1 - \mathcal B_e} \, \Gamma_{\rm SM},
\end{equation}
where the SM Higgs width is $\Gamma_{\rm SM} = 4.07 \cdot 10^{-3}$ GeV. The resulting 90\% C.L. limit on the Higgs Yukawa is
\begin{eqnarray}
\left[Y_e^\prime \right]_{\mu e, \, e \mu} < 1.7 \cdot 10^{-4}, 
\end{eqnarray}
which is a factor of 100 weaker than the bound from $\mu \rightarrow e \gamma$, in a single coupling analysis.
\\ The BR for the decay $t\to q e^+\mu^-$ is \cite{Davidson:2015zza,Cirigliano:2021img}
\begin{eqnarray}
& & \textrm{BR}(t\to q e^+\mu^-)=\frac{1}{6 \widehat \Gamma_t}\! \left(\frac{m_t}{4 \pi v}\right)^{\!2} \! \! \bigg[4\Bigl(\left| \left[C_{LQ, U}\right]_{\mu e q t}\right|^2 \!+\left| \left[C_{Lu}\right]_{\mu e q t}\right|^2 \!
+\left|\left[C_{Qe}\right]_{\mu e q t}\right|^2  \! +\left|\left[C_{e u}\right]_{\mu e q t}\right|^2 \Bigr) \nn \\
& & \qquad\quad + \left| \left[C_{LeQu}^{(1)}\right]_{\mu e q t} \right|^2
+ \left| \left[C_{LeQu}^{(1)}\right]_{e \mu t q} \right|^2 
+ 48 \left| \left[C_{LeQu}^{(3)}\right]_{\mu e q t} \right|^2 + 48 \left| \left[C_{LeQu}^{(3)}\right]_{e \mu t q} \right|^2
\bigg],
\end{eqnarray}
where we expressed the SM top width as 
\begin{equation}
\Gamma(t \rightarrow W b) = \frac{m_t^3}{16 \pi v^2} \widehat{\Gamma}_t, 
\end{equation}
where $m_t$ is the $\overline{\rm MS}$ top quark mass, $m_t = 162.5$ GeV \cite{ParticleDataGroup:2024cfk}, and $\widehat{\Gamma}_t$ a dimensionless function of $V_{tb}$, $m_t$ and $m_W$. In terms of the measured top width, $\widehat{\Gamma}_t = 1.01^{+0.14}_{-0.11}$ \cite{ParticleDataGroup:2024cfk}.
The resulting bounds on $t$-$u$ CLFV operators are 
\begin{equation}
\left[C_{LQ,U} \right]_{\mu e ut}< 2.2 \cdot 10^{-2},\qquad
\left[C^{(1)}_{LeQu} \right]_{\mu e ut}< 1.1 \cdot 10^{-2},\qquad
\left[C^{(3)}_{LeQu} \right]_{\mu e ut}<  1.6 \cdot 10^{-3},
\end{equation}
where the limit on $C_{Lu},~C_{Qe}$ and $C_{eu}$ is the same as the one on $C_{LQ,U}$.
The bounds on the $t$-$c$ operators are somewhat weaker
\begin{equation}
\left[C_{LQ,U} \right]_{\mu e ct}< 8.0 \cdot 10^{-2},\qquad
\left[C^{(1)}_{LeQu} \right]_{\mu e ct}< 4.0 \cdot 10^{-2},\qquad
\left[C^{(3)}_{LeQu} \right]_{\mu e ct}<  5.8 \cdot 10^{-3}.
\end{equation}

Semileptonic four-fermion operators containing quarks lighter than the top can be looked for
in $p p \rightarrow e \mu$, and they can be particularly enhanced in the high-invariant mass region, $m_{\mu e}\sim$ TeVs, as, in the LFV sector,  was first pointed out in Ref. \cite{Angelescu:2020uug}.
Here we recast the analysis in Ref.
\cite{CMS:2022fsw},
based on 138 fb$^{-1}$ of data at the center of mass energy $\sqrt{S}= 13$ TeV, in terms of SMEFT operators, assuming that SMEFT is valid also at high invariant mass. 
In the analysis, we use 28
$m_{\mu e}$ bins, from 50 GeV to 3 TeV.
We take the background estimates from Ref. \cite{CMS:2022fsw}. For the signal,
we generate $e\mu$ events by generalizing the SMEFT  Drell-Yan code of Ref.
\cite{Alioli:2018ljm}, 
implemented in \texttt{POWHEG}
\cite{Nason:2004rx,Alioli:2010xd,Frixione:2007vw}, to lepton-flavor-violating processes \cite{Cirigliano:2021img}. The implementation
includes NLO QCD corrections, which were found to be non-negligible at high invariant mass \cite{Alioli:2018ljm}.
Acceptance and efficiency are estimated by showering the events with
\texttt{Pythia8} \cite{Bierlich:2022pfr},
and then passing them through \texttt{Delphes} \cite{deFavereau:2013fsa}.
Schematically, the observed signal cross section in a given bin $i$ is related to the \texttt{POWHEG} cross section by
\begin{equation}
    \sigma^{\rm obs}_{i} = \sum_{j} K_{i j} \sigma^{\texttt{POWHEG}}_{j},
\end{equation}
where the response matrix depends on the SMEFT operator under consideration.
As we found the efficiency to depend mostly on the SMEFT operator flavor rather than the Lorentz structure \cite{Cirigliano:2021img},
we calculated $K$ for the purely left-handed operators $C_{LQ, U}$ and $C_{LQ, D}$ and for the left-right operator $C_{Qe}$, with all possible flavor combinations. We then use the same response matrix for all four-fermion operators. We checked that for the scalar operators $\left[C_{LeQu}\right]_{uu}$
and $\left[C_{LedQ} \right]_{bb}$ the limits obtained by using the same response matrix as for vector operators or by recalculating the response matrix differ by only a few percent.
Once we obtained the number of events in each bin, we use \texttt{pyhf} \cite{pyhf,pyhf_joss} to extract bounds on the SMEFT coefficients. Our re-analysis assumes uncorrelated background, and thus might be overly simplistic. With the assumption of uncorrelated background, the 90\% C.L. bounds are showed in Table \ref{limit_lhc}. We see that for valence quarks ($u$ and $d$) the bounds reach the $10^{-4}$ level, while being roughly a factor of ten weaker for $b$ quarks. 
The limits in Table \ref{limit_lhc} used an integrated luminosity of 138 fb$^{-1}$. The High Luminosity LHC will collect 20 times more data, and might thus improve the bounds in Table \ref{limit_lhc} by a factor of 4-5. 

The original analysis of Ref. \cite{Angelescu:2020uug} was based on 36 fb$^{-1}$
of data presented in Ref. \cite{ATLAS:2018mrn}, and was extended in 
Refs. \cite{Allwicher:2022mcg,Allwicher:2022gkm} to include the CMS dataset \cite{CMS:2022fsw}. The main difference of our analysis is the inclusion of NLO QCD corrections. As observed in Ref. 
\cite{Alioli:2018ljm} in the case of lepton-flavor-conserving processes, NLO QCD corrections can increase the SMEFT cross section by 20-30\% 
for $m_{\ell \ell^\prime} \sim 2$-$3$ TeV.  This statement depends on the quark flavor structure of the  SMEFT operators. For the operator $\left[C_{LQ, U}\right]_{uu}$, the ratio $\sigma_{\rm NLO}/\sigma_{\rm LO}$ varies between $1.22$ in the
$m_{e\mu} =  [1.55, 1.70]$ TeV  bin  to $1.33$ 
in the $[2.7, 3.0]$ TeV bin. The corrections are smaller for heavy quarks. For example, for 
$\left[C_{LQ, D}\right]_{bb}$, the same ratio  varies between $1.04$ to $1.12$ 
as $m_{e\mu}$ goes between 1.5 and 3 TeV. For valence quarks, LO bounds are thus about 10\% weaker than in Table \ref{limit_lhc}, 
while the effect is negligible for heavy quarks.

\begin{table}[t]
\centering
\begin{tabular}{||c|| c| c|| c | c || c | c |}
\hline
$C_{LQ,U}$, $C_{Lu}$, $C_{eu}$ & $uu$ & $1.6 \cdot 10^{-4}$ & $uc$ & $2.9 \cdot 10^{-4}$ & $cc$ & $1.4 \cdot 10^{-3}$  \\
$C_{LQ,D}$, $C_{Ld}$, $C_{ed}$ & $dd$ & $2.0 \cdot 10^{-4}$ & $ds$ & $3.3 \cdot 10^{-4}$ & $db$ & $5.4 \cdot 10^{-4}$   \\
           & $ss$ & $7.2 \cdot 10^{-4}$ & $sb$ & $1.4 \cdot 10^{-3}$ & $bb$ & $2.0 \cdot 10^{-3}$ \\
            \hline \hline 
            $C_{Qe}$  & $dd$&  $1.4 \cdot 10^{-4}$
            & $ds$ & $2.1 \cdot 10^{-4}$ & $db$& $5.5 \cdot 10^{-4}$ \\
          &  $ss$ & $5.1 \cdot 10^{-4}$ & $sb$ & $1.4 \cdot 10^{-3}$ & $bb$ & $2.2 \cdot 10^{-3}$
            \\
                        \hline \hline 
 $C^{(1)}_{LeQu}$ & $uu$ & $1.4 \cdot 10^{-4}$ & $uc$ & $3.3 \cdot 10^{-4}$ & $cc$ & $9.4 \cdot 10^{-4}$ \\
 $C_{LedQ}$ & $dd$ &  $1.7 \cdot 10^{-4}$                   & $ds$ & $3.8 \cdot 10^{-4}$ & $db$ & $6.3 \cdot 10^{-4}$ \\
            & $ss$  &      $6.0 \cdot 10^{-4}$               & $sb$ & $1.5 \cdot 10^{-3}$ & $bb$ & $1.7 \cdot 10^{-3}$ \\
 \hline
\end{tabular}
\caption{90\% C.L. upper limits
from CLFV Drell-Yan.
Limits on $C_{Lu}$,$C_{eu}$
and $C_{Ld}$, $C_{ed}$ are identical to $C_{LQ, U}$
and $C_{LQ, D}$, respectively
.
 }
\label{limit_lhc}
\end{table}

\section{EIC sensitivity to $e\rightarrow \mu$ transitions}\label{sec:eic}

In the landscape of future colliders, the
EIC stands out as it is scheduled to start collecting data in the early 2030s \cite{AbdulKhalek:2021gbh}. Working towards its main objective to investigate the structure of the proton, the EIC will collect a large quantity of $ep$ data,   
between 10 and 100 fb$^{-1}$ per year \cite{AbdulKhalek:2021gbh}, at center of mass energies close to the electroweak scale,  $\sqrt{S} \sim 100$ GeV.
These features can be exploited to perform sensitive searches of BSM physics
\cite{Gonderinger:2010yn,Boughezal:2020uwq,Batell:2022ogj,Boughezal:2023ooo,Davoudiasl:2023pkq,Davoudiasl:2024vje,Yan:2021htf,Liu:2021lan,Li:2021uww,Yan:2022npz,Wang:2024zns,Wen:2024cfu}.
In Ref. \cite{Cirigliano:2021img}, we studied the EIC sensitivity to $e \rightarrow \tau$ transitions. We now generalize the discussion to $e \rightarrow \mu$ operators.
For studies of $e \rightarrow \mu$ transitions at other $e p$ colliders,
we refer to Refs. \cite{Furletova:2021wyq,Antusch:2020vul,Jueid:2023fgo}.

At the EIC, $e  \rightarrow \mu$ transitions can be probed by LFV deep inelastic scattering (DIS) $e^- p \rightarrow \mu^- X$. At the parton level, the process is identical to $e \rightarrow \tau$ DIS, which, in the framework of the SMEFT, was considered in Ref. \cite{Cirigliano:2021img}.
With the normalization of Section \ref{sec:formalism}, SMEFT coefficients of $\mathcal O(1)$ (corresponding to new physics at the electroweak scale $\Lambda = v$) would induce cross sections in the range  0.1 and 10 pb, depending on the flavor structure of the operator. 
The production cross sections for different SMEFT operators, calculated at leading order in QCD, are given in Table 1 and 2 of Ref.  \cite{Cirigliano:2021img}.
From these cross sections, in the idealized situation of no background events, an EIC run at 
$\sqrt{S} = 141$ GeV, collecting 100 fb$^{-1}$ of data would be sensitive to  $\left[C_{Lu}\right]_{uu} \sim 1\cdot  10^{-3}$ and to $\left[C_{Ld}\right]_{bb} \sim 5 \cdot 10^{-3}$.
To realistically assess the sensitivity of the EIC, we need an estimate of the SM backgrounds.
We performed an analysis with \texttt{Delphes}
and \texttt{Pythia8}. We used the \texttt{Delphes} analysis card developed in Ref. \cite{Arratia:2021uqr},
assuming the muon reconstruction efficiency to be similar to electron reconstruction. Further, Ref. \cite{Arratia:2021uqr} provides the expected particle identification capabilities of the detector according to Ref. \cite{AbdulKhalek:2021gbh}. In the low $\eta$ ($|\eta| \leq 3.5$) regime, pions are misidentified as electrons (and vice versa) less than 2\% of the time, whereas all electrons are misidentified as pions in the high $\eta$ $(|\eta| > 3.5)$ or low p$_T$ $(p_T\cosh\eta < 0.05)$ regimes. In our analysis, however, we assume perfect lepton identification if the track is reconstructed. These assumptions need to be validated by a realistic detector simulation, especially in light of the fact that currently there are no plans for a muon detector at the EIC.  

We consider a potential EIC run 
at the center-of-mass energy $\sqrt{S} = 141$ GeV, which would have maximal sensitivity to SMEFT operators
\cite{Cirigliano:2021img}. We assumed an integrated luminosity of 100 fb$^{-1}$. With such luminosity, we expect billions of SM DIS events.
We generated $3\cdot 10^9$ SM neutral and charged current DIS events, with a cut on the minimum transverse momentum of the partons in the hard process, $p_T \ge 5$ GeV.  
In Ref. 
\cite{Cirigliano:2021img}, we identified semileptonic four-fermion operators as the most promising for the SMEFT signal.  In this class, we observed that the signal efficiency depended minimally on the Lorentz structure of the operators, but showed a strong dependence on the flavor content. We therefore simulated SMEFT events for two four-fermion operators, $C_{Lu}$ and $C_{Ld}$, and all possible flavor components active at the EIC, namely
$\left[C_{Lu}\right]_{i j}$ with
$ i, j \in \{u, c\}$
and $\left[C_{Ld}\right]_{i j}$ with
$ i, j \in \{d, s, b\}$. We generate $5 \cdot 10^6$ events for each SMEFT coefficients. 

\begin{figure}
\includegraphics[width=0.45\textwidth]{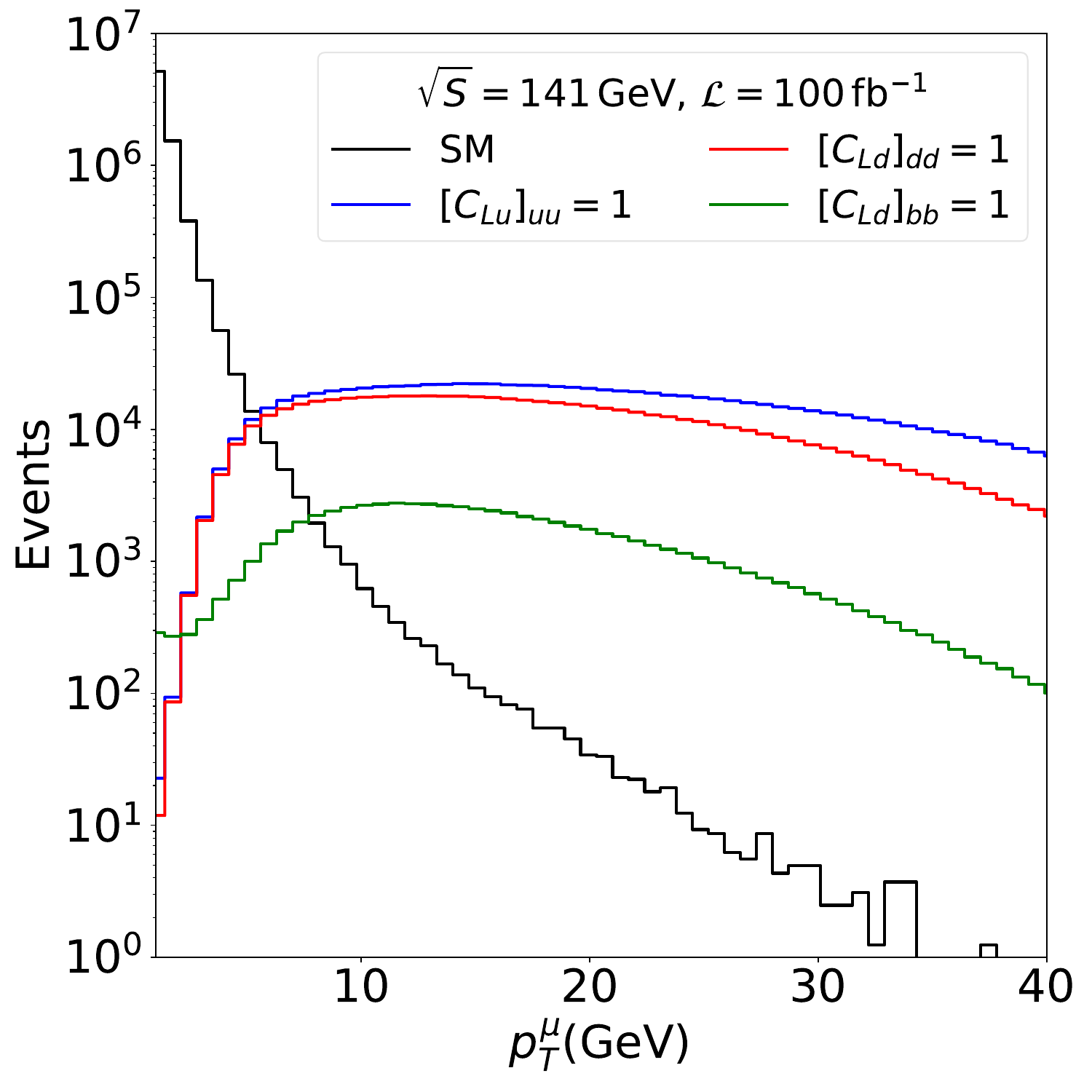}
\includegraphics[width=0.45\textwidth]{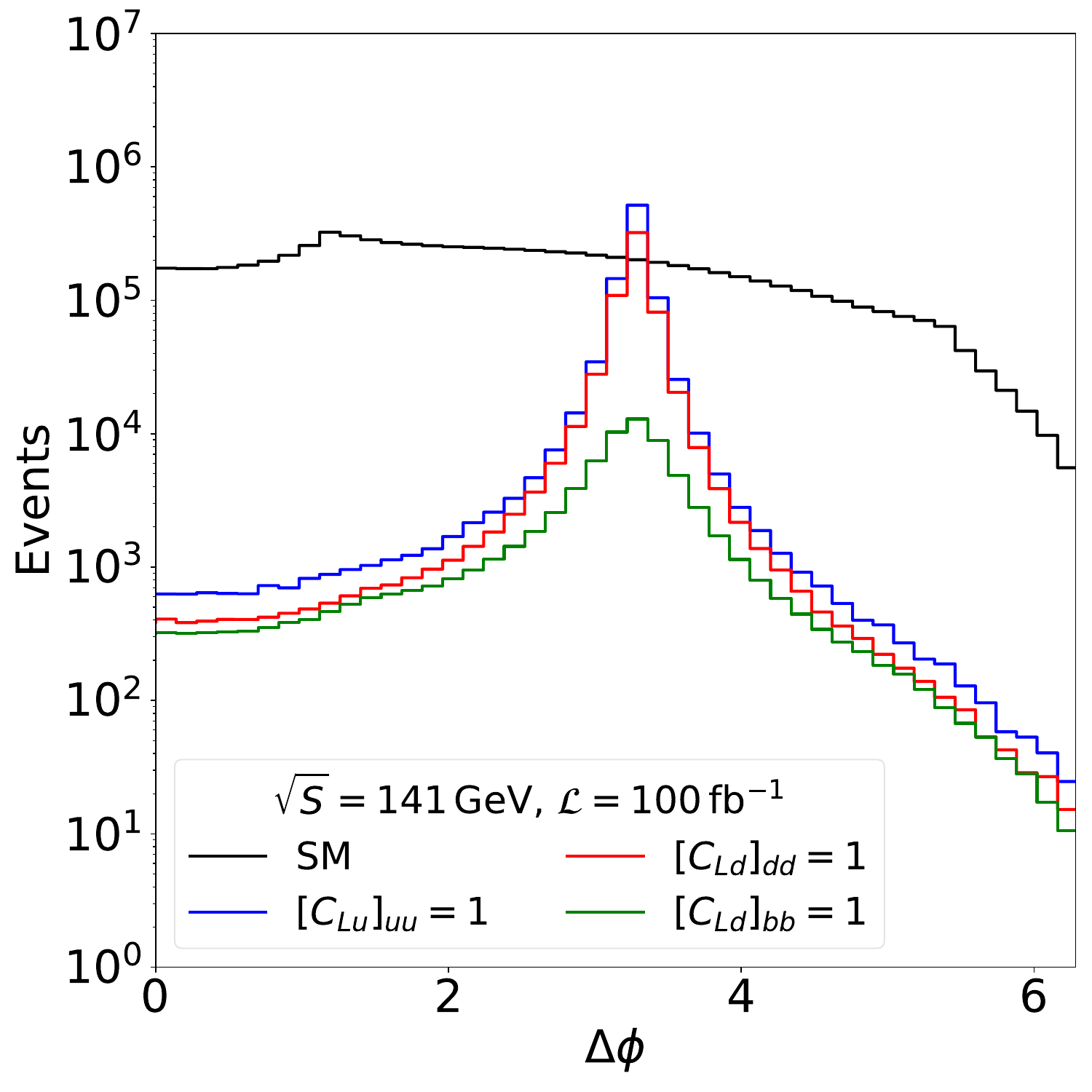}
\caption{Muon transverse momentum distribution (left) 
and azimuthal separation between the muon and the hardest jet (right) in the SM and in the presence of three LFV SMEFT operators, $\left[C_{Lu}\right]_{uu}$, 
$\left[C_{Ld}\right]_{dd}$
and $\left[C_{Ld}\right]_{bb}$, with their coefficients set to 1.}\label{fig:eic1}
\end{figure}

We studied the effects of a variety of kinematics cuts to minimize the SM background without reducing the signal from SMEFT operators.
We require events to have at least one muon with high transverse momentum, not to have any electron with high transverse momentum, and to have at least one jet. We also require the muon and the hardest jet to be almost back-to-back in the azimuthal plan.   
We found the rapidity difference between the muon and the jet to provide little discriminating power.
As an illustration of the differences between SM background and SMEFT 
signal events, in Fig. \ref{fig:eic1}, we show the distribution of the muon transverse momentum, $p_T^\mu$, and the jet-muon azimuthal separation $\Delta \phi$, for SM events (black) and for events generated by LFV SMEFT operators. We see that, in the SM, most muons have very small $p_T$. Also, SM background events show little azimuthal correlation between the muon and the hardest jet in the event, while signal events are peaked at $\Delta \phi =\pi$. 

\begin{table}[t]
\centering
\begin{tabular}[b]{||c | c | c | c | c|| c | c|| c | c  ||} 
 \hline
 cut & $p_{T, \rm cut}^\mu$ & $p_{T, \rm cut}^j$ & $p^e_{T,\rm cut}$ & $ \Delta\phi $  & $\epsilon_{\rm SM}$ & $n_{\rm bkgd}$  & $\epsilon_{uu}$ & $\epsilon_{bb}$ \\ [0.5ex] 
 \hline\hline
 1 &  5 & 5 &  10 & $[2,4]$ & $6.8 \cdot 10^{-6}$ & $1.2 \cdot 10^4$ & $0.44$ & $0.22$ \\ 
 \hline
 2 &  \ 10 &  \ 5 &  \ 10 & $[2,4]$ & $9.4 \cdot 10^{-7}$ & $1.6 \cdot 10^3$ & 0.40 & 0.19\\
 \hline
 3 &  \ 5 &  \ 10 &  \ 10 & $[2,4]$ & $4.0 \cdot 10^{-6}$
 & $6.7 \cdot 10^3$ & 0.39 & 0.18 
 \\
 \hline
 4 &  \ 5 &  \ 15 &  \ 10 & $[2,4]$ & $1.7 \cdot 10^{-6}$ & $2.9 \cdot 10^3$ & 0.32 & 0.13\\ \hline
 5 &  \ 5 &  \ 5 &  \ 5 & $[2,4]$ & $2.7 \cdot 10^{-6}$  &$4.7 \cdot 10^3$ & 0.44 & 0.22\\ 
 \hline
 6 &  \ 10 &  \ 10 &  \ 10 & $[2,4]$ & $7.9 \cdot 10^{-7}$ & $1.4 \cdot 10^3$ & 0.38 & 0.17\\
 \hline
 7 &  \ 10 &  \ 15 &  \ 10 & $[2,4]$ & $5.6 \cdot 10^{-7}$ & $9.7 \cdot 10^2$ & 0.32 & 0.13 \\
 \hline
 8 &  \ 10 &  \ 10 &  \ 5 & $[2,4]$ & $2.7 \cdot 10^{-7}$ & $4.8\cdot 10^2$ & 0.38 & 0.17\\ \hline
 9 &  \ 10 &  \ 15 &  \ 5 & $[2.5,3.5]$ & $1.8 \cdot 10^{-7}$ &  $3.1 \cdot 10^2$ & 0.32 & 0.12\\
 \hline
 10 &  \ 10 &  \ 15 &  \ 5 & $[2,4]$ & $2.0 \cdot 10^{-7}$ & $3.5 \cdot 10^2$ & 0.32 & 0.13\\ 
 \hline
 11 &  \ 5 &  \ 5 &  \ 10 & $[2.5, 3.5]$ & $4.5 \cdot 10^{-6}$ & $7.8 \cdot 10^3$ & 0.43 & 0.19\\
 \hline
 12 &  \ 15 &  \ 5 &  \ 5 & $[2,4]$ & $8.9 \cdot 10^{-8}$ & $1.5\cdot 10^2$ & 0.33 & 0.14\\  \hline
 13 &  \ 15 &  \ 15 &  \ 5 & $[2,4]$ & $7.9 \cdot 10^{-8}$ & $1.4 \cdot 10^2$ & 0.31 & 0.11\\ 
 \hline
 14 &  \ 10 &  \ 20 &  \ 5 & $[2,4]$ & $1.1 \cdot 10^{-7}$ & $1.9 \cdot 10^2$ & 0.25 & 0.08\\
 \hline
 15 &  \ 20 &  \ 20 &  \ 5 & $[2,4]$ & $2.2 \cdot 10^{-8}$ & 38 & 0.24 & 0.07 \\ [1ex] 
 \hline
\end{tabular}%
\caption{List of all cuts applied to both SM background and BSM processes. The transverse momentum, $p_T$, is given in GeV, while $\Delta\phi$ in radians. The sixth and seventh columns show the background efficiency $\epsilon_{\rm SM}$, and the number of SM background events,
assuming $\sqrt{S} = 141$ GeV and an integrated luminosity of $\mathcal L = 100$ fb$^{-1}$. The last two columns show the signal efficiency for two SMEFT operators, $\left[C_{Lu}\right]_{uu}$ and $\left[C_{Ld}\right]_{bb}$, which induce couplings to light and heavy quarks, respectively. }
\label{table:eic1}
\end{table}

The sets of cut that we impose is captured by five parameters,
$p_{T, \rm cut}^\mu$, $p_{T, \rm cut}^e$, $p_{T, \rm cut}^j$, $\Delta \phi_{\rm min}$, 
$\Delta \phi_{\rm max}$, and by the conditions
\begin{equation}
    p_T^\mu > p_{T, \rm cut}^\mu, \qquad 
    p_T^e < p_{T, \rm cut}^e, 
    \qquad
    p_T^{j} > p_{T, \rm cut}^j, \qquad  \Delta \phi_{\rm min} < \Delta \phi < \Delta \phi_{\rm max}.
\end{equation}
We studied 15 combinations of these parameters, summarized in Table \ref{table:eic1}. The table also gives the background cut efficiency $\epsilon_{\rm SM}$, defined as the ratio of SM events that pass the imposed cuts, and the expected number of background events, assuming a tree level DIS cross section of $\sigma_{\rm DIS}= 1.75 \cdot 10^4$ pb and an integrated luminosity of $100$ fb$^{-1}$. The cross section was computed in \texttt{Pythia8}, with a cut of $5$ GeV on the minimum transerve momentum of the partons in the hard process. We neglect in this study theoretical errors and the contributions of higher order corrections. 
The most important cut by far is the cut on the muon transverse momentum. Even with the hardest cuts, we see that we cannot completely reduce the SM background to zero.
The last two columns of Table \ref{table:eic1}
 show the signal efficiency for two SMEFT operators, $\left[C_{Lu}\right]_{uu}$ and 
$\left[C_{Ld}\right]_{bb}$, which induce couplings to light and heavy quarks, respectively. 
The signal efficiency does not depend too strongly on the imposed cuts.
As noticed in Ref. \cite{Cirigliano:2021img}, heavy flavor operators are more affected by hard $p_T$ cuts. We explored the possibility of softening the cuts by requiring the hardest jet to be $b$-tagged, but found no significant advantage in this strategy, at least in our naive implementation.

\begin{figure}
\includegraphics[width=\textwidth]{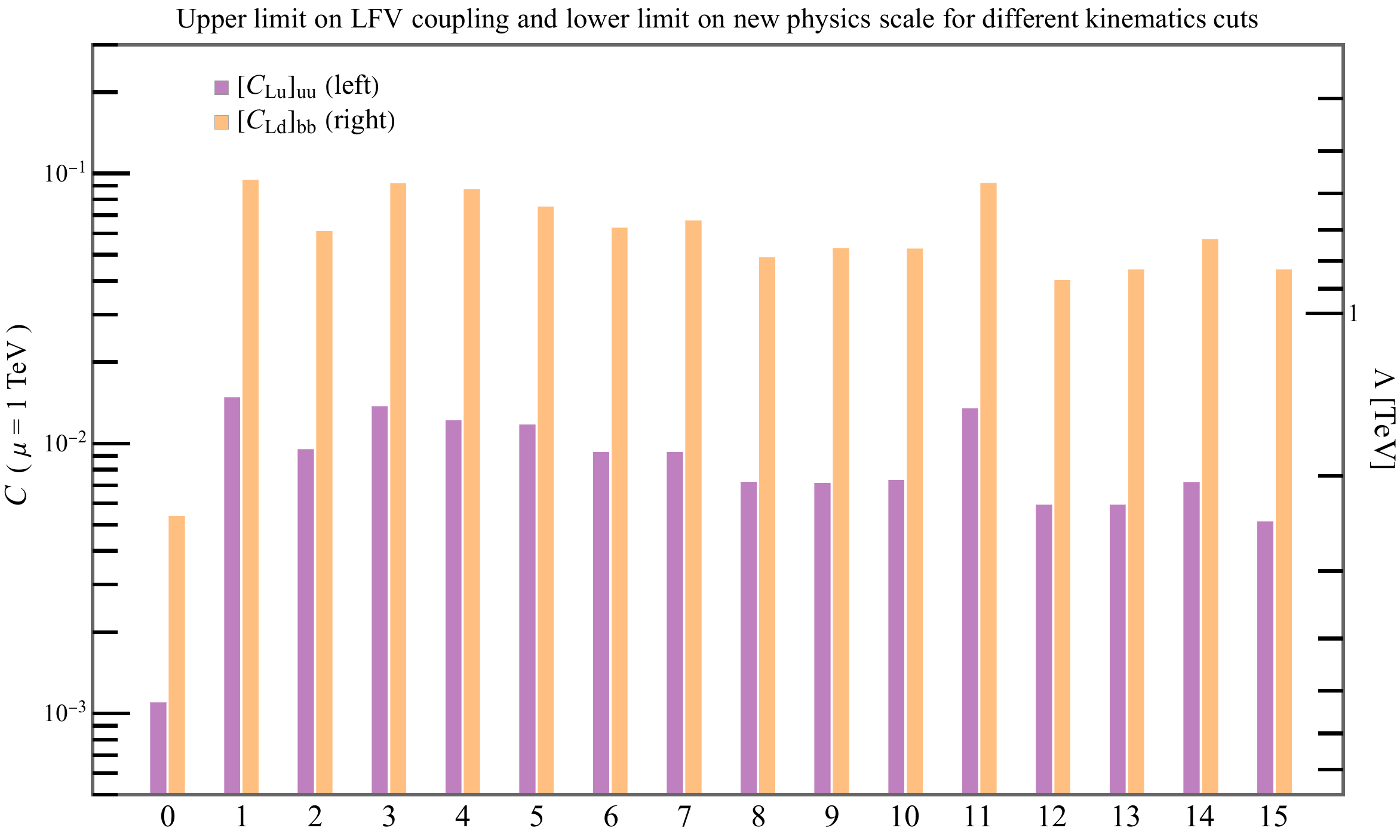}
\caption{Dependence of the bounds on the effective coefficients $\left[C_{Lu}\right]_{uu}$
and $\left[C_{Ld}\right]_{bb}$ on the kinematics cuts summarized in Table \ref{table:eic1}. The cut denoted by 0 indicates the idealized situation in which $\epsilon_{\rm SM} = 0$ and $\epsilon_{uu} = \epsilon_{bb} = 1$. 
}\label{fig:eic2}
\end{figure}

The dependence of the bounds on the EFT coefficients $\left[C_{Lu}\right]_{uu}$ and $\left[C_{Ld}\right]_{bb}$ on the kinematics cuts summarized in Table \ref{table:eic1}
is shown in Fig. \ref{fig:eic2}.
To estimate the EIC sensitivity, we 
assume that the backgrounds are known with negligible errors, and that the number of signal and background events are Poisson distributed.
The upper
limit on the CLFV coefficients at the $1-\alpha$ credibility level, when $n$ events have been
observed and $n_b$ events are expected, can be obtained by solving the equation
\begin{equation}
    1-\alpha = 1 - \frac{\Gamma(1+n, n_b + n_s)}{\Gamma(1+n, n_b)},
\end{equation}
where $n_s$ is a function of the SMEFT coefficient and of the efficiency
$\epsilon_{\rm SMEFT}$, $n_s = \mathcal L \times \sigma |C_{\rm SMEFT}|^2 \epsilon_{\rm SMEFT}$.
We solve this equation assuming $n = n_b$,
and using the cross sections given in Ref. \cite{Cirigliano:2021img}.
We neglect the theoretical error on the cross section, which can be improved by extending the calculation to next-to-leading order in QCD.
The first bar in Fig. \ref{fig:eic2}, denoted by the subscript 0, indicated the ``ideal'' sensitivity, in which one could cut all SM background without affecting the signal,
$\epsilon_{\rm SM} = 0$ and $\epsilon_{uu}=\epsilon_{bb} = 1$. The remaining bars denote the bounds we obtain after applying the cuts in Table \ref{table:eic1}. For valence quark operators,  as $\left[C_{Lu}\right]_{uu}$, the bounds improve by a factor of 3 in going from the softest to the hardest cut. The best constraint comes from cut 15 and is  $|\left[C_{Lu}\right]_{uu} | < 4.7 \cdot 10^{-3}$. Cut 12 and 13 come close, with $|\left[C_{Lu}\right]_{uu} | < 6.0 \cdot 10^{-3}$. Even with cut 15, the bound we obtain is roughly a factor of 5 weaker than the ``ideal'' sensitivity, $\left|\left[C_{Lu}\right]_{uu}\right| < 1.1 \cdot 10^{-3}$, indicating that there is room to further refine  the analysis.
For heavy quarks, the best constraint come from cut 13, $|\left[C_{Ld}\right]_{bb}| < 5.2 \cdot 10^{-2}$, with cut 12 and 15 very close. 
In this case, hardening the cut on $p^\mu_T$
beyond 15 GeV does not help, because the signal efficiency also decreases.  Since cut 15 provides the best, or very close to the best, sensitivity, we will use this cut to show projected bounds on SMEFT operators. The signal efficiencies for this cut are shown in Table \ref{table:cut15}.

\begin{table}[t]
\centering
\begin{tabular}{||c | c | c | c | c | c | c | c||} 
 \hline
  &   $\epsilon_{15}$ (\%)  &   & $\epsilon_{15}$ (\%) &   & $\epsilon_{15}$ (\%) &   & $\epsilon_{15}$ (\%) \\ [0.2ex] 
 \hline\hline
 $[C_{Lu}]_{uu}$   & 27.1  &
 $[C_{Lu}]_{uc}$   & 11.7 &
$[C_{Lu}]_{cu}$   & 28.5 &
 $[C_{Lu}]_{cc}$   & 10.1  
\\ 
 \hline
 $[C_{Ld}]_{dd}$   & 20.7 &
 $[C_{Ld}]_{ds}$   & 14.9 & 
 $[C_{Ld}]_{db}$  & 12.1 &  & \\ 
 \hline
  $[C_{Ld}]_{sd}$   & 20.1 &
  $[C_{Ld}]_{ss}$   & 11.2 &
   $[C_{Ld}]_{sb}$   & 7.8 & & \\
 \hline
 $[C_{Ld}]_{bd}$   & 19.4 &
 $[C_{Ld}]_{bs}$   & 9.6&
$[C_{Ld}]_{bb}$   & 5.4  & & \\
 \hline
\end{tabular}%
\\
\caption{Efficiencies for cut 15 for SMEFT operators with different flavor structures.}
\label{table:cut15}
\end{table}

\begin{table}[t]
\centering
\begin{tabular}{||c|| c| c|| c | c || c | c || c | c||}
\hline
$C_{Lu}$ & $uu$ & $4.7 \cdot 10^{-3}$ & $uc$ & $1.1 \cdot 10^{-2}$ & $cu$ & $5.2 \cdot 10^{-3}$ & $cc$ & $1.7 \cdot 10^{-2}$  \\
\hline
$C_{LQ,U}$, $C_{eu}$ & $uu$ & $3.3 \cdot 10^{-3}$ & $uc$ & $1.4 \cdot 10^{-2}$ & $cu$ & $3.3 \cdot 10^{-3}$ & $cc$ & $1.7 \cdot 10^{-2}$  \\
\hline
$C_{LeQu}^{(1)}$ & $uu$ & $1.1 \cdot 10^{-2}$ & $uc$ & $3.5 \cdot 10^{-2}$ & $cu$ & $1.1 \cdot 10^{-2}$ & $cc$ & $4.8 \cdot 10^{-2}$  \\
$C_{LeQu}^{(3)}$ & $uu$ & $1.0 \cdot 10^{-3}$ & $uc$ & $3.3 \cdot 10^{-3}$ & $cu$ & $1.0 \cdot 10^{-2}$ & $cc$ & $4.5 \cdot 10^{-3}$  \\
\hline \hline
 $C_{Ld}$ & $dd$ & $6.3 \cdot 10^{-3}$ & $ds$ & $9.6 \cdot 10^{-2}$ & $db$ & $1.1 \cdot 10^{-2}$ & &   \\
           & $sd$ & $8.6 \cdot 10^{-3}$ & $ss$ & $2.3 \cdot 10^{-2}$ & $sb$ & $3.0 \cdot 10^{-2}$ & &  \\
           & $bd$ & $9.2 \cdot 10^{-3}$ & $bs$ & $3.3 \cdot 10^{-2}$ & $bb$ & $5.2 \cdot 10^{-2}$ & &  \\
           \hline
           $C_{LQ,D}$, $C_{ed}$ & $dd$ & $5.1 \cdot 10^{-3}$ & $ds$ & $1.4 \cdot 10^{-2}$ & $db$ & $1.7 \cdot 10^{-2}$ & &   \\
           & $sd$ & $5.5 \cdot 10^{-3}$ & $ss$ & $2.3 \cdot 10^{-2}$ & $sb$ & $3.7 \cdot 10^{-2}$ & &  \\
           & $bd$ & $5.6 \cdot 10^{-3}$ & $bs$ & $2.7 \cdot 10^{-2}$ & $bb$ & $5.2 \cdot 10^{-2}$ & &  \\
           \hline
$C_{Qe}$ & $dd$ & $3.7 \cdot 10^{-3}$ & $ds$ & $6.9 \cdot 10^{-3}$ & $db$ & $1.1 \cdot 10^{-2}$ & &   \\
           & $sd$ & $4.3 \cdot 10^{-3}$ & $ss$ & $1.1 \cdot 10^{-2}$ & $sb$ & $3.0 \cdot 10^{-2}$ & &  \\
           & $bd$ & $9.2 \cdot 10^{-3}$ & $bs$ & $3.3 \cdot 10^{-2}$ & $bb$ & $5.2 \cdot 10^{-2}$ & &  \\
           \hline
            $C_{LedQ}$  & $dd$ & $1.6 \cdot 10^{-2}$ & $ds$ & $3.1 \cdot 10^{-2}$ & $db$ & $3.7 \cdot 10^{-2}$ & &   \\
           & $sd$ & $5.7 \cdot 10^{-2}$ & $ss$ & $6.3 \cdot 10^{-2}$ & $sb$ & $8.8 \cdot 10^{-2}$ & &  \\
           & $bd$ & $1.9 \cdot 10^{-2}$ & $bs$ & $7.9 \cdot 10^{-2}$ & $bb$ & $1.2 \cdot 10^{-1}$ & &  \\
 
 \hline
\end{tabular}
\caption{Projected 90\% C.L. upper limits on 
$u$-type and $d$-type vector, scalar and tensor
operators from CLFV DIS, assuming $\sqrt{S} = 141$ GeV and $\mathcal L = 100$ fb$^{-1}$. }
\label{tab:limit_eic}
\end{table}

In Table \ref{tab:limit_eic}, we show the projected limit on all the flavor components
of $u$-type and $d$-type vector, scalar and tensor 
four-fermion operators using cut 15,
and assuming the signal efficiencies to be independent on the Lorentz structure of the operator. For $C_{Qe}$, the bounds are obtained retaining only terms up to $\mathcal O(\lambda)$ in the CKM matrix, where $\lambda \sim V_{us} \sim 0.225$.

\section{Constraints on SMEFT operators}
\label{sec:summary}

We can now compare the bounds on CLFV operators from various high- and low-energy probes. 
In Section \ref{sec:single} we consider a scenario in which we assume that only a single operator is nonzero at the high scale $\Lambda$.
Although BSM scenarios in general predict several LFV couplings, the single operator analysis would provide a good insight into the relative sensitivity of various LFV probes. In Section \ref{sec8} we then consider a scenario with multiple couplings where, to limit the number of coefficients, we impose quark flavor symmetries on the SMEFT
operators. This scenario will illustrate the complementarity between high- and low-energy probes in less than minimal BSM scenarios.

\subsection{Single-operator dominance hypothesis}\label{sec:single}
\begin{figure}
\includegraphics[width=\textwidth]{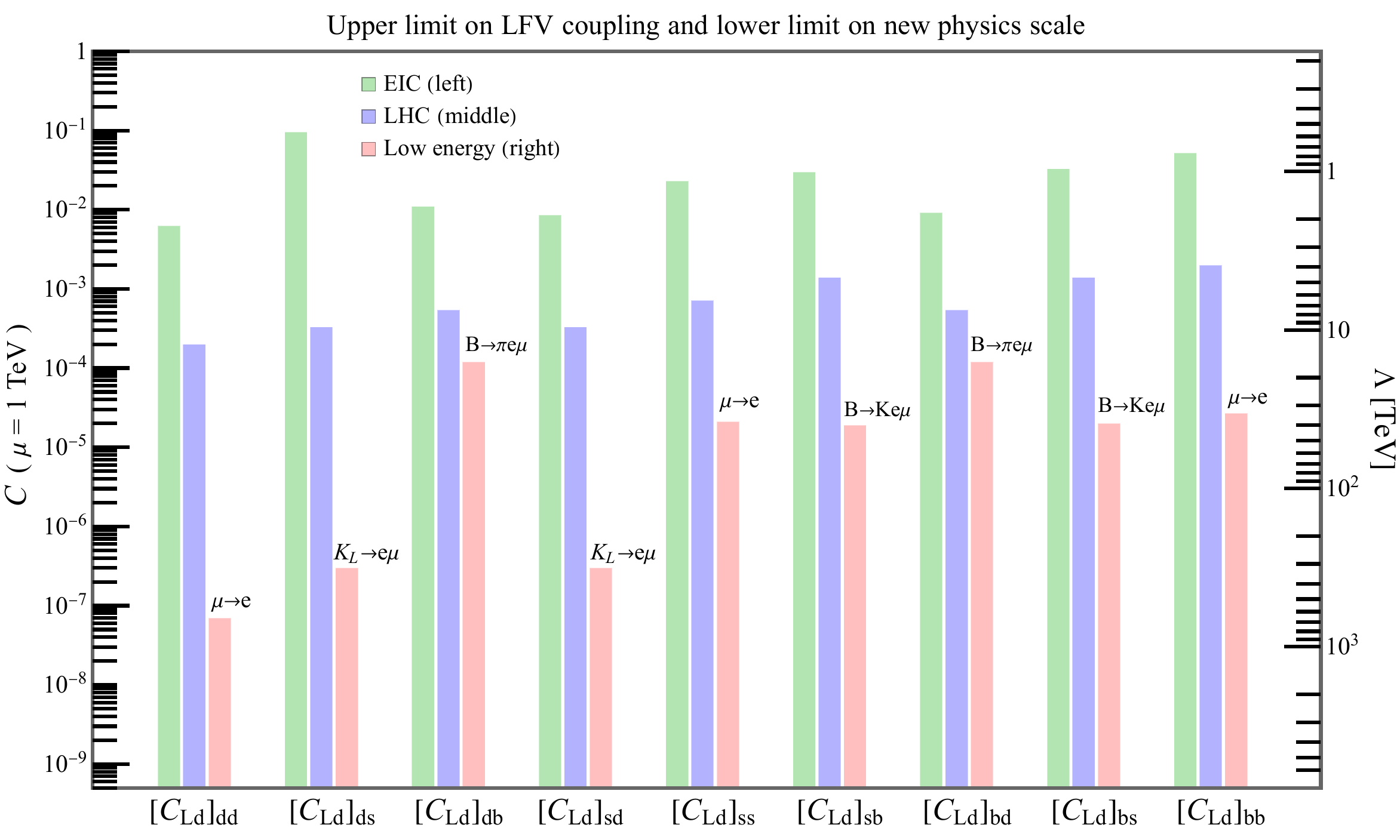}

\vspace{8ex}

\includegraphics[width=\textwidth]{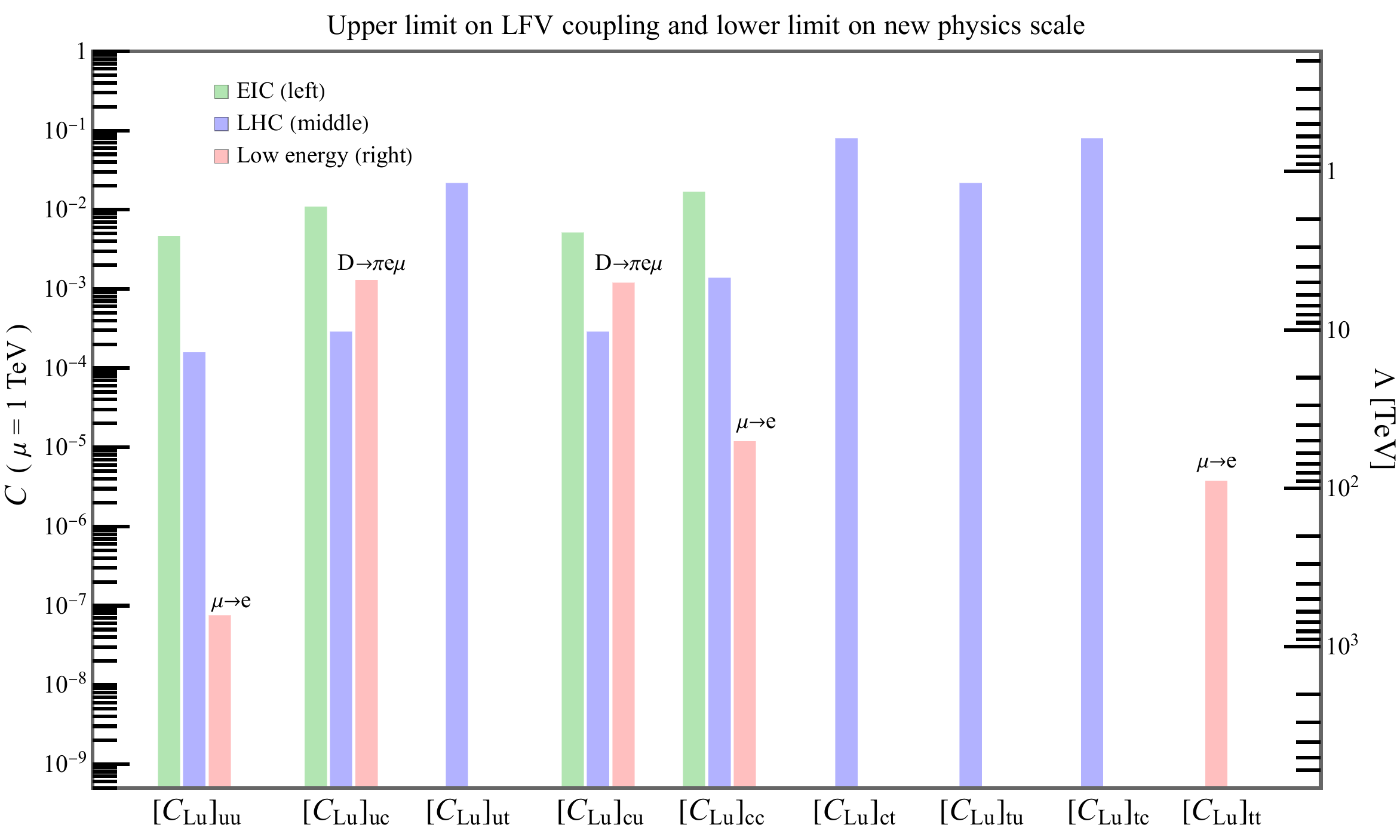}
\caption{Upper bounds (leftmost axis) on $C_{Ld}$ (top) and $C_{Lu}$ (bottom) and lower bounds on new physics scale $\Lambda$ (rightmost axis) from the EIC (left), LHC (middle) and low-energy observables (right). 
}
\label{fig:barcharts}
\end{figure}

In this section, we compare the limits on CLFV SMEFT operators from low-energy probes, discussed in Sections \ref{sec:mu2e}
and \ref{sec:Flavor}, with the LHC and EIC sensitivities of Sections \ref{sec:lhc} and \ref{sec:eic}, in the assumption that BSM physics only induces one operator at the scale $\Lambda$.
Starting from lepton bilinear operators, the photon dipoles $[\Gamma^e_\gamma]_{e\mu,\, \mu e}$ are overwhelmingly constrained by $\mu\rightarrow e \gamma$. 
The $Z$ dipoles, $[\Gamma^e_Z]_{e\mu,\, \mu e}$, receive a direct constraint from $Z \rightarrow e \mu$,
but this is six orders of magnitude weaker than what can be inferred from $\mu \rightarrow e \gamma$.
The strongest low-energy bounds on the $Z$ couplings $c^{(1)}_{L\varphi}$,
$c^{(3)}_{L\varphi}$ and $c_{e\varphi}$ arise from $\mu \rightarrow e$ conversion in nuclei. The single operator bounds in Table \ref{limit_dipole_yukawa_Z}
can be converted into a new physics scale $\Lambda \sim 265$ TeV, where we define $\Lambda$ by $4 G_F C/\sqrt{2} = 1/\Lambda^2$. In comparison, the LHC bound on the $Z \rightarrow \mu e$ branching ratio can be converted into a scale of $4$ TeV. 
$\mu\rightarrow e \gamma$ and $\mu \rightarrow e$ conversion thus rule out the direct observation of $Z \rightarrow \mu e$ at the LHC. Similar conclusions for future $e^+ e^-$ colliders were reached in Ref.
\cite{Calibbi:2021pyh}. 
Low- and high-energy bounds are closer in the case of Higgs couplings, 
where $\mu \rightarrow e \gamma$ arises  dominantly at two loops and $\mu \rightarrow e$ conversion is suppressed by the small Yukawa couplings of light quarks.
Low-energy bounds are nonetheless a factor of 100 stronger than direct LHC bounds, requiring an improvement of four orders of magnitude in the measurement of the $H \rightarrow e \mu$ branching ratio for the latter to become competitive. 

We next consider semileptonic four-fermion operators. We discuss in detail the two operators  $C_{Ld}$ and $C_{Lu}$, for which we obtained bounds in Sections \ref{sec:mu2e}, \ref{sec:Flavor}, \ref{sec:lhc}, and \ref{sec:eic}. Other vector-like operators, such as $C_{LQ, U}$ or $C_{LQ, D}$, present very similar patterns.
Again, we assume that a single operator is nonzero at the high scale $\Lambda$.
The bar charts in Fig.~\ref{fig:barcharts} present upper bounds on the LFV coefficient $C~(\mu=1~{\rm TeV})$ on the leftmost axis, converted in a lower bound on $\Lambda$ on the rightmost axis. The latter is obtained by assuming, as before, $4G_FC_{Ld/Lu}/\sqrt{2}=1/\Lambda^2$. The blue (middle) and pink (right) bars correspond to current limits from the LHC and low-energy observables. The decay channel that gives the dominant low-energy bound is labeled right above each pink bar. On the other hand, the green bars (left) represent expected sensitivities from the future EIC with the assumption of $\sqrt{S} = 141$ GeV and ${\cal L}=100~{\rm fb}^{-1}$. Overall, low-energy and LHC bounds are stronger than those expected from the EIC. 
On flavor diagonal couplings,
the bound on $\Lambda$ from $\mu\to e$ conversion almost reaches $1000$ TeV for $[C_{Ld}]_{dd}$ and $[C_{Lu}]_{uu}$. The bounds from CLFV Drell-Yan, while approaching 10 TeV, are three/four orders of magnitude weaker.  
It is interesting to notice that the next strongest bound 
is on $[C_{Lu}]_{tt}$, with scale approaching 100 TeV. As commented earlier, this is due to the strong running of top operators onto $Z$ couplings, mediated by the large top Yukawa. The $\mu \rightarrow e$ conversion bounds of $[C_{Ld}]_{ss}$, $[C_{Ld}]_{bb}$ and $[C_{Lu}]_{cc}$ are somewhat weaker, but still about a factor of 100 stronger than the LHC or EIC. In Section \ref{sec8} we will discuss how these conclusions are affected by turning on multiple couplings.    
In the case of quark flavor changing operators, kaon decays probe scales very close to $\mu \rightarrow e$ conversion. $B$ decays are weaker (see also Ref.~\cite{Ali:2023kua}).
In particular, the bounds on the
$bd$ and $db$ components arising from $B \rightarrow \pi e \mu$ are only a factor of 4 stronger than the LHC. 
$D$ meson decays are even weaker, and in this case CLFV Drell-Yan outperforms low-energy searches, while the expected sensitivity at the EIC is roughly a factor of 10 weaker. 
Currently, the only constraints on flavor-changing $tc$ and $tu$ operators are from top decay searches at the LHC,
which yield weak bounds 
in the 
${\cal O} (10^{-1})$-${\cal O} (10^{-2})$ range. In principle, the operators can radiatively generate others that can be constrained by low-energy probes. Such contributions are however suppressed by CKM and Yukawa factors, and are not competitive. 
The discussion for the $[C_{Lu}]$ and 
$[C_{Ld}]$ operators is closely mirrored by the other vector-like operators, as can be seen from Tables \ref{limit_semileptonic}, \ref{Table:UpTypeLimits_flavor},
\ref{Table:DownTypeLimits_flavor},
\ref{limit_lhc} and \ref{tab:limit_eic}.
In the case of scalar and tensor operators, the low-energy limits tend to be even stronger, because of the enhancement of pseudoscalar contributions to leptonic meson decays
and of the strong running of tensor operators into dipoles. The only exception are the flavor-changing tensor couplings $[C^{(3)}_{LeQu}]_{uc,\, cu}$, which are better constrained by Drell-Yan.
Next-generation $\mu\to e$ searches at Fermilab (Mu2e) and J-PARC (COMET) aim to improve the experimental sensitivity by four orders of magnitude, from which we expect that the bounds can be more stringent by a factor of 100. Since the bound from collider probes scale as $1/\sqrt{\cal L}$, the High-Luminosity LHC with $3000~{\rm fb}^{-1}$ would be able to achieve a factor of 5 improvement.  Finally, it should be noted that the CLFV operators of interest can also generate pseudoscalars and quarkonium decays such as $\pi^0\to\mu e$ and $J/\psi\to \mu e$. These channels, however, yield weaker bounds compared to $\mu\to e$ transition searches; see discussions in Refs. \cite {Hazard:2016fnc, Hoferichter:2022mna, Calibbi:2022ddo}.

\subsection{Multiple couplings fit}
\label{sec8}

\begin{figure}[t]
\centering
\includegraphics[width=6.7cm]{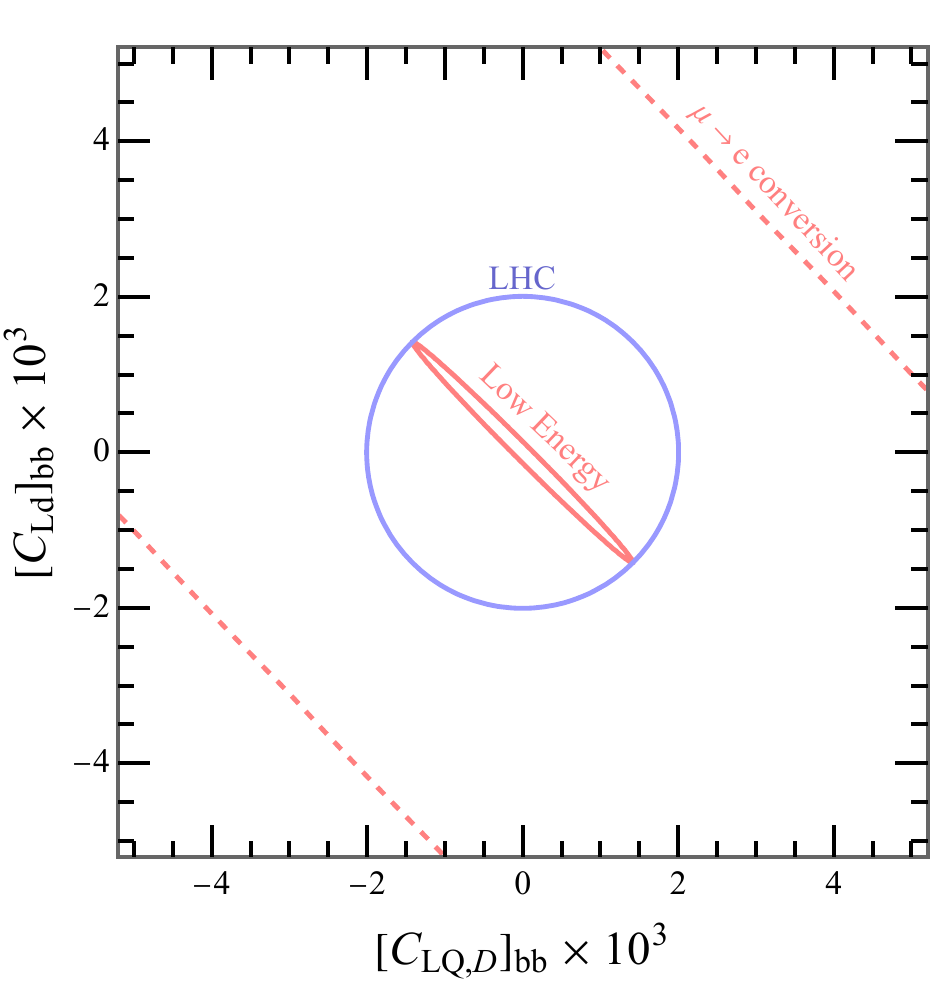}
\hspace{4ex}
\includegraphics[width=6.7cm]{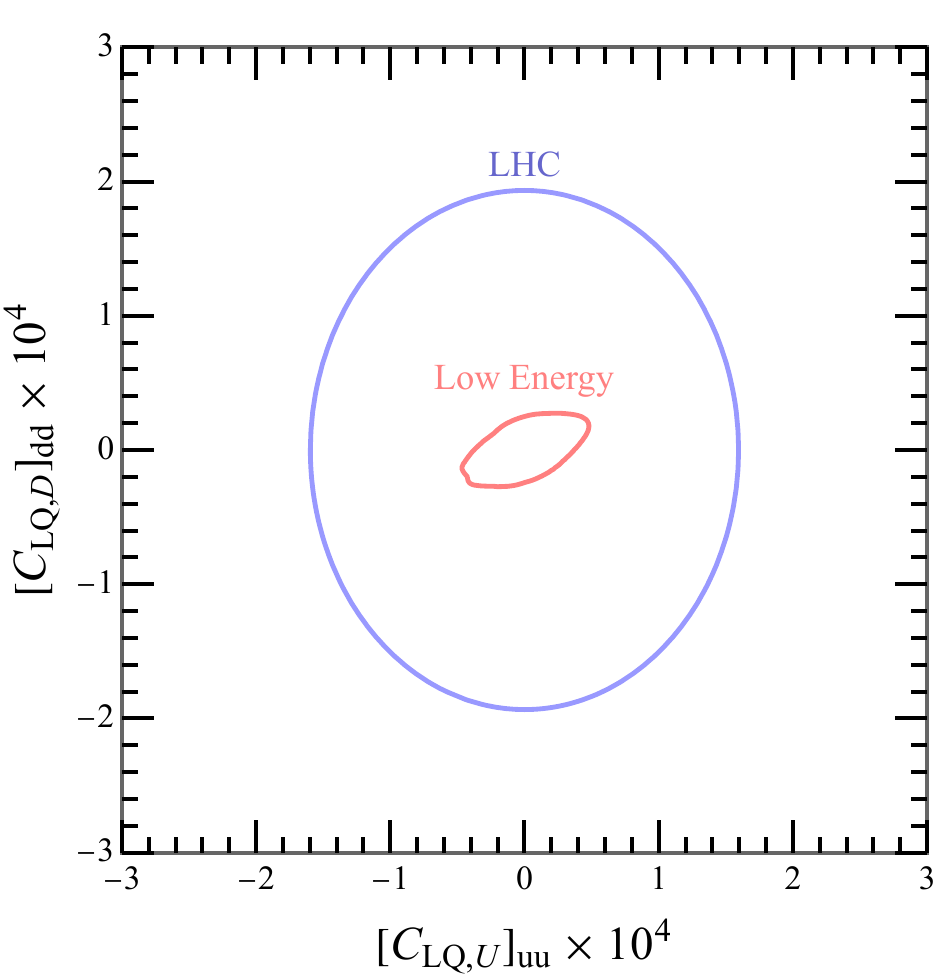}
\caption{Bounds on $[C_{LQ,D}]_{bb}$ and $[C_{Ld}]_{bb}$ (left) and $[C_{LQ,U}]_{uu}$ and $[C_{LQ,D}]_{dd}$ (right) at $90\%$ C.L. 
The region inside the pink lines is allowed by low-energy processes ($\mu N \to e N$ and $\mu\to 3e$), while the blue contour depicts the current LHC limits.}
\label{fig:U2flavor}
\end{figure}

The results in Section \ref{sec:single} confirm that, in a single coupling scenario, $\mu \rightarrow e \gamma$, $\mu \rightarrow e$ conversion in nuclei and meson decays provide, in most cases, the strongest constraints on CLFV new physics. In this section, we explore the robustness of these conclusions by turning on multiple couplings, as is typically the case in BSM models. 
We consider here a simplified scenario in which the SMEFT couplings satisfy an approximate $U(2)_Q\times U(2)_{u}\times U(2)_d$ flavor symmetry \cite{Faroughy:2020ina, Barbieri:2011ci, Barbieri:2012uh, Blankenburg:2012nx}, that can distinguish the first two generations of quark fields from the third one. Focusing on vector operators for simplicity, one can see that the exact symmetry allows the following operators 
\begin{align}
{\cal L}=&-\frac{4G_F}{\sqrt{2}}\Bigg\{C_{LQ,D}~\bar{e}_L\gamma^{\mu}\mu_L\bar{D}_L\gamma_{\mu}D_L + \left[C_{LQ,D}\right]_{bb}\bar{e}_L\gamma^{\mu}\mu_L\bar{b}_L\gamma_{\mu}b_L \nonumber\\
&\hspace{1.5cm}+C_{ed}~\bar{e}_R\gamma^{\mu}\mu_R \bar{D}_R\gamma_{\mu}D_R+\left[C_{ed}\right]_{bb}\bar{e}_R\gamma^{\mu}\mu_R\bar{b}_R\gamma_{\mu}b_R\nonumber\\
&\hspace{1.5cm}+C_{Ld}~\bar{e}_L\gamma^{\mu}\mu_L \bar{D}_R\gamma_{\mu}D_R+\left[C_{Ld}\right]_{bb}\bar{e}_L\gamma^{\mu}\mu_L\bar{b}_R\gamma_{\mu}b_R\nonumber\\
&\hspace{1.5cm}+C_{Qe}~\bar{e}_R\gamma^{\mu}\mu_R\bar{D}_L\gamma_{\mu}D_L
+\left[C_{Qe}\right]_{bb}\bar{e}_R\gamma^{\mu}\mu_R\bar{b}_L\gamma_{\mu}b_L\nonumber\\
&-\frac{4G_F}{\sqrt{2}}\Bigg\{ C_{LQ,U}~\bar{e}_L\gamma^{\mu}\mu_L\bar{U}_L\gamma_{\mu}U_L+ \left[C_{LQ,U}\right]_{tt}\bar{e}_L\gamma^{\mu}\mu_L\bar{t}_L\gamma_{\mu}t_L \nonumber\\
&\hspace{1.5cm}+C_{eu}~\bar{e}_R\gamma^{\mu}\mu_R \bar{U}_R\gamma_{\mu}U_R+\left[C_{eu}\right]_{tt}\bar{e}_R\gamma^{\mu}\mu_R\bar{t}_R\gamma_{\mu}t_R\nonumber\\
&\hspace{1.5cm}+C_{Lu}~\bar{e}_L\gamma^{\mu}\mu_L \bar{U}_R\gamma_{\mu}U_R+\left[C_{Ld}\right]_{tt}\bar{e}_L\gamma^{\mu}\mu_L\bar{t}_R\gamma_{\mu}t_R
\nonumber \\
 & \hspace{1.5cm}  
+C_{Qe}~\bar{e}_R\gamma^{\mu}\mu_R\bar{U}_L\gamma_{\mu}U_L
+\left[C_{Qe}\right]_{bb}\bar{e}_R\gamma^{\mu}\mu_R\bar{t}_L\gamma_{\mu}t_L
  \Bigg\},
 \label{U2_operators}
\end{align}
with $D_{L/R}=(d_{L/R},s_{L/R})^T$ and $U_{L/R}=(u_{L/R},c_{L/R})^T$. The Wilson coefficients without flavor indices are given by
\begin{align}
C_{LQ,D}&=\left[C_{LQ,D}\right]_{dd}=\left[C_{LQ,D}\right]_{ss},\\
C_{ed}&=\left[C_{ed}\right]_{dd}=\left[C_{ed}\right]_{ss},\\
C_{Ld}&=\left[C_{Ld}\right]_{dd}=\left[C_{Ld}\right]_{ss},\\
C_{Qe}&=\left[C_{Qe}\right]_{dd}=\left[C_{Qe}\right]_{ss},
\end{align}
and
\begin{align}
C_{LQ,U}&=\left[C_{LQ,U}\right]_{uu}=\left[C_{LQ,U}\right]_{cc},\\
C_{eu}&=\left[C_{eu}\right]_{uu}=\left[C_{eu}\right]_{cc}, \\
C_{Lu}&=\left[C_{Lu}\right]_{uu}=\left[C_{Lu}\right]_{cc}.
\end{align}
Based on this flavor symmetry, we consider multiple couplings fits to LFV couplings in the following two multi-operator scenarios:
\begin{center}
\begin{itemize}
\item[1.]  Down-type quark operator case~(8 couplings, {\it left} plot in Fig. \ref{fig:U2flavor}),\\
\vspace{-2ex}
\begin{align}
[C_{LQ,D}],~[C_{Ld}],~[C_{ed}],~[C_{Qe}],~[C_{LQ,D}]_{bb},~[C_{Ld}]_{bb},~[C_{ed}]_{bb},~[C_{Qe}]_{bb},
\end{align}
    \item[2.]  Light-quark operator case~(7 couplings, {\it right} plot in Fig. \ref{fig:U2flavor}),\\
    \vspace{-2ex}
\begin{align}
[C_{LQ,D}],~[C_{Ld}],~[C_{ed}],~[C_{Qe}],~[C_{LQ,U}],~[C_{Lu}],~[C_{eu}].
\end{align}
\end{itemize}
\end{center}
Figure \ref{fig:U2flavor} presents $90\%$ C.L. limits on two Wilson coefficients obtained by marginalizing over remaining couplings in each case. Here, the blue and pink contours are the latest LHC and low-energy bounds arising from $\mu\to e$ conversion and $\mu\to 3e$, respectively. Note that the region inside the contours is allowed. The left plot corresponds to the down-type operator case and presents the bounds on two bottom-quark operators. As can be seen from Table \ref{topRGE} and was also observed in the case of $\tau \rightarrow e$ transitions \cite{Cirigliano:2021img}, the axial component $[C_{LQ,D}-C_{Ld}]_{bb}$ contribution
to $\mu \rightarrow e$ conversion and $\mu \rightarrow 3 e$ is suppressed by at least a factor of ten compared to the vector. Coupled with the fact that we now have multiple vector couplings, this lead to a direction unconstrained by $\mu\to e$ conversion (pink dashed lines), which closes only when adding  $\mu\to 3e$. The 
resulting bound is very close to the current LHC sensitivity.
The right plot represents the case with light-quark operators. It is clearly seen that the bounds on $[C_{ LQ,U}]_{uu}$ and $[C_{LQ,D}]_{dd}$ are much weaker than those from single-operator analysis. This is because there are enough couplings to engineer cancellations in low-energy observables, dominantly $\mu\to e$ conversion. For example, if $C_{LQ,U}\sim -C_{Lu}$, their contributions to the SI $\mu\to e$ process are suppressed, while the axial-vector combination $C_{LQ,U} -C_{Lu}$ can still be constrained by the SD process. This results in much milder bounds as seen in Eq. (\ref{SD_AVuu}), in reach of the LHC. Although our analysis of $\mu\to e$ conversion focuses on the case with titanium targets, whose nuclear response functions are available, utilizing different
nuclear targets in conversion searches could help to further constrain multiple couplings as discussed in Refs. \cite{Haxton:2022piv, Davidson:2017nrp, Fuyuto:2024skf}. For example, naturally occurring  titanium is predominantly composed of the $^{48}$Ti isotope $(\sim 74\%)$ with nuclear spin zero. The isotopes $^{47}$Ti and $^{49}$Ti carry nonzero spin and their abundance are less than $10\%$, leading to a suppression of SD contributions.
On the other hand, the aluminum target that will be used in the next-generation experiments, Mu2e and COMET, has a $100\%$ natural abundance of the isotope $^{27}$Al with nuclear spin $J = 5/2$. This property would be helpful to investigate further bounds on the axial-vector couplings. 
The fits showed in Fig. \ref{fig:U2flavor} neglect the theoretical errors due to parton distributions and missing perturbative QCD higher order corrections, nucleon and nuclear matrix elements.
These can play a role, especially in assessing the discriminating power of $\mu \rightarrow e$ conversion with different targets \cite{Noel:2024led,Noel:2024swe}.  
Notice that  gauge invariance links the $e\mu$ components of the vector operators $C_{LQ, U}$, $C_{LQ, D}$,
$C_{Lu}$, $C_{Ld}$ and of the scalar and tensor operators to charged-current and neutrino operators \cite{Cirigliano:2021img}, which can be constrained by precision $\beta$ decays, meson charged-current decays, searches for rare processes such as $K \rightarrow \pi \nu \nu$ \cite{Cirigliano:2021img}, or searches for neutrino non-standard interactions \cite{Coloma:2024ict}. In a complete global analysis, it will be important to consider such constraints.

We finally comment on the complementarity between LHC and EIC. 
In the framework of the SMEFT, and neglecting small interference effects proportional to the quark and lepton masses, the CLFV Drell-Yan and DIS 
cross sections both take the form
\begin{equation}
    \sigma^{\rm DIS, DY}  = \sum  \hat{\sigma}^{\rm DIS, DY}_i |C_i|^2,  
\end{equation}
where the sum is extended over the set of SMEFT operators. Therefore, there can be no cancellation between operators with different chiralities or with different flavor content, and the one-at-a-time bounds in Tables \ref{limit_lhc}
 and \ref{tab:limit_eic}
 hold even in multiple coupling scenarios. If the SMEFT formalism is applicable at both the EIC and LHC, the latter will thus in general provide stronger bounds.  The limits from CLFV Drell-Yan, however, are dominated by high invariant mass bins, with $m_{\mu e} \sim 2-3$ TeV. These limits can be evaded in models with relatively light new physics which is not resonantly produced in Drell-Yan. An example could be leptoquarks that are exchanged in the $t$ channel in Drell-Yan but in $s$ channel at the EIC. The exchange of a $\sim 1$ TeV particle in the $t$-channel would suppress the cross section at high invariant mass  \cite{Cirigliano:2021img}, thus weakening the LHC bounds. A discovery of $e\rightarrow \mu$ transitions at the EIC would thus point to the interesting possibility of light BSM physics.


\section{Conclusions}\label{sec:conclusions}
We have analyzed experimental constraints on charged lepton flavor violating (CLFV) $\mu-e$ interactions from a variety of low- and high-energy searches, mainly $\mu\to e\gamma$, $\mu\to e$ conversion, $\mu \rightarrow 3e$, CLFV meson decays, the Large Hadron Collider (LHC), and the Electron-Ion Collider (EIC). We construct  $\mu-e$ operators with a generic quark flavor structure in the formalism of the Standard Model Effective Field Theory (SMEFT), which can be applied to any concrete UV models where CLFV interactions are induced by new particles heavier than electroweak scale. Low-energy probes can be categorized into two classes of observables depending on the quark-flavor structure: $\mu\to e$ transitions and meson decays. The former mainly consists of $\mu\to e\gamma$, $\mu\to e$ conversion, and  $\mu\to 3e$, and receive contributions from quark-flavor conserving interactions,  while the latter includes CLFV $K,D,B$ meson decays and top quark decays. In the case of quark-flavor conserving operators, we consider scale running effects based on the SMEFT renormalization group equations. Our main findings for the low-energy probes are:
\begin{itemize}
    \item The photon dipole operator receives the most severe constraint, $\Gamma^e_{\gamma}<6.0\times 10^{-11}$, from $\mu\to e \gamma$, which can also strongly constrain top-quark scalar and tensor operators, $\left[C_{LeQu}^{(1)}\right]_{tt}<9.3\times 10^{-8}$ and $\left[C_{LeQu}^{(3)}\right]_{tt}<1.3\times 10^{-10}$, due to their large mixing with the photon dipole.  
    \item $\mu \rightarrow e \gamma$ also provides the strongest constraint on $Z$ dipoles and CLFV Yukawa interactions,  
    while vector and axial CLFV $Z$ couplings are constrained by $\mu \rightarrow e$ conversion.
    \item Four-fermion operators are bounded by $\mu\to e$ conversion searches, even in the case of non-valence $s$, $c$, $b$ and $t$ quarks. For these, contributions to $\mu \rightarrow e$ arise because of operator mixing and renormalization group evolution. 
    Strange-quark vector operators are well constrained 
    via their mixing with $u$ and $d$ operators, even though their direct contribution to $\mu \rightarrow e$ is suppressed by the small strange nucleon axial and vector form factors.
The same apply to $c$, $b$ and $t$ quarks.
    \item Four-fermion operators with quark-flavor violation are constrained by $D$ meson decays for the $uc/cu$ components and $K$ and $B$ meson decays for the $ds/sd$, $db/bd$, and $sb/bs$ components. Among them, the $K_L \to e^{\mp}\mu^{\pm}$ channel  gives the most stringent limit $\left[C_{LedQ}\right]_{ds/sd}<1.3\times 10^{-8}$. 
\end{itemize}
The LHC has also been conducting searches for the CLFV decays of the Higgs, $Z$ boson, and top quark, and for the production of  $e\mu$ pairs in $pp$ collisions. We translate upper bounds on branching ratios of CLFV decays into bounds on CLFV operators, while for the Drell-Yan process we recast the experimental analysis and obtain bounds on SMEFT semileptonic four-fermion operators. In addition to the existing searches at the LHC, we also explored the sensitivity of the future EIC to CLFV deep inelastic scattering. 
We find that currently only the search for CLFV top decays can constrain quark-flavor-changing four-fermion operators involving top quarks, resulting in a bound at the level of $10^{-(2-3)}$. 
The limit on vector operators with $uc/cu$ indices by the Drell-Yan process is roughly a factor of 4 stronger than that from $D^+\to\pi^+e^{\pm}\mu^{\mp}$. For all remaining operators, in the simplified scenario of single operator dominance, LHC bounds are weaker than either $\mu \rightarrow e$ transitions or $K$ and $B$ meson decays. LHC bounds are however usually stronger than the projected EIC sensitivity, in the assumption that the SMEFT framework can be adopted at both colliders. 

Finally we analyzed multi-operator scenarios, which are naturally realized in various BSM models. We assumed the SMEFT operators to obey a $U(2)_Q\times U(2)_u\times U(2)_d$ flavor symmetry, and further divided the operators into two groups with 8 or 7 operators, due to the complexity of handling a ${\cal O}(10)$ dimensional parameter space. The analysis demonstrated that in each case it is easy to evade the strongest bounds from spin-independent $\mu \rightarrow e$ conversion, and that the bounds from spin dependent 
$\mu \rightarrow e$ or $\mu \rightarrow 3e$ become comparable to the LHC reach. This illustrates the importance of considering multiple probes of CLFV interactions.

\acknowledgments
We thank Evan Rule for helpful discussions on numerical calculations of nuclear response functions and for comments on the manuscript. We thank B.~Yan for his help on the analysis of the EIC sensitivity, and acknowledge fruitful discussions with V.~Cirigliano, M.~Hoferichter, C.~Lee, S.~Mantry and D.~Pefkou. 
F.~D. acknowledges the support of the Los Alamos National Laboratory Student Program Office and  Undergraduate Student Internship Program. 
F.~D, K.~F and E.~M. were supported by the US Department of Energy Office and by the Laboratory Directed Research and Development (LDRD) program of Los Alamos National Laboratory under project numbers 20250164ER and 20210190ER. Los Alamos National Laboratory is operated by Triad National Security, LLC, for the National Nuclear Security Administration of the U.S. Department of Energy (Contract No. 89233218CNA000001).
The work of S.~G-S.~is supported by MICIU/AEI/10.13039/501100011033 and 
by FEDER UE through grants PID2023-147112NB-C21; and 
through the ``Unit of Excellence Mar\'ia de Maeztu 2020-202'' 
award to the Institute of Cosmos Sciences, grant CEX2019-000918-M. 
Additional support is provided by the Generalitat de Catalunya (AGAUR) 
through grant 2021SGR01095. 
S.~G-S.~is a Serra H\'{u}nter Fellow. 

\appendix

\section{SMEFT and LEFT bases}\label{app:basis}

\subsection{SMEFT basis}

We follow here the conventions of Ref. \cite{Cirigliano:2021img}, which adopts a notation similar to the standard SMEFT notation \cite{Grzadkowski:2010es}, with some normalization factors inserted so to have dimensionless coefficients.
The theory contains the right-handed quarks, $u_R$ and $d_R$, and charged leptons, $e_R$, which are singlets under $SU(2)_L$, while the  
left-handed quarks and leptons and the scalar field $\varphi$
transform as doublets under $SU(2)_L$
\begin{equation}
q_L  = \Biggl( \begin{array}{c}
u_L \\
d_L
\end{array}
\Biggr), \qquad \ell_L = \Biggl( \begin{array}{c}
\nu_L \\
e_L
\end{array}
\Biggr), \qquad
\varphi = \frac{v}{\sqrt{2}} U(x) \Biggl( 
\begin{array}{c}
0 \\
1 +  \frac{h}{v}
\end{array}\Biggr),
\end{equation}
where the last expression holds in the unitary gauge, 
$v=246$ GeV is the scalar vacuum expectation value (vev), $h$ is the physical Higgs field and $U(x)$ is a unitary matrix that encodes the Goldstone bosons. $\tilde \vp$ denotes the combination $\tilde \vp = i\tau_2 \vp^*$. 
The gauge covariant derivative is given by
\begin{equation}
D_\mu =  \partial_\mu + i g_1 {\rm y} B_\mu + i  \frac{g_2}{2} \tau^I  W^I_\mu   + i g_s G^a_\mu t^a  
\end{equation}
where $B_\mu$, $W^I_{\mu}$ and $G^a_{\mu}$ are the $U(1)_Y$, $SU(2)_L$ and $SU(3)_c$ gauge fields, respectively, and $g_1$, $g_2$, and $g_s$ are their gauge couplings. Furthermore, 
$\tau^I/2$ and $t^a$ are the $SU(2)_L$ and $SU(3)_c$ generators, in the representation of the field on which the derivative acts.
In the SM, the gauge couplings $g_1$ and $g_2$ are related to the electric charge and the Weinberg angle by $g_2 s_w = g_1 c_w = e$, where $e > 0$ is the charge of the positron
and $s_w = \sin\theta_W$, $c_w = \cos\theta_W$. SMEFT corrections to these relations are subleading for the processes considered here, which have no SM background.

At dimension-6 in the SMEFT, there are two kind of LFV operators, lepton bilinears and four-fermion operators. The bilinears are given by 
\begin{eqnarray}
\mathcal L_{\psi^2 \varphi^2 D} &=&   -  \frac{\varphi^{\dagger} i \DLR_{\mu} \varphi}{v^2} \, \left(   \bar \ell_L   \gamma^{\mu} \, c^{(1)}_{L\varphi} \ell_L +   
  \bar e_R  \gamma^{\mu}\, c_{e\varphi}  e_R\right)  -  \frac{ \varphi^{\dagger}  i \DLR^I_{\mu} \varphi}{v^2}\,  \bar \ell_L \tau^I  \gamma^{\mu} c^{(3)}_{L\varphi} \ell_L, \label{eq:Z}\\
\mathcal L_{\psi^2 X \varphi} &=&  -\frac{1}{\sqrt{2}}\bar \ell_L \sigma^{\mu\nu}(g_1\Gamma_{B}^e B_{\mu\nu}+g_2\Gamma_{W}^e {\tau}^I  {W}^I_{\mu\nu})\frac{\varphi}{v^2}  e_R + \mathrm{h.c.}\ , \label{eq:dipole}\\
\mathcal L_{\psi^2 \varphi^3} &=& - \sqrt{2} \frac{\varphi^{\dagger} \varphi}{v^2} \bar \ell_L Y^\prime_e \varphi e_R  + \textrm{h.c.}, \label{eq:Y}
\end{eqnarray}
where $\DLR_{\mu}=  D_\mu-\DL_\mu$, $\DLR^I_{\mu}= \tau^I D_\mu-\DL_\mu \tau^I$.\footnote{Here, $\varphi^{\dagger}\overleftarrow D_{\mu}\varphi \equiv \left(D_{\mu}\varphi \right)^{\dagger}\varphi$.}
The couplings $c_{L\vp}^{(1)}$, $c_{L\vp}^{(3)}$, $c_{e\vp}$ are hermitian, 3 $\times$ 3 matrices in lepton-flavor space.
$\Gamma_W^{e}$ and $\Gamma_B^{e}$ in \eq{dipole} are generic $3\times 3$ matrices in flavor space,  which we find convenient to trade for dipole couplings to the $Z$ and photon field
\begin{equation}
\label{eq:gZdipole}
\Gamma^e_\gamma = \Gamma^e_B  - \Gamma^e_W, \qquad  \Gamma^e_Z = -c_w^2 \Gamma^e_W - s_w^2 \Gamma^{e}_B.
\end{equation}
$Y_e^\prime$ in \eqref{eq:Y} denotes a dimension-six correction to the Yukawa coupling, to be added to the dimension-four SM Yukawa
\begin{equation}
\mathcal L_{\psi^2 \varphi} = - \sqrt{2}  \bar \ell_L Y^{(0)}_e \varphi e_R + \textrm{h.c.}
\end{equation}
We use the same normalization for quark Yukawa interactions.
After the Higgs gets its vev, we can write
\begin{equation}
\label{eq:hYukawa}
\mathcal L_{\rm yuk} = -  v \bar e_L Y_e  e_R  \left(1 + \frac{h}{v}\right) - \bar e_L Y_e^\prime e_R h + \ldots + \textrm{h.c.}, \qquad Y_e = Y_e^{(0)} + \frac{1}{2} Y_e^\prime,
\end{equation}
where the dots denote higher-order terms in $h$.
We diagonalize the first term, so that the charged lepton masses are given by $M_e = v Y_e$. $Y_e^\prime$ is, in general,  off-diagonal.
For both quark and lepton SM Yukawa couplings we will use the convention $M_f = v Y_f$.

$\mathcal L_{\psi^4}$ includes four-fermion operators. The most relevant for collider searches are semileptonic four-fermion operators,
\begin{align}\label{eq:fourfermion}
&\mathcal L_{\psi^4} = -\frac{4 G_F}{\sqrt{2}}  \bigg\{ 
C^{(1)}_{LQ}\, \bar \ell_L \gamma^\mu \ell_L \, \bar q_L \gamma_\mu q_L + C^{(3)}_{LQ} \, \bar \ell_L \tau^I \gamma^\mu \ell_L \, \bar q_L  \tau^I \gamma_\mu q_L \\
& \qquad\qquad\qquad + C_{eu} \, \bar e_R \gamma^\mu e_R \, \bar u_R \gamma_\mu u_R  +\
 C_{ed} \, \bar e_R \gamma^\mu e_R \, \bar d_R \gamma_\mu d_R \nn \\
 & \qquad\qquad\qquad + C_{Lu}\, \bar \ell_L \gamma^\mu \ell_L \, \bar u_R \gamma_\mu u_R +  C_{Ld}\, \bar \ell_L \gamma^\mu \ell_L \, \bar d_R \gamma_\mu d_R  +\ C_{Qe}  \, \bar e_R \gamma^\mu e_R \, \bar q_L \gamma_\mu q_L   \bigg\}  \nn \\
&\quad -\frac{4 G_F}{\sqrt{2}} \bigg\{  C_{LedQ}\, \bar \ell^i_L e_R\, \bar d_R q_L^i + C^{(1)}_{LeQu}\, \varepsilon^{ij} \bar \ell^i_L e_R\, \bar q_L^j u_R +  C^{(3)}_{LeQu}\, \varepsilon^{ij} \bar \ell^i_L \sigma^{\mu \nu} e_R\, \bar q_L^j \sigma_{\mu \nu} u_R \, + \textrm{h.c.}
  \bigg\}. \nonumber
\end{align}
Here, $i,j$ represent $SU(2)_L$ indices.
The couplings are, in general, four-index tensors in flavor. We allow the operators to have a generic structure in quark flavor. We follow the flavor conventions of Ref.~\cite{Alioli:2018ljm} and assign operator labels to the neutral current components
with charged leptons, after rotating to the $u$ and $d$ quark mass basis. 
For most operators, the rotation to the mass basis leads to an inconsequential relabelling of the Wilson coefficients. The exceptions are the operators $C_{LQ}^{(1,3)}$
and $C_{Qe}$. For $C_{LQ}^{(1,3)}$, the rotation induces  factors of the SM CKM matrix $V_{\rm CKM}$ in the charged-current and in the neutral current neutrino components. Introducing 
\begin{equation}
C_{LQ, U} = \left(U^u\right)_L^{\dagger} \left( C_{LQ}^{(1)} - C_{LQ}^{(3)} \right) U_L^{u}, \qquad  C_{LQ, D} = \left(U^d_L\right)^{\dagger} \left( C_{LQ}^{(1)} + C_{LQ}^{(3)} \right) U^d_L,
\end{equation}
where $U_{L,R}^{u,d}$ are unitary matrices that diagonalize the quark mass matrices, the first two terms in the four-fermion Lagrangian \eq{fourfermion} become
\begin{eqnarray}
\mathcal L &=&
-\frac{4 G_F}{\sqrt{2}}  \bigg\{ 
   \left[ C^{}_{LQ, U} \right]_{p r s t}\, \bar e^p_L \gamma^\mu e^r_L \, \bar u^s_L \gamma_\mu u^t_L 
+  \left[C^{}_{LQ, D} \right]_{pr st}\, \bar e^p_L \gamma^\mu e^r_L \, \bar d^s_L \gamma_\mu d^t_L \nonumber \\
& & +\left[ V_{\rm CKM} C_{LQ, D} V^{\dagger}_{\rm CKM } \right]_{p r s t} \, \bar \nu^{p}_L \gamma^\mu \nu^r_L \, \bar u^s_L \gamma_\mu u^t_L 
+ \left[V_{\rm CKM}^{\dagger} C^{}_{LQ, U} V_{\rm CKM} \right]_{p r s t}\, \bar \nu^p_L \gamma^\mu \nu^r_L \, \bar d^s_L \gamma_\mu d^t_L  \nonumber\\
& &+ \left( \left[C_{LQ, D} V^{\dagger}_{\rm CKM} - V_{\rm CKM}^{\dagger} C^{}_{LQ, U} \right]_{p r s t} \bar \nu^p_L \gamma^\mu e^r_L  \bar d^s_L \gamma_\mu u^t_L  + \textrm{h.c.}\right)
\bigg\},
\end{eqnarray}
where $p$, $r$, $s$, $t$ are flavor indices in the quark/lepton mass bases.
For $C_{Qe}$, we define it as
\begin{equation}\label{CQe}
\mathcal L = - \frac{4 G_F}{\sqrt{2}}    \, \bar e_R \gamma^\mu e_R \, \left(  \bar d_L C_{Qe} \gamma_\mu d_L +  \bar u_L V_{\rm CKM} C_{Qe} V^{\dagger}_{\rm CKM}  \gamma_\mu u_L \right) .
\end{equation}
The renormalization group evolution of the operators in Eq. \eqref{eq:fourfermion} also induces  purely leptonic operators
\begin{equation}\label{eq:fourleptons}
\mathcal L_{\psi^4} = -\frac{4 G_F}{\sqrt{2}}  \bigg\{ 
C^{}_{LL}\, \bar \ell_L \gamma^\mu \ell_L \, \bar \ell_L \gamma_\mu \ell_L  + C_{ee} \, \bar e_R \gamma^\mu e_R \, \bar e_R \gamma_\mu e_R  
 + C_{Le}\, \bar \ell_L \gamma^\mu \ell_L \, \bar e_R \gamma_\mu e_R  \bigg\}.
\end{equation}

\subsection{LEFT basis}

Below the electroweak scale,
we map the SMEFT LFV operators onto 
a low-energy $SU(3)_c \times U(1)_{\rm em}$ EFT (LEFT). The matching can be done more immediately in the basis of Ref. \cite{Jenkins:2017jig,Jenkins:2017dyc,Dekens:2019ept}, from which we differ only in the fact that we factorize dimensionful parameters so that the Wilson coefficients of the LEFT operators become dimensionless.

At dimension five, we consider leptonic dipole operators \begin{equation}
\mathcal L_5 = - \frac{e}{2 v}  \bar e_{L}^{p} \, \sigma^{\mu \nu} \left[ \Gamma^{e}_\gamma\right]_{p r} {e_{R}^{r}} F_{\mu \nu}  + {\rm h.c.},
\end{equation}
where $p,r$ are leptonic flavor indices.

At dimension six, there are several semileptonic four-fermion operators.
Those relevant for direct LFV probes have two charged leptons. There are eight vector-type operators, denoted by the subscript $VIJ$, with $I$, $J \in \{L, R\}$, four scalar and two tensor operators, denoted by the subscripts  $S$ or $T$  
\begin{align}\label{eq:fourfermion2}
\mathcal L_{6} = -\frac{4 G_F}{\sqrt{2}} & \bigg( 
 C^{eu}_{\rm VLL}\, \bar e_L \gamma^\mu e_L \, \bar u_L \gamma_\mu u_L
+ C^{ed}_{\rm VLL}\, \bar e_L \gamma^\mu e_L \, \bar d_L \gamma_\mu d_L 
+ C^{eu}_{\rm VRR}\, \bar e_R \gamma^\mu e_R \, \bar u_R \gamma_\mu u_R \nn \\
&  + C^{ed}_{\rm VRR}\, \bar e_R \gamma^\mu e_R \, \bar d_R \gamma_\mu d_R 
+     C^{ue}_{\rm VLR}\, \bar e_R \gamma^\mu e_R \, \bar u_L \gamma_\mu u_L
+ C^{de}_{\rm VLR}\, \bar e_R \gamma^\mu e_R \, \bar d_L \gamma_\mu d_L \nonumber \\ 
&  + C^{eu}_{\rm VLR}\, \bar e_L \gamma^\mu e_L \, \bar u_R \gamma_\mu u_R 
 + C^{ed}_{\rm VLR}\, \bar e_L \gamma^\mu e_L \, \bar d_R \gamma_\mu d_R 
\bigg) \nonumber \\
  - \frac{4 G_F}{\sqrt{2}}  & \bigg( 
  C^{eu}_{\rm SRR}\, \bar e_L  e_R \, \bar u_L  u_R
+ C^{ed}_{\rm SRR}\, \bar e_L  e_R \, \bar d_L  d_R 
+ C^{eu}_{\rm TRR}\, \bar e_L \sigma^{\mu\nu} e_R \, \bar u_L \sigma_{\mu \nu} u_R \nonumber \\
& + C^{ed}_{\rm TRR}\, \bar e_L \sigma^{\mu\nu} e_R \, \bar d_L \sigma_{\mu \nu} d_R 
+     C^{eu}_{\rm SRL}\, \bar e_L  e_R \, \bar u_R  u_L
+ C^{ed}_{\rm SRL}\, \bar e_L  e_R \, \bar d_R d_L  + \textrm{h.c.}
\bigg).
\end{align}
There are in addition four purely leptonic operators
\begin{align}
\label{eq:fourlepton}
{\cal L}_{6}=-\frac{4G_F}{\sqrt{2}} & \Bigl[C_{\rm VLL}^{ee}\bar{e}_L\gamma^{\mu}e_L\bar{e}_L\gamma_{\mu}e_L +C_{\rm VRR}^{ee}\bar{e}_R\gamma^{\mu}e_R\bar{e}_R\gamma_{\mu}e_R +C_{\rm VLR}^{ee}\bar{e}_L\gamma^{\mu}e_L\bar{e}_R\gamma_{\mu}e_R  \nonumber\\
&
+\bigl(C_{\rm SRR}^{ee}\bar{e}_Le_R\bar{e}_Le_R +{\rm h.c.} \bigr) \Bigr].
\end{align}
Finally, scalar operators with heavy quarks induce dimension-7 gluonic operators at low energy
\begin{align}
\label{eq:Lgluonic}
\mathcal L_g = C_{GG}  \frac{1}{v^3} \frac{\alpha_s}{4\pi} \left( G^a_{\mu \nu} G^{a \mu \nu} \right)   \bar e_L  e_R  & + C_{G\widetilde G} \frac{1}{v^3} \frac{\alpha_s}{4\pi} \left( G^a_{\mu \nu} \widetilde G^{a\,\mu \nu} \right)  \bar e_L  e_R   + \textrm{h.c.}.
\end{align}
where
$\widetilde G^{a\mu\nu} = \frac{1}{2} \epsilon^{\mu\nu\alpha\beta} G^{a}_{\alpha\beta}$.

\subsection{Tree level matching}
\label{sec:tree}

The coefficients of the operators in Eqs.~\eqref{eq:fourfermion2} and \eqref{eq:fourlepton} are not all independent, if one matches the SMEFT onto LEFT.
For example, the four-fermion contributions to the semileptonic vector operators with charged leptons in \eq{fourfermion2} are given by:
\begin{subequations}
\begin{eqnarray}
\Big[ C^{eu}_{\rm VLL} \Big]_{\mu e j i} &=& \Big[ C_{LQ, U}\Big]_{\mu e j i} + \delta_{i j} \Big[c^{(1)}_{L\varphi} + c^{(3)}_{L\varphi}\Big]_{\mu e} z_{u_L},  \label{eq:match1}\\
\Big[ C^{ed}_{\rm VLL} \Big]_{\mu e j i} &=& \Big[ C_{LQ, D}\Big]_{\mu e j i} + \delta_{i j} \Big[c^{(1)}_{L\varphi} + c^{(3)}_{L\varphi}\Big]_{\mu e} z_{d_L}, \\ 
\Big[ C^{eu}_{\rm VRR} \Big]_{\mu e j i} &=& \Big[ C_{eu}\Big]_{\mu e j i}    + \delta_{i j} \Big[c_{e\varphi}\Big]_{\mu e} z_{u_R}, \\  
\Big[ C^{ed}_{\rm VRR} \Big]_{\mu e j i} &=& \Big[ C_{ed}\Big]_{\mu e j i}    + \delta_{i j} \Big[c_{e\varphi}\Big]_{\mu e} z_{d_R}, \\
\Big[ C^{eu}_{\rm VLR} \Big]_{\mu e j i} &=& \Big[ C_{Lu}\Big]_{\mu e j i}    + \delta_{i j} \Big[c^{(1)}_{L\varphi} + c^{(3)}_{L\varphi}\Big]_{\mu e} z_{u_R}, \\
\Big[ C^{ed}_{\rm VLR} \Big]_{\mu e j i} &=& \Big[ C_{Ld}\Big]_{\mu e j i}    + \delta_{i j} \Big[c^{(1)}_{L\varphi} + c^{(3)}_{L\varphi}\Big]_{\mu e} z_{d_R}, \\  
\Big[ C^{ue}_{\rm VLR} \Big]_{\mu e j i} &=& \Big[ V_{\rm CKM} C_{Qe} V^{\dagger}_{\rm CKM}\Big]_{\mu e j i} + \delta_{i j} \Big[ c_{e\varphi}\Big]_{\mu e} z_{u_L},  \label{eq:cqe1}\\ 
\Big[ C^{de}_{\rm VLR} \Big]_{\mu e j i} &=& \Big[ C_{Qe} \Big]_{\mu e j i}   + \delta_{i j} \Big[ c_{e\varphi}\Big]_{\mu e} z_{d_L}. \label{eq:cqe2}
\end{eqnarray}
\end{subequations}
Here the SM couplings of the $Z$ bosons to fermions are given by
\begin{equation}
    z_{f_L} =  T_{3 f} - Q_f s_w^2, \qquad z_{f_R} =  - Q_f s_w^2.
\end{equation}
The coefficients of the leptonic operators in \eq{fourlepton} are given by:
\begin{subequations}
\begin{eqnarray}
\Big[C_{\rm VLL}^{ee} \Big]_{p r st}&=&   \left[C_{LL}\right]_{p r s t}   + 
\frac{z_{e_L}}{4} \left[ \left(c^{(1)}_{L\varphi}+c^{(3)}_{L\varphi} \right)_{pr}\delta_{st} + \left(c^{(1)}_{L\varphi}+c^{(3)}_{L\varphi} \right)_{p t}\delta_{sr} \right]  \\
& & + \frac{z_{e_L}}{4} \left[ \left(c^{(1)}_{L\varphi}+c^{(3)}_{L\varphi} \right)_{st}\delta_{pr} + \left(c^{(1)}_{L\varphi}+c^{(3)}_{L\varphi} \right)_{s r}\delta_{pt } \right] \,, \nn\\
\Big[C_{\rm VRR}^{ee} \Big]_{p r st}&=& \left[C_{ee}\right]_{p r s t} +   \frac{z_{e_R}}{4} \left[
\left(c_{e\varphi}\right)_{p r}\delta_{st} + \left(c_{e\varphi}\right)_{p t}\delta_{s r}  \right]  \\
& &  +   \frac{z_{e_R}}{4} \left[
\left(c_{e\varphi}\right)_{s t}\delta_{p r} + \left(c_{e\varphi}\right)_{s r}\delta_{p t}  \right]  \nn \\
\Big[C_{\rm VLR}^{ee} \Big]_{prst}&=& \left[C_{Le}\right]_{pr st}
+
z_{e_R}\left(c^{(1)}_{L\varphi}+c^{(3)}_{L\varphi}\right)_{pr}\delta_{st}+z_{e_L}\left(c_{e\varphi}\right)_{st}\delta_{pr}
,\\
\Big[C_{\rm SRR}^{ee} \Big]_{p r st}&=&- \frac{v^2}{2m^2_H}\left( \left(Y_e^{\prime} \right)_{pr }\left(Y_e \right)_{st}\delta_{st}
+  \left(Y_e^{\prime} \right)_{s t}\left(Y_e \right)_{p r}\delta_{pr}
\right).
\end{eqnarray}
\end{subequations}
The scalar and tensor operators, $C^{(1)}_{LeQu}$, $C^{(3)}_{LeQu}$ and $C^{}_{LedQ}$, 
and the LFV Yukawa $Y^\prime_e$ match onto scalar and tensor operators at low energy. In the neutral current sector one finds
\begin{subequations}
\begin{eqnarray}
\Big[ C^{eu}_{\rm SRR} \Big]_{\mu e j i} &=& - \Big[C^{(1)}_{LeQu}\Big]_{\mu e j i} - \delta_{i j} \frac{v^2}{2 m_H^2} \Big[Y^{\prime}_e\Big]_{\mu e} Y_u , \\
\Big[ C^{ed}_{\rm SRR} \Big]_{\mu e j i} &=& - \delta_{i j} \frac{v^2}{2 m_H^2} \Big[Y^{\prime}_e\Big]_{\mu e} Y_d, \\
\Big[ C^{eu}_{\rm SRL} \Big]_{\mu e j i} &=& - \delta_{i j} \frac{v^2}{2 m_H^2} \Big[Y^{\prime}_e\Big]_{\mu e} Y_u \\
\Big[ C^{ed}_{\rm SRL} \Big]_{\mu e j i} &=& + \Big[C_{LedQ}\Big]_{\mu e j i} - \delta_{i j} \frac{v^2}{2 m_H^2} \Big[Y^{\prime}_e\Big]_{\mu e} Y_d \\
\Big[ C^{eu}_{\rm TRR} \Big]_{\mu e j i} &=& - \Big[ C^{(3)}_{LeQu}\Big]_{\mu e j i}, \\
\Big[ C^{ed}_{\rm TRR} \Big]_{\mu e j i} &=& 0,
\end{eqnarray}
\end{subequations}
At the $b$ and $c$ thresholds, the scalar operators also induce corrections to the gluonic operators in \eq{Lgluonic}, yielding
\begin{subequations}
\begin{align}
\left[C_{GG} \right]_{\mu e} \!&= \frac{1}{3} \! \sum_{q = b, c} \frac{v}{m_q} \!\left[ C^{eq}_{\rm SRR} + C^{eq}_{\rm SRL} \right]_{\!\mu e qq}, \
\left[C_{GG} \right]_{e \mu}\! =  \frac{1}{3}\! \sum_{q = b, c}   \frac{v}{m_q} \!\left[ C^{eq}_{\rm SRR} + C^{eq}_{\rm SRL} \right]_{\!e \mu qq}\label{matching_GGb}\\
\left[C_{G\widetilde G} \right]_{\!\mu e} \!&= \frac{i}{2} \!\sum_{q = b, c}  \frac{v}{m_q} \!\left[  C^{eq}_{\rm SRR} - C^{eq}_{\rm SRL} \right]_{\!\mu e qq}, \
\left[C_{G\widetilde G} \right]_{e \mu} \!=  \frac{i}{2} \!\sum_{q = b, c}  \frac{v}{m_q}\! \left[  C^{eq}_{\rm SRR} -C^{eq}_{\rm SRL} \right]_{\! e \mu qq}  \label{matching_GGtildeb}
\end{align}
\end{subequations}

\section{$\mu \rightarrow e$ conversion in nuclei}\label{App:mutoe}

{\renewcommand{\arraystretch}{1.3}\begin{table}[t]
\centering
\begin{tabular}{|c|| c c c c c c c |}
\hline
{\rm Target} & $W_{MM}^{00}$ & $W_{MM}^{11}$ & $W_{MM}^{01, 10}$  & $W_{\Sigma^{\prime\prime}\Sigma^{\prime\prime}}^{00}$  & $W_{\Sigma^{\prime\prime}\Sigma^{\prime\prime}}^{11}$ &  $W_{\Sigma^{\prime\prime}\Sigma^{\prime\prime}}^{01, 10}$ &  \\
\hline\hline
{\rm Al} & $2.6\cdot 10^2$ & $0.28$ & $-8.24$ & $0.11$ & $9.2\cdot 10^{-2}$ & $0.1$ & \\
{\rm Ti} & $5.2\cdot 10^2$ & $1.6$& $-28.8$ & $8.8\cdot 10^{-3}$ & $7.1\cdot 10^{-3}$ & $-7.9\cdot 10^{-3}$ & \\
\hline 
\hline
{\rm Target} & $W_{\Sigma^{\prime}\Sigma^{\prime}}^{00}$ & $W_{\Sigma^{\prime}\Sigma^{\prime}}^{11}$  & $W_{\Sigma^{\prime}\Sigma^{\prime}}^{01,10}$ & $W_{\Phi^{\prime\prime}M}^{00}$ & $W_{\Phi^{\prime\prime}M}^{11}$ & $W_{\Phi^{\prime\prime}M}^{01}$ & $W_{\Phi^{\prime\prime}M}^{10}$ \\
\hline\hline
{\rm Al} & $9.1\cdot 10^{-2}$ & $5.4\cdot 10^{-2}$ & $7.2\cdot 10^{-2}$ & $1.1\cdot 10^2$ & $0.17$ & $-0.41$  & $-0.82$ \\
{\rm Ti} & $3.7\cdot 10^{-3}$ & $2.5\cdot 10^{-3}$ & $-3.1\cdot 10^{-3}$ &  $1.6\cdot 10^2$ & $4.5$ & $-1.1$ & $-9.4$\\
\hline 
\end{tabular}
\caption{Nuclear response functions for Al and Ti taken from \cite{Haxton:2022piv} and calculated using the Mathematica script \cite{MathematicaScript}.}
\label{Table:response}
\end{table}}


In this appendix, we provide more details on the calculation of the $\mu \rightarrow e$ branching ratio. Following the formalism of Refs. \cite{Haxton:2022piv, Haxton:2024lyc},
the conversion rate is given by \cite{Haxton:2022piv, Haxton:2024lyc}
\begin{align}
\Gamma(\mu\to e)&=~\frac{(\alpha_{\rm em}Z_{\rm eff})^3}{2\pi^2}\frac{q^2_{\rm eff}}{1+q/M_T}\left(\frac{m_{\mu}M_T}{m_{\mu}+M_T} \right)^3\nonumber\\
&\times\sum_{\tau,\tau^{\prime}=0,1} \Bigg[R^{\tau\tau^{\prime}}_{MM}W^{\tau\tau^{\prime}}_{MM} + R^{\tau\tau^{\prime}}_{\Sigma^{\prime\prime}\Sigma^{\prime\prime}}W^{\tau\tau^{\prime}}_{\Sigma^{\prime\prime}\Sigma^{\prime\prime}}
+ R^{\tau\tau^{\prime}}_{\Sigma^{\prime}\Sigma^{\prime}}W^{\tau\tau^{\prime}}_{\Sigma^{\prime}\Sigma^{\prime}}-\frac{2q_{\rm eff}}{m_N} R^{\tau\tau^{\prime}}_{\Phi^{\prime\prime}M}W^{\tau\tau^{\prime}}_{\Phi^{\prime\prime}M} \Bigg], \label{rate_mutoe}
\end{align}
where the first term with the nuclear response function $W_{MM}^{\tau\tau^{\prime}}$ in the square brackets corresponds to the SI contribution while the second $(W_{\Sigma^{\prime\prime}\Sigma^{\prime\prime}}^{\tau\tau^{\prime}})$ and third $(W_{\Sigma^{\prime}\Sigma^{\prime}}^{\tau\tau^{\prime}})$ terms originate from SD processes. The last term with $W^{\tau\tau^{\prime}}_{\Phi^{\prime\prime}M}$ is the interference of the longitudinal projection of the nuclear spin-velocity current with the charge operator in the multipole expansion. The values of these response functions for Al and Ti are shown in Table \ref{Table:response}.
$R_{XX}^{\tau\tau^{\prime}}$ depends on couplings $c_i$s
\begin{align}
R^{\tau\tau^{\prime}}_{MM}&=c_1^{\tau}c_1^{\tau^{\prime}*}+c_{11}^{\tau}c_{11}^{\tau^{\prime}*},\\
R^{\tau\tau^{\prime}}_{\Sigma^{\prime\prime}\Sigma^{\prime\prime}}&=(c_4^{\tau}-c_6^{\tau})(c_4^{\tau^{\prime}*}-c_6^{\tau^{\prime}*})+c_{10}^{\tau}c_{10}^{\tau^{\prime}*},\\
R^{\tau\tau^{\prime}}_{\Sigma^{\prime}\Sigma^{\prime}}&=c_4^{\tau}c_4^{\tau^{\prime}*}+c_9^{\tau}c_{9}^{\tau^{\prime}*},\\
R^{\tau\tau^{\prime}}_{\Phi^{\prime\prime}M}&={\rm Re}\left[c^{\tau *}_3c^{\tau^{\prime}}_1+ic^{\tau *}_{12}c^{\tau^{\prime}}_{11}  \right].
\end{align}
{\renewcommand{\arraystretch}{1.3}\begin{table}[t]
\centering
\begin{tabular}{|c|| c c c c c  |}
\hline
{\rm Target} & $Z_{\rm eff}$ & $q_{\rm eff}$ [MeV] & $q$ [MeV] & $M_T$ [MeV] & $\Gamma_{\rm cap}$~[GeV]   \\
\hline\hline
{\rm Al} & $11.3086$ & $110.81$ & $104.98$ & $25133.1$ & $0.46\cdot 10^{-18}$   \\
{\rm Ti} & $16.6562$ & $112.43$  & $104.28$ & $44587.6$  & $1.71\cdot 10^{-18}~$  \\
\hline 
\end{tabular}
\caption{Input parameters for decay rate of $\mu\to e $ conversion in Eq. (\ref{rate_mutoe}) and capture rate.}
\label{Table:input_mutoe}
\end{table}}
Here $\tau$ and $\tau^{\prime}$ represent isoscalar and isovector indices defined by $c^0_i=(c^p_i+c^n_i)/2$ and $c^1_i=(c^p_i-c^n_i)/2$. 
Note that we use a slightly different notation in $R^{\tau\tau^{\prime}}_{\Phi^{\prime\prime}M}$ from that in Ref. \cite{Haxton:2024lyc}. In particular, we use
\begin{equation}
c^{\tau}_{11} = \left[i c^{\tau *}_{11}\right]([32]),
\qquad 
c^{\tau}_{12} = \left[c^{\tau *}_{12}\right]([32]).
\end{equation}
The couplings for SI contributions are expressed by LEFT operators as follows:
\begin{align}
c_1^N=&~\frac{2\pi\alpha_{\rm em}}{qv}\left(\sum_{q=u,d,s}Q_qF^{q/N}_1 \right)\left[\left(\Gamma^{e*}_{\gamma} \right)_{e\mu} +\left(\Gamma^{e}_{\gamma} \right)_{\mu e} \right]\nonumber\\
&-\frac{G_F}{\sqrt{2}}\sum_{q=u,d,s}F^{q/N}_1\Bigg(C^{eq}_{\substack{\rm VLL \\ \mu e}}+C^{q e}_{\substack{\rm VLR \\ \mu e}}+C^{eq}_{\substack{\rm VRR \\ \mu e}}+C^{eq}_{\substack{\rm VLR \\ \mu e}} \Bigg) \nonumber\\
&-\frac{G_F}{\sqrt{2}}\sum_{q=u,d,s} \frac{1}{m_q} F^{q/N}_S \Bigg(C^{eq*}_{\substack{\rm SRR \\ e \mu}}+C^{eq}_{\substack{\rm SRR \\ \mu e}} +C^{eq*}_{\substack{\rm SRL \\ e\mu}} + C^{eq}_{\substack{\rm SRL \\ \mu e}} \Bigg) \nonumber\\
&+\frac{2G_F}{\sqrt{2}}\frac{q}{m_N}\sum_{q=u,d,s}\Bigg(\hat{F}^{q/N}_{T,0}-\hat{F}^{q/N}_{T,1}+ 4\hat{F}^{q/N}_{T,2} \Bigg)\Bigg(C^{eq*}_{\substack{{\rm TRR} \\ e\mu}}+ C^{eq}_{\substack{{\rm TRR} \\ \mu e}} \Bigg)\nonumber\\
&+\frac{3}{2v^3}F^N_G\Bigg(C^*_{\substack{GG\\ e\mu}}+C_{\substack{GG\\ \mu e}} \Bigg), 
\end{align}
\begin{align}
c_{11}^N=&~\frac{2\pi\alpha_{\rm em}}{qv}\left(\sum_{q=u,d,s}Q_qF^{q/N}_1 \right)\left[\left(\Gamma^{e*}_{\gamma} \right)_{e\mu} -\left(\Gamma^{e}_{\gamma} \right)_{\mu e} \right]\nonumber\\
&-\frac{G_F}{\sqrt{2}}\sum_{q=u,d,s}F^{q/N}_1\Bigg(C^{eq}_{\substack{\rm VLL \\ \mu e}}-C^{qe}_{\substack{\rm VLR \\ \mu e}}-C^{eq}_{\substack{\rm VRR \\ \mu e}}+C^{eq}_{\substack{\rm VLR \\ \mu e}} \Bigg) \nonumber\\
&-\frac{G_F}{\sqrt{2}}\sum_{q=u,d,s} \frac{1}{m_q} F^{q/N}_S \Bigg(C^{eq*}_{\substack{\rm SRR \\ e\mu}}-C^{eq }_{\substack{\rm SRR \\ \mu e}} +C^{eq *}_{\substack{\rm SRL \\ e\mu}} - C^{eq}_{\substack{\rm SRL \\ \mu e}} \Bigg) \nonumber\\
&+\frac{2G_F}{\sqrt{2}}\frac{q}{m_N}\sum_{q=u,d,s}\Bigg(\hat{F}^{q/N}_{T,0}-\hat{F}^{q/N}_{T,1}+ 4\hat{F}^{q/N}_{T,2} \Bigg)\Bigg(C^{eq *}_{\substack{{\rm TRR} \\ e\mu}}- C^{eq}_{\substack{{\rm TRR} \\ \mu e}} \Bigg)\nonumber\\
&+\frac{3}{2v^3}F^N_G\Bigg(C^*_{\substack{GG\\ e\mu}}-C_{\substack{GG\\ \mu e}} \Bigg).
\end{align}
It should be noted that tensor operators can contribute to coherent process with a suppression factor $\sim m_{\mu}/m_N$ arising from nonzero nucleon velocity \cite{Haxton:2022piv, Haxton:2024lyc}. The couplings for SD contributions are given by 
\begin{align}
c^N_{4}=&~\frac{G_F}{\sqrt{2}}\sum_{q=u,d,s}F^{q/N}_A\Bigg(C^{eq}_{\substack{{\rm VLL}\\ \mu e}}-C^{qe}_{\substack{{\rm VLR}\\ \mu e}} + C^{eq}_{\substack{{\rm VRR}\\ \mu e}} - C^{eq}_{\substack{{\rm VLR}\\ \mu e}} \Bigg)\nonumber\\
&-\frac{4G_F}{\sqrt{2}}\sum_{q=u,d,s}\hat{F}^{q/N}_{T,0}\Bigg(C^{eq *}_{\substack{{\rm TRR} \\ e\mu}} + C^{eq}_{\substack{{\rm TRR}\\ \mu e}} \Bigg),\\
c^N_6=&~\frac{qm_+}{4m^2_N}\frac{G_F}{\sqrt{2}}\sum_{q-u,d,s}F^{q/N}_{P^{\prime}}\Bigg(C^{eq}_{\substack{{\rm VLL}\\ \mu e}}-C^{qe}_{\substack{{\rm VLR}\\ \mu e}} + C^{eq}_{\substack{{\rm VRR}\\ \mu e}} - C^{eq}_{\substack{{\rm VLR}\\ \mu e}} \Bigg)\nonumber\\
&-\frac{q}{2m_N}\frac{G_F}{\sqrt{2}}\sum_{q=u,d,s}\frac{1}{m_q}F^{q/N}_{P}\Bigg(C^{eq*}_{\substack{{\rm SRR}\\ e\mu}}-C^{eq}_{\substack{{\rm SRL}\\ \mu e}} - C^{eq*}_{\substack{{\rm SRL}\\ e\mu}} + C^{eq}_{\substack{{\rm SRR}\\ \mu e}} \Bigg)\nonumber\\
&+i\frac{q}{2m_N}\frac{1}{v^3}F^N_{\tilde G}\Bigg(C^*_{\substack{G\tilde{G} \\ e \mu}}-C_{\substack{G\tilde{G} \\ \mu e}} \Bigg),\\
c^N_9=&-\frac{G_F}{\sqrt{2}}\sum_{q=u,d,s}F^{q/N}_A\Bigg(C^{eq}_{\substack{{\rm VLL}\\ \mu e}} +C^{qe}_{\substack{{\rm VLR}\\ \mu e}} -C^{eq}_{\substack{{\rm VRR}\\ \mu e}}-C^{eq}_{\substack{{\rm VLR}\\ \mu e}}\Bigg)\nonumber\\
&+\frac{4G_F}{\sqrt{2}}\sum_{q=u,d,s}\hat{F}^{q/N}_{T,0}\Bigg(C^{eq *}_{\substack{{\rm TRR} \\ e\mu}} - C^{eq}_{\substack{{\rm TRR}\\ \mu e}} \Bigg),\\
c^N_{10}=&~-i\frac{G_F}{\sqrt{2}}\sum_{q=u,d,s}\Bigg(F^{q/N}_A-\frac{qm_-}{4m^2_N}F^{q/N}_{P^{\prime}} \Bigg)\Bigg(C^{eq}_{\substack{{\rm VLL}\\ \mu e}}+C^{qe}_{\substack{{\rm VLR}\\ \mu e}} -C^{eq}_{\substack{{\rm VRR}\\ \mu e}}-C^{eq}_{\substack{{\rm VLR}\\ \mu e}} \Bigg)\nonumber\\
&-i\frac{q}{2m_N}\frac{G_F}{\sqrt{2}}\sum_{q=u,d,s}\frac{1}{m_q}F^{q/N}_P\Bigg(C^{eq*}_{\substack{{\rm SRR}\\ e\mu}}+C^{eq}_{\substack{{\rm SRL}\\ \mu e}} -C^{eq*}_{\substack{{\rm SRL}\\ e\mu}}-C^{eq}_{\substack{{\rm SRR}\\ \mu e}} \Bigg)\nonumber\\
&+i\frac{4G_F}{\sqrt{2}}\sum_{q=u,d,s}\hat{F}^{q/N}_{T,0}\Bigg(C^{eq *}_{\substack{{\rm TRR} \\ e\mu}} - C^{eq}_{\substack{{\rm TRR}\\ \mu e}} \Bigg)
-\frac{q}{2m_N}\frac{1}{v^3}F^N_{\tilde G}\Bigg(C^*_{\substack{G\tilde{G} \\ e\mu}}+ C_{\substack{G\tilde{G} \\ \mu e}} \Bigg),
\end{align}
with $m_{\pm}=m_{\mu}\pm m_e$. The above expressions include axial-vector, pseudoscalar, tensor and CP-odd gluonic currents while the interactions that contribute to SI process are omitted since their non-coherent contributions are suppressed by $m_{\mu}/m_N$ and nuclear response functions. Finally, for the interference terms, 
\begin{align}
c^N_3=&~\frac{4G_F}{\sqrt{2}}\sum_{q=u,d,s}\hat{F}^{q/N}_{T,0}\Bigg(C^{eq*}_{\substack{{\rm TRR}\\ e\mu}}+ C^{eq}_{\substack{{\rm TRR}\\ \mu e}} \Bigg),\\
c^N_{12}=&~i\frac{4G_F}{\sqrt{2}}\sum_{q=u,d,s}\hat{F}^{q/N}_{T,0}\Bigg(C^{eq*}_{\substack{{\rm TRR}\\ e\mu}}- C^{eq}_{\substack{{\rm TRR}\\ \mu e}} \Bigg).
\end{align}
Combining with nuclear response functions in Table \ref{Table:response}, one can naively see that the above tensor contributions can be the same order of magnitude as the one included in the SI contribution.  $F_i^{q/N}$ represents nucleon form factors and the values are given in Table \ref{Table:FF}. Since the momentum transfer in Al and Ti is close, we adopt those for Al in \cite{Haxton:2024lyc}. 
Notice that for the scalar form factors $F_S^{q/p}$ and $F_S^{q/n}$ we use the values at zero momentum transfer,
as there is some discrepancy between the momentum dependence derived in Refs. \cite{Haxton:2024lyc}
and \cite{Hoferichter:2016nvd}.
The branching ratio of $\mu\to e$ conversion is normalized by muon capture rate, ${\rm BR}(\mu\to e)\equiv \Gamma(\mu\to e)/\Gamma_{\rm capt}$, which is experimentally measured \cite{Suzuki:1987jf}.

{\renewcommand{\arraystretch}{1.3}
\begin{table}[t]
\centering
\begin{tabular}{|c || c c c c c c |}
\hline
{\rm Quark} ($q$) & $F_1^{q/N}$ &  $F_A^{q/N}$ & $F_{P^{\prime}}^{q/N}$ & $F^{q/p}_S$ [MeV] & $F^{q/n}_S$ [MeV] &  \\
\hline\hline
$u$ & $1.94$  & $0.82$ & $69$  &  $16.3$ & $14.5$  &\\
$d$ & $0.97$   & $-0.43$ & $-80$  & $30.6$ & $34.5$ &\\
$s$ & $1.1\cdot 10^{-4}$  & $-4.4\cdot 10^{-2}$  &  $-1.5$  & $43.3$ & $43.3$  &\\
\hline 
\hline
{\rm Quark} ($q$) & $F^{q/N}_P$ [GeV] & ${\hat F}^{q/N}_{T,0}$ & ${\hat F}^{q/N}_{T,1}$ & ${\hat F}^{q/N}_{T,2}$ & $F^N_G$ [MeV] & $F^N_{\tilde G}$ [MeV]   \\
\hline\hline
$u$ & $0.16$ & $0.78$   & $-2.8$  & $-0.08$   &  \multirow{3}{*}{$-48.7$} & \multirow{3}{*}{$-0.3\cdot 10^3$}  \\
$d$ & $-0.55$  & $-0.2$  & $0.9$ & $0.57$  &   & \\
$s$ & $-0.28$  & $-2.7\cdot 10^{-3}$  & $1.7\cdot 10^{-2}$  & $3.8\cdot 10^{-3}$   &   & \\
\hline 
\end{tabular}
\caption{Nucleon form factors where isospin limit of $F^{u,d,s/p}_i=F^{d,u,s/n}_i$ is taken except scalar form factors \cite{Haxton:2024lyc}.}
\label{Table:FF}
\end{table}}

\section{Exclusive $K,D$ and $B$ meson decays}\label{App:BRandFF}

In this appendix we provide the expressions for the observables used in our analysis of Section~\ref{sec:Flavor}.
The $\rm{BR}$ for the up-type LFV $D^{0}\to e\mu$ leptonic decay is given by~\cite{Cirigliano:2021img}:
\begin{eqnarray}
{\rm{BR}}(D^{0}\to\mu^{-}e^{+})&=&\tau_{D^{0}}\frac{G_{F}^{2}f_{D}^{2}}{16\pi m_{D^{0}}}\lambda^{1/2}\left(1,\frac{m_{e}^{2}}{m^{2}_{D^{0}}},\frac{m_{\mu}^{2}}{m^{2}_{D^{0}}}\right)\nonumber\\[1ex]
&\times&\Bigg[\left(m^{2}_{D^{0}}-(m_{e}+m_{\mu})^{2}\right)\left|(m_{\mu}-m_{e})A+\frac{m^{2}_{D^{0}}}{m_{c}+m_{u}}C\right|^{2}\nonumber\\[1ex]
&+&\left(m^{2}_{D^{0}}-(m_{e}-m_{\mu})^{2}\right)\left|(m_{\mu}+m_{e})B+\frac{m^{2}_{D^{0}}}{m_{c}+m_{u}}D\right|^{2}\Bigg]\,,
\label{Eq:BRleptonic}
\end{eqnarray}
where $\lambda(x,y,z)=x^{2}+y^{2}+z^{2}-2(xy+xz+yz)$, and\footnote{We have noted a typo in Ref.~\cite{Cirigliano:2021img} concerning the last line of Eq.~(\ref{Eq:CouplingsRedefinition}). 
The sign in front of the coupling $(C_{\rm{SRR}}^{eu*})_{e\mu uc}$ is ``+" rather than ``-".}
\begin{eqnarray}
\begin{aligned}
A&=(C_{\rm{VLR}}^{eu})_{\mu ecu}-(C_{\rm{VLL}}^{eu})_{\mu ecu}+(C_{\rm{VRR}}^{eu})_{\mu ecu}-(C_{\rm{VLR}}^{ue})_{cu\mu e}\,,\\[1ex]
B&=-(C_{\rm{VLR}}^{eu})_{\mu ecu}+(C_{\rm{VLL}}^{eu})_{\tau ecu}+(C_{\rm{VRR}}^{eu})_{\mu ecu}-(C_{\rm{VLR}}^{ue})_{cu\mu e}\,,\\[1ex]
C&=(C_{\rm{SRR}}^{eu})_{\mu ecu}-(C_{\rm{SRL}}^{eu})_{\mu ecu}+(C_{\rm{SRL}}^{eu*})_{e\mu uc}-(C_{\rm{SRR}}^{eu*})_{e\mu uc}\,,\\[1ex]
D&=(C_{\rm{SRR}}^{eu})_{\mu ecu}-(C_{\rm{SRL}}^{eu})_{\mu ecu}-(C_{\rm{SRL}}^{eu*})_{e\mu uc}+(C_{\rm{SRR}}^{eu*})_{e\mu uc}\,.\\[1ex]
\end{aligned}
\label{Eq:CouplingsRedefinition}
\end{eqnarray}
The input parameters relevant for the calculation of Eq.~(\ref{Eq:BRleptonic}) are taken from the PDG~\cite{Workman:2022ynf} with $f_{D}=209.0(2.4)$ MeV~\cite{FlavourLatticeAveragingGroupFLAG:2021npn}.
The description of the down-type LFV transitions $K_{L}^{0}\to e^{\pm}\mu^{\mp},B^{0}\to e^{\pm}\mu^{\mp},B_{s}\to e^{\pm}\mu^{\mp}$ is formally identical to the one above, with the replacements of the corresponding input parameters.

The up-type LFV $D^{+}\to\pi^{+}e\mu$ semileptonic decay distribution is given by:\footnote{We note that the decay distribution contains interference terms between vector- and scalar-like couplings, between vector and tensor couplings, and between the tensor coupling and its complex-conjugate that we do not show in Eq.~(\ref{Eq:BRSemileptonic}) due to their length.}
\begin{eqnarray}
&&\frac{d\Gamma(D^{+}\to\pi^{+}e^{+}\mu^{-})}{dq^{2}}=\frac{G_{F}^{2}}{768\pi^{3}m_{D}^{3}}\frac{\sqrt{\lambda(m_{e}^{2},m_{\mu}^{2},q^{2})}\sqrt{\lambda(m_{D}^{2},m_{\pi}^{2},q^{2})}}{q^{4}}\nonumber\\[1ex]
&\times&\bigg\{\left|\left(C_{\rm{VLL}}^{eu}+C_{\rm{VLR}}^{eu}-C_{\rm{VRR}}^{eu}\right)_{\mu ecu}-(C_{\rm{VLR}}^{ue})_{cu\mu e}\right|^{2}\nonumber\\[1ex]
&\times&\bigg[\frac{\lambda(m_{D}^{2},m_{\pi}^{2},q^{2})}{q^{2}}\left(q^{2}-m_{e}^{2}-m_{\mu}^{2}-2m_{e}m_{\mu}\right)\left(m_{e}^{2}+m_{\mu}^{2}-2m_{e}m_{\mu}+2q^{2}\right)\left|f_{+}(q^{2})\right|^{2}\nonumber\\[1ex]
&+&3\left(m_{e}+m_{\mu}\right)^{2}\left(q^{2}-m_{e}^{2}-m_{\mu}^{2}+2m_{e}m_{\mu}\right)\left(\frac{m_{D}^{2}-m_{\pi}^{2}}{q}\right)^{2}\left|f_{0}(q^{2})\right|^{2}\bigg]\nonumber\\[1ex]
&+&\left|\left(C_{\rm{VLL}}^{eu}+C_{\rm{VLR}}^{eu}+C_{\rm{VRR}}^{eu}\right)_{\mu ecu}+(C_{\rm{VLR}}^{ue})_{cu\mu e}\right|^{2}\nonumber\\[1ex]
&\times&\bigg[\frac{\lambda(m_{D}^{2},m_{\pi}^{2},q^{2})}{q^{2}}\left(q^{2}-m_{e}^{2}-m_{\mu}^{2}+2m_{e}m_{\mu}\right)\left(m_{e}^{2}+m_{\mu}^{2}+2m_{e}m_{\mu}+2q^{2}\right)\left|f_{+}(q^{2})\right|^{2}\nonumber\\[1ex]
&+&3\left(m_{e}-m_{\mu}\right)^{2}\left(q^{2}-m_{e}^{2}-m_{\mu}^{2}-2m_{e}m_{\mu}\right)\left(\frac{m_{D}^{2}-m_{\pi}^{2}}{q}\right)^{2}\left|f_{0}(q^{2})\right|^{2}\bigg]\nonumber\\[1ex]
&+&3q^{2}\left(\frac{m_{D}^{2}-m_{\pi}^{2}}{m_{c}}\right)^{2}\bigg[\left(q^{2}-(m_{e}-m_{\mu})^{2}\right)\left|\left(C_{\rm{SRL}}^{eu*}+C_{\rm{SRR}}^{eu*}\right)_{e\mu uc}-(C_{\rm{SRL}}^{eu}+C_{\rm{SRR}}^{eu})_{\mu ecu}\right|^{2}\nonumber\\[1ex]
&+&\left(q^{2}-(m_{e}+m_{\mu})^{2}\right)\left|\left(C_{\rm{SRL}}^{eu*}+C_{\rm{SRR}}^{eu*}\right)_{e\mu uc}+(C_{\rm{SRL}}^{eu}+C_{\rm{SRR}}^{eu})_{\mu ecu}\right|^{2}\bigg]\left|f_{0}(q^{2})\right|^{2}\nonumber\\[1ex]
&+&\frac{4\lambda(m_{D}^{2},m^{2}_{\pi},q^{2})}{(m_{D}+m_{\pi})^{2}q^{2}}\bigg[\left(q^{2}-(m_{e}-m_{\mu})^{2}\right)\left(2(m_{e}+m_{\mu})^{2}+q^{2}\right)\left|\left(C_{\rm{TRR}}^{eu}\right)_{\mu ecu}\right|^{2}\nonumber\\[1ex]
&+&\left(q^{2}-(m_{e}+m_{\mu})^{2}\right)\left(2(m_{e}-m_{\mu})^{2}+q^{2}\right)\left|\left(C_{\rm{TRR}}^{eu*}\right)_{e\mu uc}\right|^{2}\bigg]\left|f_{T}(q^{2})\right|^{2}\bigg\}\,,
\label{Eq:BRSemileptonic}
\end{eqnarray}
where $f_{+,0,T}(q^{2})$ are, respectively, the $D\to\pi$ vector, scalar and tensor form factors.

The expressions for the decays $K^{+}\to\pi^{+}e\mu,D_{s}\to K^{+}e\mu,B^{+}\to\pi^{+}e\mu,B^{+}\to K^{+} e\mu$ are similar to Eq.~(\ref{Eq:BRSemileptonic}) with the corresponding replacements of the input parameters and form factors.
In the following we provide the related form factors.

\begin{figure}
\centering\includegraphics[scale=0.35]{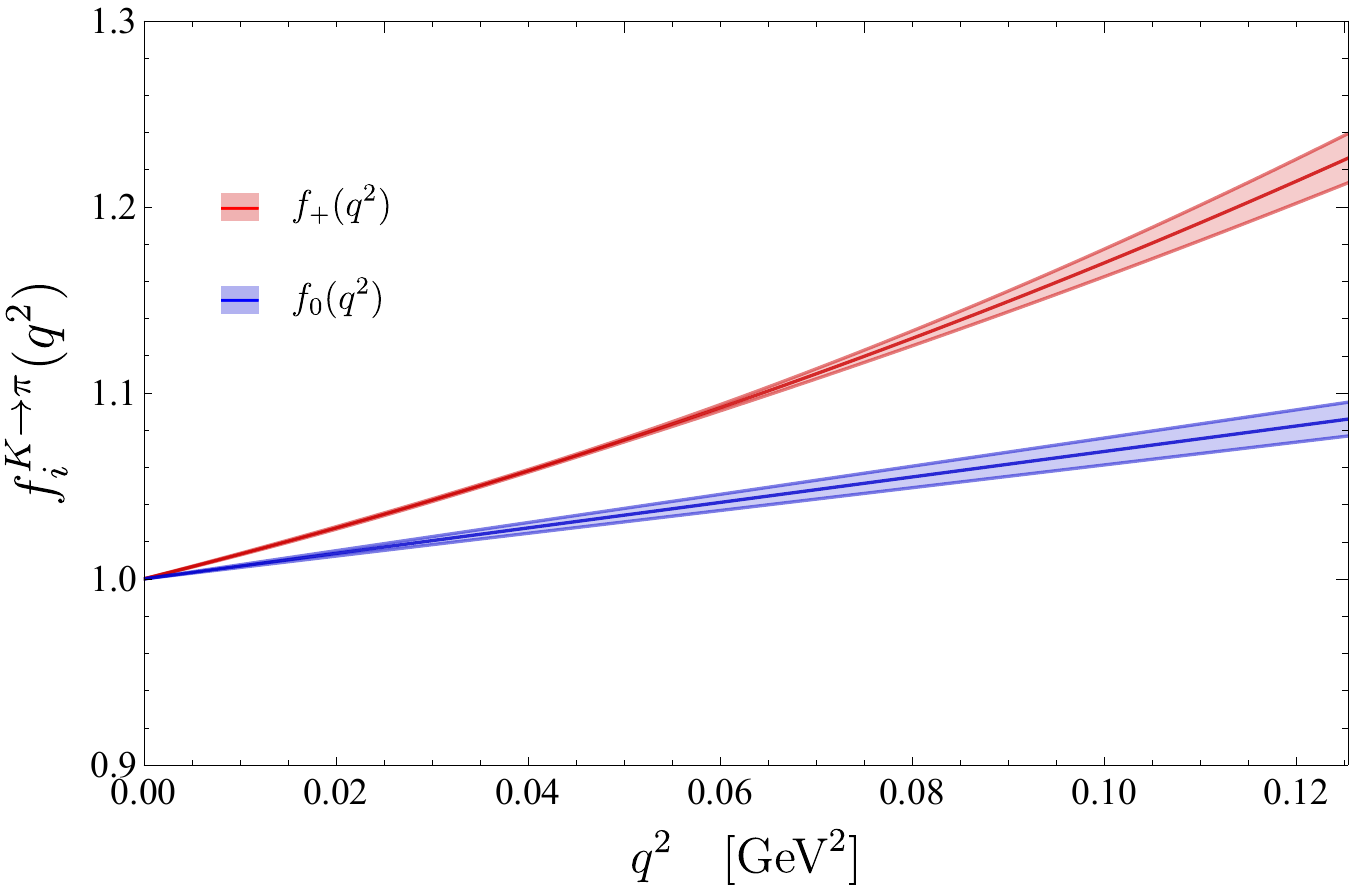}
\caption{\label{Fig:KFormFactors}$K\to\pi$ form factors from FlaviaNet~\cite{FlaviaNetWorkingGrouponKaonDecays:2010lot}.}
\end{figure}

Among the various methods proposed in the literature to describe the parameterization of the $q^{2}$ dependence of the $K\to\pi$ form factors, we rely on the conventional one given by the Taylor expansion,
\begin{equation}
\tilde{f}_{+,0}(q^{2})=1+\lambda^{\prime}_{+,0}\frac{q^{2}}{m_{\pi}^{2}}+\frac{1}{2}\lambda^{\prime\prime}_{+,0}\left(\frac{q^{2}}{m_{\pi}^{2}}\right)^{2}+\cdots\,,
\end{equation}
where the $\lambda^{\prime}_{+,0}$ and $\lambda^{\prime\prime}_{+,0}$ are, respectively, the slope and the curvature of the form factors, and the higher order terms are negligible in the kinematic range of the decay.
For the purpose of our work, it is sufficient to use the expansion parameters from the FlaviaNet working group~\cite{FlaviaNetWorkingGrouponKaonDecays:2010lot}, and the $N_{f}=2+1$ FLAG-average $f_{+,0}(0)=0.9677(27)$~\cite{FlavourLatticeAveragingGroupFLAG:2021npn}.
The form factors $\tilde{f}_{+,0}(q^{2})$ are shown in Fig.~\ref{Fig:KFormFactors}.

\begin{figure}
\includegraphics[scale=0.325]{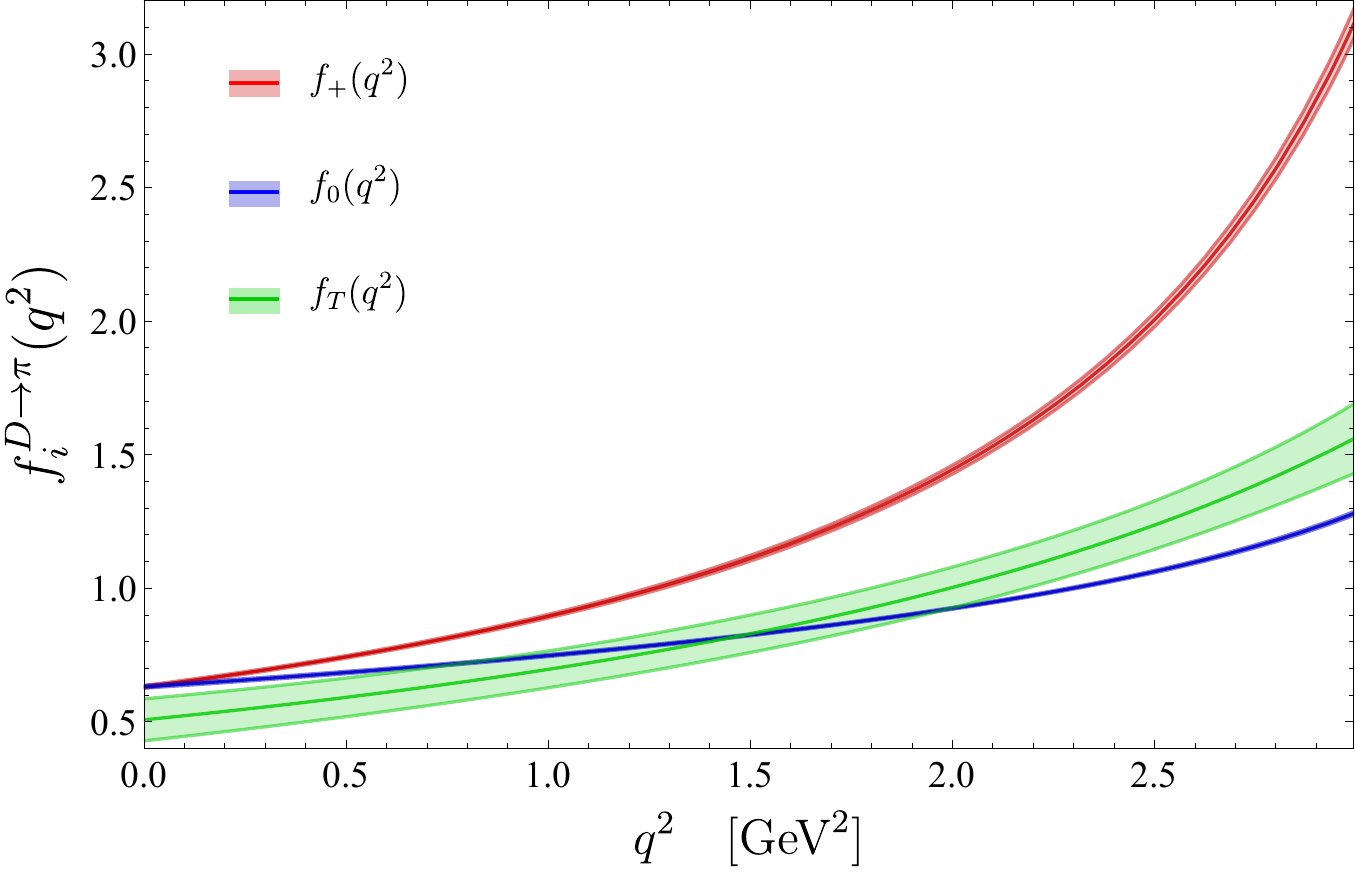}\quad\includegraphics[scale=0.325]{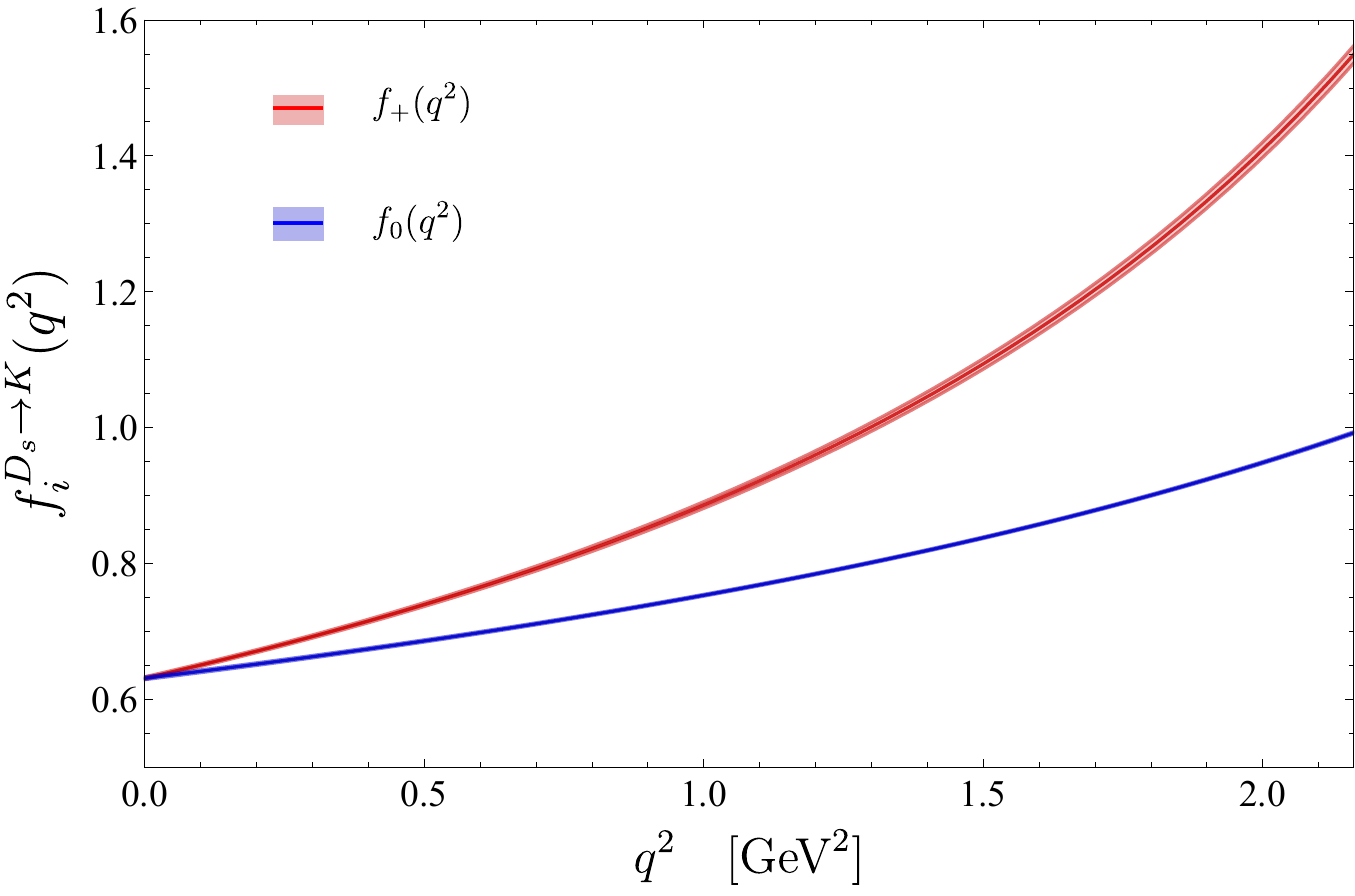}
\caption{\label{Fig:DFormFactors}$D\to\pi$ (left) and $D_{s}\to K$ (right) $f_{+,0}(q^{2})$ form factors from the FNAL/MILC collaboration~\cite{FermilabLattice:2022gku}. The tensor form factors $f_{T}(q^{2})$ are from the ETM collaboration~\cite{Lubicz:2018rfs}.}
\end{figure}

\begin{figure}
\includegraphics[scale=0.325]{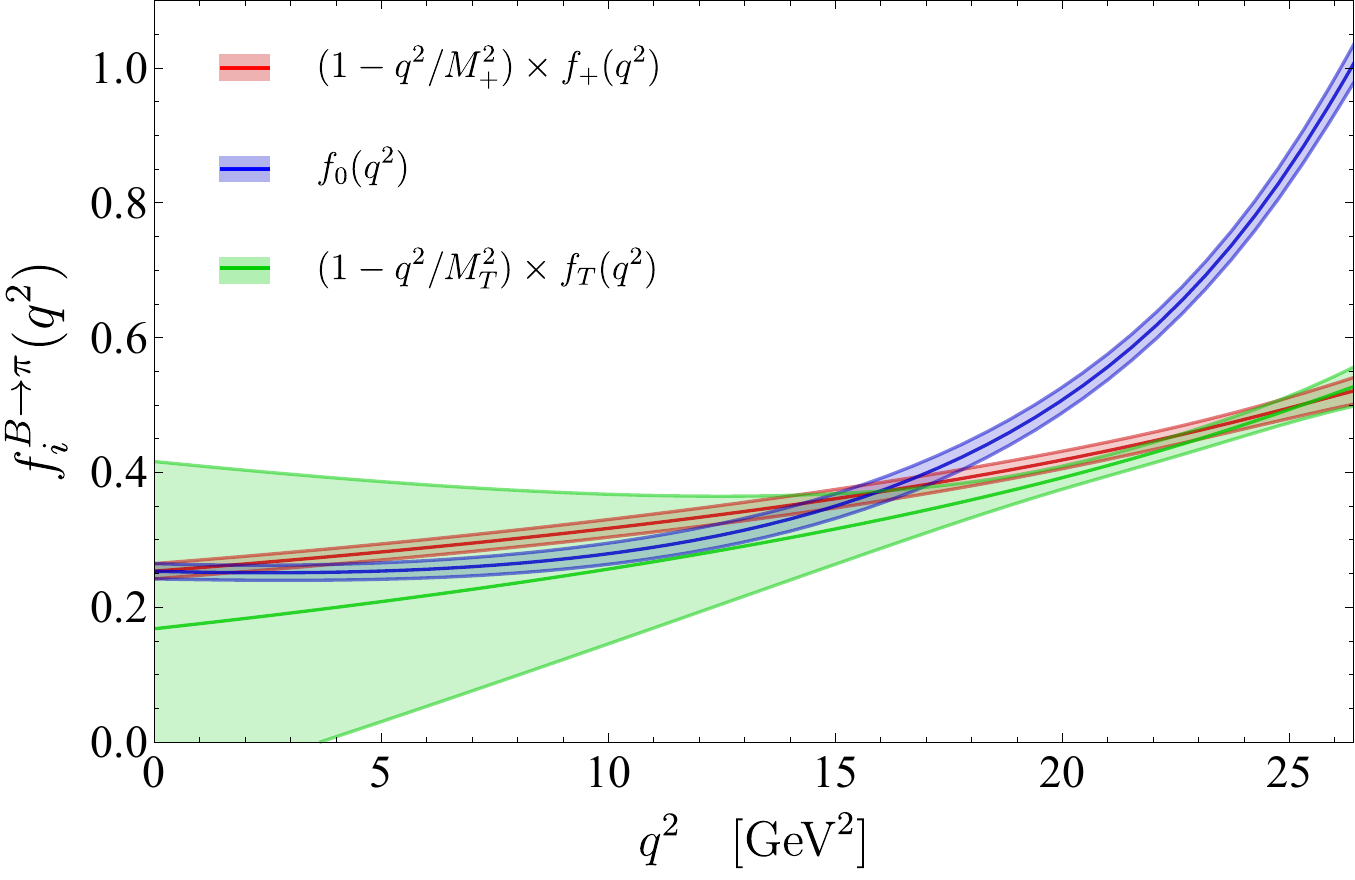}\quad\includegraphics[scale=0.325]{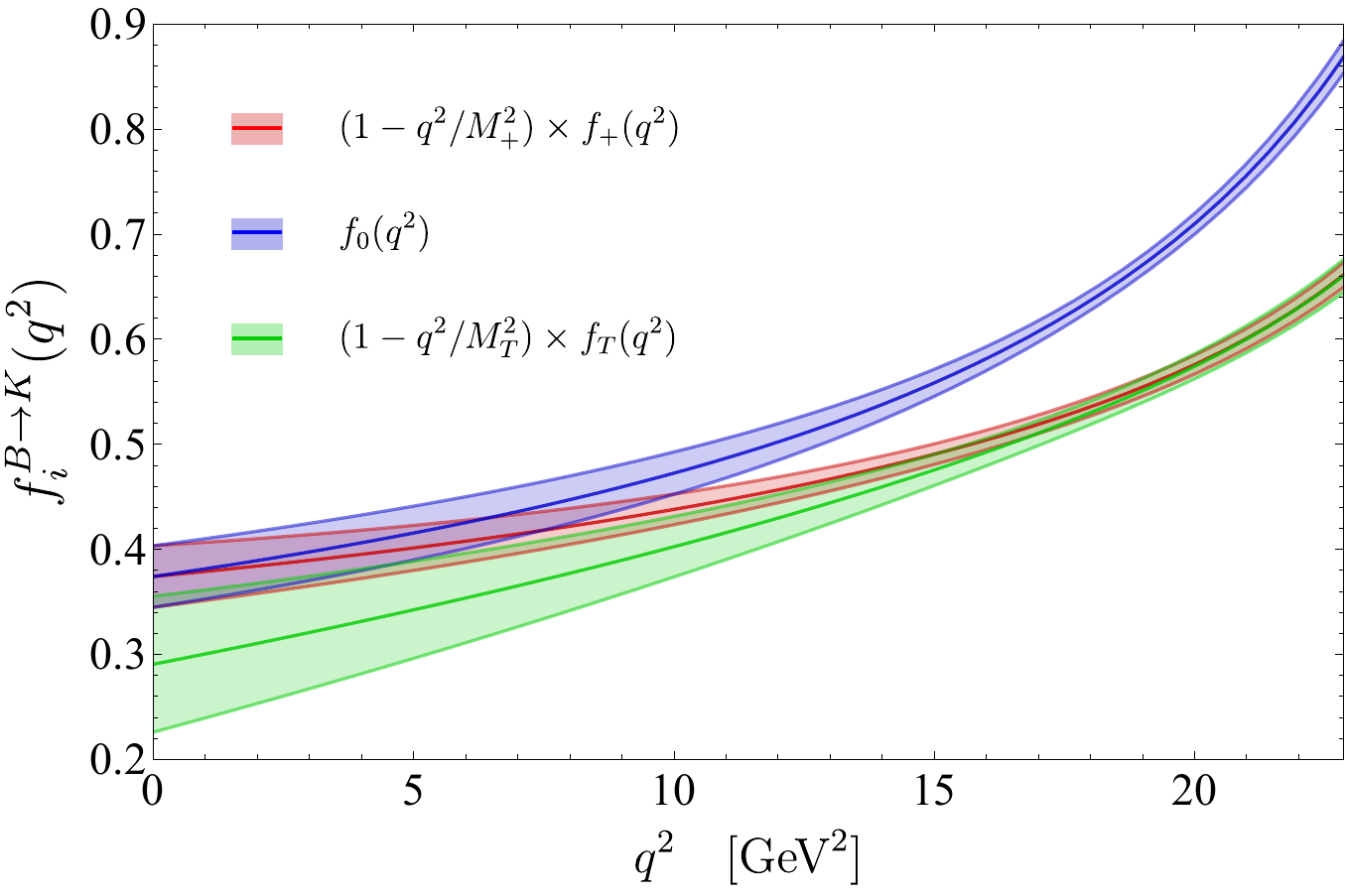}
\caption{\label{Fig:BFormFactors}$B\to\pi$ (left)~\cite{FermilabLattice:2015mwy} and $B\to K$ (right)~\cite{Bailey:2015dka} $f_{+,0,T}(q^{2})$ form factors from the FNAL/MILC collaboration~\cite{FermilabLattice:2022gku}.}
\end{figure}

For the $D_{(s)}\to\pi(K)$ and $B\to\pi,B\to K$ form factors we use Lattice-QCD results 
using the BCL representation~\cite{Bourrely:2008za}:\footnote{Correlations between form factors for different processes are neglected.
See Ref.~\cite{Gonzalez-Solis:2021pyh} 
for the first results
of the $B\to\pi$ vector and scalar form factors with a stamp of correlation with the $B_{s}\to K$ ones.}
\begin{eqnarray}
\begin{aligned}
f_{+,T}(q^2)&=\frac{1}{1-q^2/M_{+,T}^2}\sum_{n=0}^{N-1} a^{+,T}_{n}\left[z(q^2,t_{0})^{n}-(-1)^{n-N}\frac{n}{N}z(q^2,t_{0})^{N}\right]\,,\\[1ex]
f_{0}(q^2)&=\frac{1}{1-q^2/M_{0}^2}\sum_{n=0}^{N-1} a^{0}_{n}z(q^2,t_{0})^{n}\,,
\label{Eq:zexpansion}
\end{aligned}
\end{eqnarray}
where the new variable $z$ is given by
\begin{equation}
z(q^{2},t_{0})=\frac{\sqrt{t_{+}-q^{2}}-\sqrt{t_{+}-t_{0}}}{\sqrt{t_{+}-q^{2}}+\sqrt{t_{+}-t_{0}}}\,.
\end{equation}
In these expressions, $M_{+,0,T}$ refers to possible sub-threshold poles, $t_{+}$ denotes the start of the multiparticle cut, and $t_{0}\in[0,t_{+}]$ is a free parameter that can be chosen to optimize the fit.
Here we provide a graphical account of the form factors output from the FNAL/MILC collaboration that we use in our analysis (see Figs.~\ref{Fig:DFormFactors} and \ref{Fig:BFormFactors}), and refer to the Refs.~\cite{FermilabLattice:2015mwy,Bailey:2015dka,FermilabLattice:2022gku} for the numerical values of the expansion coefficients $a_{n}^{+,0,T}$, $M_{+,0,T}$ and $t_{+,0}$. 
For $f^{D\to\pi}_{T}(q^{2})$ we use the results from the ETM collaboration~\cite{Lubicz:2018rfs}.
The form factors for the $D^{0}\to\pi^{0}e^{\pm}\mu^{\mp}$ transition are scaled as $f^{D\to\pi}_{+,0,T}\to f^{D\to\pi}_{+,0,T}/\sqrt{2}$ by isospin symmetry.

\begin{figure}[h!]
\centering\includegraphics[scale=0.5]{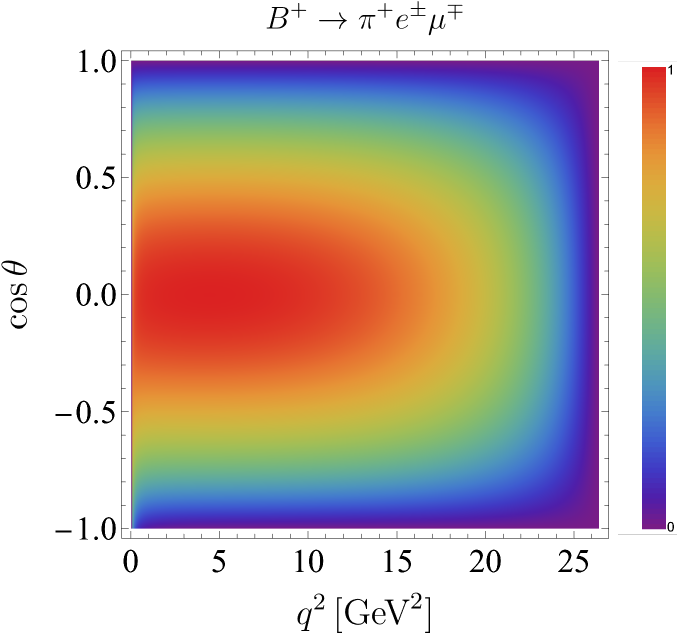}\quad\includegraphics[scale=0.5]{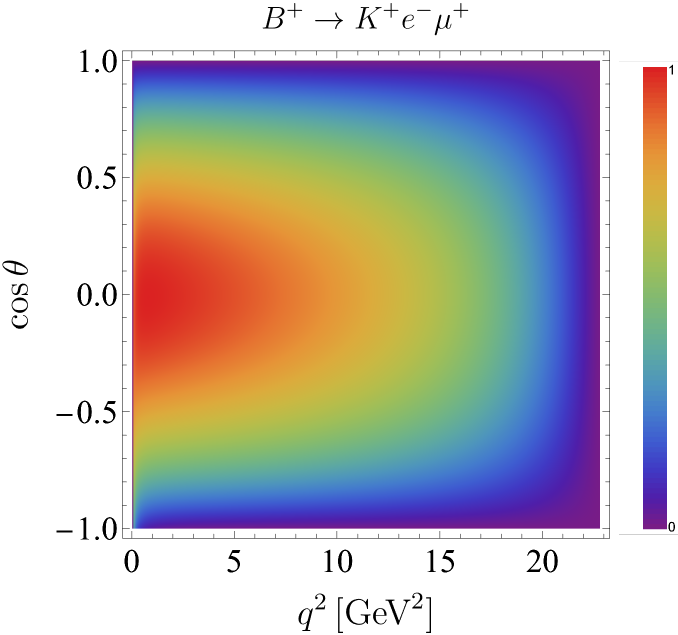}
\caption{\label{Fig:DalitzPlotBdecays}Dalitz plot distribution (in arbitrary units) for the decays $B^{+}\to\pi^{+}e^{\pm}\mu^{\mp}$ (left), $B^{+}\to K^{+}e^{-}\mu^{+}$ (right) in the $(q^{2},\cos\theta)$ variables assuming one single vector operator from Table \ref{Table:DownTypeLimits_flavor} is turned on at a time.}
\end{figure}

Fig. \ref{Fig:DalitzPlotBdecays} follows from Section \ref{sec:Flavor} showing the Dalitz plot distributions for the decays $B^{+}\to\pi^{+}e^{\pm}\mu^{\mp}$ (left) and $B^{+}\to K^{+}e^{-}\mu^{+}$ (right). 
As can be observed, the region near $\cos\theta=0$ is the most densely populated, and the distributions are found to be symmetric with respect to this variable.

\begin{figure}[h!]
\includegraphics[scale=0.45]{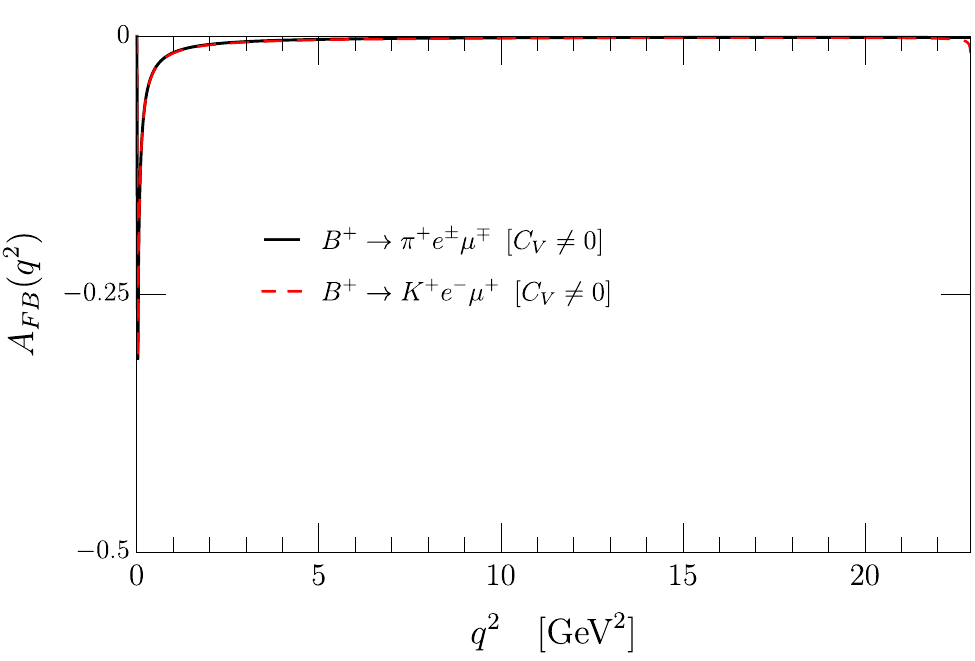}\quad\includegraphics[scale=0.45]{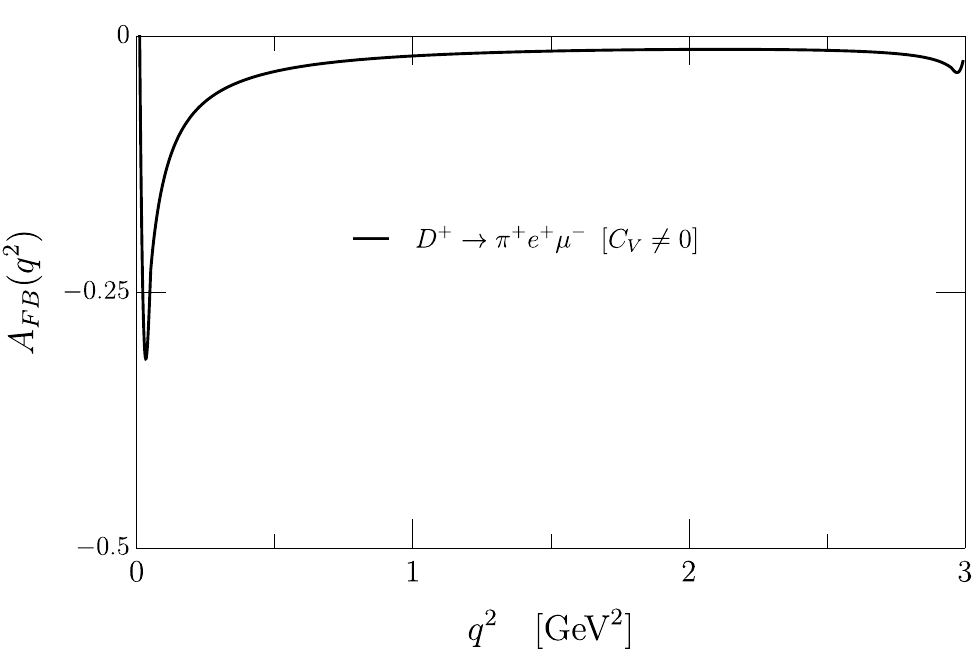}
\caption{\label{Fig:AFB}Left: Variation of the forward-backward asymmetry for the decays $B^{+}\to\pi^{+}e^{\pm}\mu^{\mp}$ (solid black) and $B^{+}\to K^{+}e^{-}\mu^{+}$ (dashed red) as a function of the dilepton invariant mass ($q^{2}$) assuming one single vector operator is turned on at a time. Right: same but for the decay $D^{+}\to\pi^{+}e^{+}\mu^{-}$.}
\end{figure}

Finally, we define the forward-backward asymmetry ($A_{FB}$) for the leptons by integrating over $\cos\theta$ as:
\begin{equation}
A_{FB}(q^{2})=\frac{\int_{0}^{1}d\cos\theta\frac{d^{2}\Gamma}{dq^{2}d\cos\theta}-\int_{-1}^{0}d\cos\theta\frac{d^{2}\Gamma}{dq^{2}d\cos\theta}}{\int_{0}^{1}d\cos\theta\frac{d^{2}\Gamma}{dq^{2}d\cos\theta}+\int_{-1}^{0}d\cos\theta\frac{d^{2}\Gamma}{dq^{2}d\cos\theta}}\,.
\end{equation}
In Fig. (\ref{Fig:AFB}) we provide a graphical account of the results of the $q^{2}$ variation of $A_{FB}$.
The possible effects of $A_{FB}$ are negative and appear to be close to the $e\mu$ threshold, then $A_{FB}$ becomes a decreasing function until the kinematical upper limit of $q^{2}$.

\section{Two-loop mixing of vector and dipole operators}\label{app:twoloop}

Refs. \cite{Crivellin:2017rmk, Ardu:2021koz} pointed out that the next-to-leading logarithmic mixing of vector into dipole operators can play an important role in constraining leptonic and semileptonic LFV four-fermion operators. In the main body of the paper we keep our analysis at the leading logarithmic level, and thus neglect this mixing. We relax here this assumption, and provide an estimate of its impact.

Neglecting the electron mass, the two-loop mixing of vector operators onto dipoles is given, in the 't Hooft-Veltman (HV) scheme, by \cite{Crivellin:2017rmk,Ciuchini:1993fk}
\begin{align}
 \frac{d}{d\log\mu}   \left[\Gamma^{e}_{\gamma}\right]_{\mu e} &= -\frac{4 m_\mu}{v} \frac{\alpha}{4\pi}\frac{1}{(4\pi)^2} N_c \sum_q \Bigg[  \left( 6 Q_q^2 Q_e + \frac{4}{9} Q_q Q_e^2\right) C^{eq}_{\rm VRR}  \nonumber \\ & +  \left( - 6 Q_q^2 Q_e + \frac{4}{9} Q_q Q_e^2\right) C^{qe}_{\rm VLR} \Bigg]   \\
 \frac{d}{d\log\mu}   \left[\Gamma^{e}_{\gamma}\right]_{e \mu} &= -\frac{4 m_\mu}{v} \frac{\alpha}{4\pi}\frac{1}{(4\pi)^2} N_c \sum_q \Bigg[  \left( 6 Q_q^2 Q_e + \frac{4}{9} Q_q Q_e^2\right) C^{eq}_{\rm VLL}  \nonumber \\ & +  \left( - 6 Q_q^2 Q_e + \frac{4}{9} Q_q Q_e^2\right) C^{eq}_{\rm VLR} \Bigg] . 
\end{align}
The two-loop anomalous dimension is scheme dependent \cite{Ciuchini:1993fk}, and the scheme dependence is absorbed by the finite part of the matching coefficient. The HV scheme is convenient since the finite part of the matching coefficient vanishes \cite{Crivellin:2017rmk}.

The solution of the RGEs in the case of operators with left-handed leptons is
\begin{align}
  & \left[ \Gamma^{e}_\gamma\right]_{e\mu}( \mu = 2\, {\rm GeV} ) = \sum_{q=u,c} \left[ -1.7 \cdot 10^{-7} C^{eq}_{\rm VLL}+ 2.1 \cdot 10^{-7} C^{eq}_{\rm VLR}\right] \nonumber \\
&   + \sum_{q=d,s} \left[ -5.9 \cdot 10^{-8} C^{eq}_{\rm VLL} + 3.8 \cdot 10^{-8} C^{eq}_{\rm VLR}\right]
+ \left[ -4.7 \cdot 10^{-8} C^{eb}_{\rm VLL} + 3.0 \cdot 10^{-8} C^{eb}_{\rm VLR}\right] \label{eq:twoloop2}
,
\end{align}
where the coefficients on the r.h.s of Eq. \eqref{eq:twoloop2} are evaluated at $\mu_0 = m_W$. Similar expressions hold for operators with right-handed leptons. 
Using the limit $|\left[\Gamma^e_\gamma\right]_{e\mu }| < 6.0 \cdot 10^{-11}$,
the limits on the individual LEFT coefficients at the electroweak scale are at the level of $10^{-3}$-$10^{-4}$, which implies 
\begin{align}
   |\left[C_{LQ, U}]_{uu,\, cc}\right| &< 3.4 \cdot 10^{-4}, &\qquad   |\left[C_{Lu}]_{uu,\, cc}\right| &< 2.8 \cdot 10^{-4}, \\
    |\left[C_{LQ, D}]_{dd,\, ss}\right| &< 1.0 \cdot 10^{-3}, &\qquad   |\left[C_{Ld}]_{dd,\, ss}\right| &< 1.6 \cdot 10^{-3}, \\
    |\left[C_{LQ, D}]_{bb}\right| &< 1.3 \cdot 10^{-3}, &\qquad   |\left[C_{Ld}]_{bb}\right| &< 2.0 \cdot 10^{-3}, 
\end{align}
where the SMEFT operators are also evaluated at the electroweak scale. 
For all quark generations, including the $s$, $c$ and $b$ quarks, these limits are much weaker than the single coupling bounds shown in Table \ref{limit_semileptonic},
which are dominated by spin-independent $\mu \rightarrow e$ conversions. In the case of $u$, $d$
and $s$ quarks, they are also weaker than the bounds from spin-dependent $\mu \rightarrow e$ conversion, shown in Eqs. \eqref{SD_AVuu}, \eqref{SD_AVdd} and \eqref{SD_AVss}. 
The bounds on $b$ quark operators are comparable to the LHC reach, and thus the two-loop vector-dipole mixing could provide additional constraints  in scenarios that contain enough couplings to cancel both spin-independent and spin-dependent contributions to $\mu \rightarrow e$ conversion, 
as the 8 coupling fit illustrated in Fig. \ref{fig:U2flavor}.

\bibliographystyle{JHEP} 
\bibliography{bibliography}

\end{document}